\documentclass[12pt]{article}
\usepackage{graphicx,a4}
\usepackage{amssymb}

\def\al{\alpha}  
\def\be{\beta} 

\def\de{\delta}

\def\th{\theta}

\def\ka{\kappa}
\def\la{\lambda}

\def\si{\sigma}

\def\up{\upsilon}

\def\om{\omega}

\def\Ga{\Gamma}
\def\Th{\Theta}

\def\K{{\mathcal{K}}}
\def\L{{\mathcal{L}}}

\newcommand{\ben}{\begin{equation}}
\newcommand{\een}{\end{equation}}
\newcommand{\bea}{\begin{eqnarray}}
\newcommand{\eea}{\end{eqnarray}}
\newcommand{\ba}{\begin{array}}
\newcommand{\ea}{\end{array}}
\newcommand{\bi}{\begin{itemize}}
\newcommand{\ei}{\end{itemize}}

\def\math{\mathsurround 0pt}
\def\oversim#1#2{\lower.5pt\vbox{\baselineskip0pt \lineskip-.5pt
        \ialign{$\math#1\hfil##\hfil$\crcr#2\crcr{\scriptstyle\sim}\crcr}}}

\def\gap{\mathrel{\mathpalette\oversim {\scriptstyle >}}}

\def\pa{\partial}

\def\Re{\mathrm{Re}}
\def\Im{\mathrm{Im}}

\begin{document}

\title{Sphalerons in Two Higgs Doublet Theories}
\author{Jackie Grant\thanks{jackieg@pact.cpes.susx.ac.uk}
and Mark Hindmarsh\thanks{m.b.hindmarsh@sussex.ac.uk}\\
Centre for Theoretical Physics\\
University of Sussex\\
Falmer\\
Brighton BN1 9QJ\\
U.K.}
\maketitle
\begin{abstract}
We undertake a comprehensive
investigation of the properties of the sphaleron
in electroweak theories with two Higgs doublets.  We do this
in as model-independent a way as possible: by
exploring the physical parameter space described by the masses and
mixing angles of the Higgs particles.  If there is a large split in the masses
of the neutral Higgs particles, there can be several sphaleron solutions, distinguished by their
properties under parity and the behaviour of the Higgs field at the origin.
In general, these solutions appear in parity conjugate
pairs and are not spherically symmetric, although the
departure from spherical symmetry is small.
Including CP violation in the Higgs potential can
change the energy of the sphaleron by up to
14 percent for a
given set of Higgs masses, with significant implications for the baryogenesis bound on the
mass of the lightest Higgs.
\end{abstract}
%\pacs{PACS numbers:
%\hfill SUSX-TH-00-nnn
%\hfill hep-ph/0101120
%}
%\vskip.5pc
%]
%%%%%%%%%%%%%%%%%%%%%%%%%%%%%%%%%%%%%%%%%%%%%%%%%%%%%%
{PACS numbers:  11.15.-q, 11.27.+d, 11.30.-j
\hfill SUSX-TH-00-003
}

\newpage

%%%%%%%%%%%%%%%%%%%%%%%%%%%%%%%%%%%%%%%%%%%%%%%%%%%%%%%%%%%%%%%%%%%%%%%%%%%%%%%%%%%%%
%%                                                                                 %%
%%                                                                                 %%
%%  			INTRODUCTION    				           %%
%%                                                                                 %%
%%%%%%%%%%%%%%%%%%%%%%%%%%%%%%%%%%%%%%%%%%%%%%%%%%%%%%%%%%%%%%%%%%%%%%%%%%%%%%%%%%%%%

\section{Introduction}
\label{s:introduction}

One of the major unsolved problems in particle cosmology is to account for the
baryon asymmetry of the Universe. This asymmetry is 
usually expressed in terms of the parameter
$\eta$, defined as the ratio between the baryon number density $n_B$ and the
entropy density $s$: $\eta = n_B/s \sim 10^{-10}$.  Sakharov 
\cite{Sakharov:1967dj} laid
down the framework for any explanation: the theory of baryogenesis 
must contain baryon number ($B$) violation;
charge conjugation ($C$)
violation; combined charge conjugation and
parity ($CP$) violation; and a departure from thermal equilibrium. 
The Standard Model is naturally $C$ and $P$
violating, and violates $CP$ through 
the couplings of fermionic charged currents to the $W^{\pm}$ (the CKM matrix).
It was also known to violate the combination $B+L$ 
(where $L$ is lepton number) non-perturbatively \cite{'tHooft:1976up}, 
and the realisation that this rate is large at high temperature, and 
that the Standard Model could depart from equilibrium at 
a first order phase transition \cite{Kuzmin:1985mm} led to
considerable optimism that the origin of the baryon asymmetry could be found in
known physics. 

However, the
Standard Model does not have a first order phase transition for Higgs
masses above about 75 GeV \cite{Kajantie:1996kf,Kajantie:1997qd}, and in any
case is not thought to have enough $CP$ violation.
Current attention is focused on the Minimal Supersymmetric Standard Model
(MSSM), where there are many sources of $CP$ violation over and above the CKM
matrix \cite{Huet:1996sh,Carena:1997gx,Cline:2000nw},
and the phase transition can be first order for Higgs masses up to 120
GeV, provided the right-handed stop is very light and the
left-handed stop very massive \cite{deCarlos:1997ru,Laine:1998qk,Cline:1998hy}.

The currently accepted picture for the way these elements fit together was
developed by Cohen, Kaplan, and Nelson \cite{Cohen:1991iu} (see also
\cite{Rubakov:1997sr,Riotto:1999yt,Trodden:1999ym} for reviews). A first order 
transition proceeds by nucleation of bubbles of the new, stable, phase. The
bubbles grow and merge until the new phase has taken over. The effect of 
$CP$ violation in the theory is to make the fermion reflection
coefficients off the wall chirally asymmetric, which results in a chiral
asymmetry building up in front of the advancing wall in the fermion species
which couple most strongly to the wall and have the largest $CP$ violating
couplings. This chiral asymmetry is turned into a baryon asymmetry by the
action of symmetric-phase sphalerons.

As the wall sweeps by, the rate of baryon number violation by sphalerons
drops as the sphaleron mass increases sharply. The formation of a
sphaleron is a thermal activation process and the rate can be estimated
to go as $\Ga_s \simeq \exp(-E_s(T)/T)$, where $E_s(T)$ is the energy of the
sphaleron at temperature $T$.  This rate must not be so large that the baryon
asymmetry is removed behind the bubble wall by sphaleron processes in thermal
equilibrium, and this condition can be
translated into a lower bound on the sphaleron mass 
\cite{Shaposhnikov:1986jp,Shaposhnikov:1987tw,Bochkarev:1990fx}
\ben
\label{e:SphEneBou}
E_s(T_c)/T_c \gap 45.
\een
Thus it is clear that any theory of baryogenesis requires a careful
calculation of the sphaleron mass. For example, it turns out that  
condition (\ref{e:SphEneBou}) is not satisfied for any value of Higgs mass
in the Standard Model \cite{Kajantie:1996kf}.

It has been known for a long time that spherically symmetric solutions exist in
SU(2) gauge theory with a single fundamental Higgs 
\cite{Dashen:1975xh,Soni:1980ps,Burzlaff:1984jb},
which is the bosonic sector of the Standard Model at zero Weinberg angle. 
However, it was Klinkhamer and Manton \cite{Klinkhamer:1984di} 
who realised that they were
unstable, with a single unstable mode, and that the formation and decay of
a sphaleron results in a simultaneous change of both $B$ and $L$ 
number by $N_f$ (the number
of fermion families).  They calculated numerically both the mass and the Chern-Simons number,
finding the mass to be 3.7 (4.2) $M_W/\al_W$ at a Higgs mass of 72 (227) GeV,
where $\al_W = g_W^2/4\pi$ and $M_W$ is the mass of the $W^{\pm}$ particle;
and the Chern-Simons number to be exactly 1/2.

At $M_h \gap 12M_W$ new solutions appear \cite{Kunz:1989sx,Yaffe:1989ms},
which have different
boundary conditions at the origin: the Higgs field does not vanish.  These
spontaneously violate parity and
occur in $P$ conjugate pairs with slightly lower energy than the original
sphaleron, which correspondingly develops a second negative eigenvalue. These
are termed {\em deformed sphalerons} or {\em bisphalerons}.

Several authors have considered models with two Higgs doublets.  Kastening,
Peccei, and Zhang (KPZ) \cite{Kastening:1991nw} studied
models with $CP$ violation, but did not use the most general spherically symmetric
ansatz, limiting themselves to a parity conserving form.
Bachas, Tinyakov, and Tomaras (BTT) \cite{Bachas:1996ap}
on the other hand, considered a two-doublet theory with no explicit $CP$
violation, used a $C$ conserving ansatz, chose the masses
of the pseudoscalar ($M_A$) and the charged Higgs ($M_{H^\pm}$) to be zero, and
chose the mixing between the two scalar Higgses to be zero. They found new
$P$ violating solutions, specific to multi-doublet models, 
at $M_H \gap 5M_W$, where $M_H$ is the mass of the
second $CP$ even Higgs.  They did not calculate the
Chern-Simons number, but we show that these solutions appear in $P$ conjugate pairs and are in fact sphalerons, 
in that they have Chern-Simons number near 1/2, and one unstable mode. In view
of the difference in behaviour of the two Higgs
fields as the origin is approached, we call them
{\em relative winding (RW) sphalerons}.
More recently, Kleihaus \cite{Kleihaus:1999bh} looked at the bisphalerons in 
a restricted two-doublet Higgs model.

Sphalerons in the MSSM have were studied by Moreno, Oaknin, and Quiros (MOQ)
\cite{Moreno:1997zm}, who included one-loop corrections, both quantum and
thermal.  However, they again did not allow for $P$ violating
bisphalerons or RW sphalerons, and did not consider
the effect of $CP$ violation either,
which can appear in the guise of complex values of the soft SUSY breaking terms
in the potential.

All of the above work was carried out at zero Weinberg angle with a
spherically symmetric ansatz: there have been 
several studies of sphalerons in the Standard Model in the full 
SU(2)$\times$U(1) theory \cite{Kleihaus:1991ks,James:1992re,Klinkhamer:1992fi},
where one is forced to adopt the more complicated axially 
symmetric ansatz:  Ref.\ \cite{Kleihaus:1991ks} used the axially
symmetric ansatz in a numerical computaton, 
\cite{James:1992re} expanded in powers of $g'/g$ using a 
partial wave decomposition,
and \cite{Klinkhamer:1992fi} estimated the energy by constructing a 
non-contractible loop in field configuration space which was sensitive to 
$\theta_W$.
The upshot of this work is that working at the physical 
value of the Weinberg angle changes the energy of the sphaleron by about 
10\%. It is interesting to note that the SU(2)$\times$U(1) theory also
contains charged sphaleron solutions \cite{Saffin:1998ae}. 

Here we report on work on sphalerons in the two-doublet Higgs model (2DHM) in
which we study the properties of sphalerons in as general a set of
realistic models as possible, although we do use the zero Weinberg angle 
approximation and a spherically symmetric ansatz.
We try to express parameter space in
terms of physical quantities: Higgs masses and mixing angles, which helps us
avoid regions of parameter space which have already been ruled out by LEP, or
where the vacuum is unstable. It also means one can take into account
ultraviolet radiative corrections by using the 
1-loop corrected values for the masses and mixing angles. 

We are interested in the energy, the Chern-Simons number, the symmetry
properties, and the eigenvalues of the normal modes of the various sphaleron
solutions in the theory, as functions of the physical parameters. 
From the point of view of the computation of the rate of baryon number 
violation,
the mass is certainly the most important quantity, followed by the number and
magnitude of negative eigenvalues of the fluctuation operator in the sphaleron
background: the largest contribution to the baryon number violation rate comes
from the sphaleron with lowest energy and hence only one negative eigenvalue.
The Chern-Simons number and the
symmetry properties under $C$, $P$, and spatial rotations, 
are also interesting as they help classify the
solutions.

We firstly check our results against the existing literature, principally 
Yaffe \cite{Yaffe:1989ms} and BTT
\cite{Bachas:1996ap}, and then reexamine the sphaleron in a more realistic part
of parameter space, where $M_A$ and $M_{H^\pm}$ are above their experimental
bounds.  We find that in large regions of parameter space, particularly when 
one of the neutral Higgses is heavy (above about 6 $M_W$), the 
RW sphaleron is the lowest energy sphaleron.  When there is $CP$ violation in the
Higgs sector, the would-be pseudoscalar Higgs can play the role of the heavy
Higgs, and the other two Higgses can remain relatively light.  The
fractional energy difference 
between the RW and the ordinary (Klinkhamer-Manton) sphaleron is small, about 
1\% in the parameter ranges we explored. 

We encounter a problem with $P$ violating sphalerons when either 
$M_A-M_{H^\pm}$, or the amount of $CP$ violation
is non-zero: there is a departure from
spherical symmetry in the energy density, signalling an inconsistency in the
ansatz for the field profiles.  However, the energy density in the
non-spherically symmetric terms is small, at most about 0.2\% of the dominant
spherically symmetric terms, so it is a good approximation to ignore them.

We also looked at the sphaleron in the restricted parameter space afforded by
the (tree level) MSSM, confirming the results of \cite{Moreno:1997zm} that the
sphaleron energy depends mainly on the mass of the lightest Higgs and on $\tan
\beta$, and finding no RW or bisphaleron solutions.

Finally, we amplify the
point made in \cite{Grant:1999ci} 
that introducing $CP$ violation makes a significant
difference to the sphaleron mass, and may significantly change bounds on the
Higgs mass from electroweak baryogenesis.

We do not explicitly compute quantum or thermal corrections
\cite{Bochkarev:1990fx,Arnold:1987mh,Carson:1990rf,Carson:1990jm,Baacke:1994aj,Diakonov:1996xz,Moore:1996jv}
as they are model-dependent. However, if particle masses
are expressed in units of $M_W$, a reasonable
approximation to the 1-loop sphaleron mass (in units of $M_W/\alpha_W$) 
can be obtained by interpreting the
masses and mixing angles as loop-corrected quantities evaluated at an
energy scale $M_W$ \cite{Moore:1996jv}.  This approximation justifiably ignores
small corrections 
due to radiatively induced operators of dimension higher than 4, but 
does not take into account the cubic term in the effective potential. 
This means our calculations are less accurate  
near the phase transition. However, as the error is in the Higgs potential,
which generally contributes
less than 10\% to the energy, the resulting uncertainty is
not large.

The plan
of the paper is as follows.  In Section \ref{s:2Higgs_ew} we describe the
bosonic sector of the two Higgs doublet SU(2) electroweak theory.
We discuss the various parametrizations of the scalar potential, and provide
translation tables in Appendix \ref{a:param_pot}. We show how we use physical
masses and mixing angles as independent parameters of the theory.  Although in
this approach the stability of the vacuum is automatic, as one chooses the
masses of the physical particles to be real, there are still the problems of
boundedness and global minimisation to be overcome.  We solve the boundedness
problem straight forwardly, but with two Higgs doublets,
finding the global minimum of the potential is non-trivial, and we are forced
to use numerical methods.

In Section \ref{s:ansatz} we discuss the sphaleron solutions and their symmetry
properties.  In Section \ref{s:finding_solns} we descibe 
the numerical method we use to find
the solutions: although the Newton method 
has been used before \cite{Yaffe:1989ms,Bachas:1996ap} there
are some difficulties associated with the boundary
conditions that were not highlighted by previous authors.
In Section \ref{s:results} we present our results.  Section 
\ref{s:conclusions} contains discussions and conclusions.

Throughout this paper we use $\hbar = c = k_B = 1$, a metric with signature
$(+,-,-,-)$, and $M_W= 80.4$ GeV.

%%%%%%%%%%%%%%%%%%%%%%%%%%%%%%%%%%%%%%%%%%%%%%%%%%%%%%%%%%%%%%%%%%%%%%%%%%%%%%%%%%%%%
%%                                                                                 %%
%%                                                                                 %%
%%  			TWO DOUBLET ELECTROWEAK THEORY			           %%
%%                                                                                 %%
%%%%%%%%%%%%%%%%%%%%%%%%%%%%%%%%%%%%%%%%%%%%%%%%%%%%%%%%%%%%%%%%%%%%%%%%%%%%%%%%%%%%%

\section{Two Higgs doublet electroweak theory}
\label{s:2Higgs_ew}

We shall be working with an SU(2) theory with two
Higgs doublets $\phi_\alpha$, with subscript $\al = 1,2$. Although we should strictly
work with the full SU(2)$\times$U(1) theory, neglecting the 
 U(1) coupling is
a reasonable approximation to make when studying the sphaleron.

The relevant Lagrangian is
\ben
\label{e:lag}
{\cal L}=-\frac{1}{4}F^{a}_{\mu\nu}F^{a\mu\nu}
           +(D_\mu\phi_{\alpha})^\dagger (D^\mu\phi_{\alpha})
         -V(\phi_{1},\phi_{2}).
\een
Here, the covariant derivative $D_\mu\phi_{\alpha}
= \pa_\mu\phi_\al + gW^a_\mu t^a\phi_\al$ with
antihermitian generators $t^a = \si^a/2i$.

This Lagrangian may have discrete symmetries, including parity, charge
conjugation invariance, and $CP$ \cite{Jarlskog:1989bm}.  These transformations are realised
on the Higgs fields by
\bea
\label{e:Pxm}
P : & & \phi_{\alpha}(t,x^j) \to \phi_{\alpha}(t,-x^j), \\
\label{e:Cxm}
C: & & \phi_{\alpha}(t,x^j) \to
{-i}\sigma_{2}e^{-i2\theta_{\alpha}}\phi_{\alpha}^{\ast}(t,x^j), \\
\label{e:CPxm}
CP: & & 
\phi_{\alpha}(t,x^j) \to
{-i}\sigma_{2}e^{-i2\theta_{\alpha}}\phi_{\alpha}^{\ast}(t,-x^j),
\eea
where $\theta_\al$ are phase factors that can only be determined by reference
to the complete theory.  The transformations
on the gauge fields are
\bea
\label{e:Pxmg}
P : & & W_{\mu}(t,x^j) \to W^{\mu}(t,-x^j), \\
\label{e:Cxmg}
C: & & W_{\mu}(t,x^j) \to
( {-i}\sigma_{2})W_{\mu}^{\ast}(t,x^j)({-i}\sigma_{2})^{\dagger},
\\
\label{e:CPxmg}
CP: & & 
W_{\mu}(t,x^j) \to
({-i}\sigma_{2}) W^{\mu\ast}(t,-x^j)({-i}\sigma_{2})^{\dagger}.
\eea
With these transformations the only place a departure from $C$, $P$, 
or $CP$ invariance can occur in Lagrangian (\ref{e:lag}) is in the 
Higgs potential 
term $V(\phi_{1},\phi_{2})$.

\subsection{The Higgs potential}
\label{s:higgs_pot}
The most general two Higgs doublets potential has 14 real parameters,
assuming that the energy density at the minimum is zero.
We shall consider one with a
discreet symmetry imposed on dimension four terms, 
$\phi_{1} \rightarrow \phi_{1}$, $\phi_{2} \rightarrow -\phi_{2}$, which 
suppresses flavour changing neutral currents \cite{Gunion:1989we}, and results
in a potential with 10 real parameters.  One of these parameters may be removed
by a phase redefinition of the fields we detail in Appendix
\ref{a:param_pot}, and the potential may be written 
\bea
\label{e:pot}
V(\phi_{1},\phi_{2})& = &(\lambda_{1}+\lambda_{3})\left(\phi_{1}^{\dagger}\phi_{1}-\frac{\upsilon_{1}^2}{2}\right)^{2}
                        +(\lambda_{2}+\lambda_{3})\left(\phi_{2}^{\dagger}\phi_{2}-\frac{\upsilon_{2}^2}{2}\right)^{2} \nonumber\\
                        & &  +2\lambda_{3}\left(\phi_{1}^{\dagger}\phi_{1}-\frac{\upsilon_{1}^2}{2}\right)
                                   \left(\phi_{2}^{\dagger}\phi_{2}-\frac{\upsilon_{2}^2}{2}\right) \nonumber\\
                       & &  +\lambda_{4}\left[\phi_{1}^{\dagger}\phi_{1}\phi_{2}^{\dagger}\phi_{2} 
                                     -\Re^2( \phi_{1}^{\dagger}\phi_{2})      
                                     -\Im^2( \phi_{1}^{\dagger}\phi_{2})\right] \nonumber\\
   & &  +(\lambda_{+}+\chi_{1})\left(\Re( \phi_{1}^{\dagger}\phi_{2})-\frac{\upsilon_{1}\upsilon_{2}}{2}\right)^{2}
+(\lambda_{+}-\chi_{1})\Im^2( \phi_{1}^{\dagger}\phi_{2})\nonumber\\
  & &  +2\chi_{2}\left(\Re( \phi_{1}^{\dagger}\phi_{2})-\frac{\upsilon_{1}\upsilon_{2}}{2}\right)
\Im(\phi_{1}^{\dagger}\phi_{2}).
\eea 
This form of the potential is convenient as the vacuum configuration, 
which we take as the zero of the potential is
entirely real: 
\ben
\label{e:minimum}
\phi^{vac}_{\alpha}=\frac{\upsilon_{\alpha}}{\sqrt{2}}\left[\ba{c}0
\\ 1 \ea\right].
\een
This form also makes clear what are the sources of $CP$ violation in
the theory. Ignoring couplings to other fields, it
can be seen that when $\chi_2 = 0$ there is a discrete symmetry
\ben
\label{e:Cxm_phi}
\phi_{\alpha}\rightarrow{-i}\sigma_{2}\phi_{\alpha}^{\ast},
\een
which sends $\Im(\phi_{1}^{\dagger}\phi_{2})\to
-\Im(\phi_{1}^{\dagger}\phi_{2})$.  This can be identified as charge
conjugation invariance.  Thus $\chi_2$ is a $C$ breaking
parameter. 
%and we shall see that its physical effect is to mix the $CP$ odd and
%$CP$ even neutral Higgses. 
%Although
%$\chi_2$
%is a $C$ breaking parameter and when considering only the theory described by 
%Lagrangian \ref{e:lag} (with $\chi_2 = 0$) 
%
%Note the Higgs fields are either $C$ odd or $C$ even, (while being $P$ even), 
In the presence of fermions, $C$ and $P$ are not separately conserved, and 
%they behave as $P$ odd
%or $P$ even, (while being $C$ even) \cite{Gunion:1989we}. It is for this reason that 
we generally refer to the field properties according to their
behavior under $CP$, and to $\chi_2$ as a $CP$ violating parameter, giving 
rise to a mixing between the $CP$ odd and
$CP$ even neutral Higgses. When one includes the other fields of the full
theory one can find further sources of $CP$ violation, such as the phases in the
CKM matrices of the quarks and, if neutrinos are massive, 
leptons. 

In Appendix\ \ref{a:param_pot} we write down how the 
nine parameters of Eq.\ \ref{e:pot} relate to the parameters of the two more
usual forms of this potential.  
 
It is useful to determine as many as possible of the nine parameters in the
potential from physical ones. The physical parameters at hand are the four
masses of the Higgs particles, the three mixing angles of the neutral Higgses,
and the vacuum expectation value ($\upsilon$) 
of the Higgs (which is determined from $M_W$,
and the SU(2) gauge coupling $g$). This
leaves one undetermined parameter which may be chosen in various ways.

In the absence of $CP$ violation, we automatically have $\chi_2=0$, and 
our input parameters are; $\upsilon$, 
$M_h$ and $M_H$ (the masses of the $CP$ even scalars),
$M_A$ (the mass of the $CP$ odd scalar), $M_{H^\pm}$ (the mass of the charged scalar), 
$\phi$ (the mixing angle between the $CP$ even scalars), 
$\tan\beta$ and $\lambda_3$, (the only parameter we choose by hand). This gives
non-zero values for the other eight of our nine parameters.

In the presence of $CP$ violation our input parameters again include 
$\upsilon$, $M_h$, $M_H$, $M_A$, $M_{H^\pm}$, $\phi$, and $\lambda_3$. 
However, now we also have $\th_{CP}$ (the mixing angle 
between the $CP$ even and the $CP$ odd neutral Higgs sector which is entirely responsible
for the $\chi_{2}$ term), 
and the third mixing angle 
$\psi$. For a non-zero $\th_{CP}$, $\tan\beta$ is determined by 
the masses and mixings, and although we still denote the three 
neutral Higgs masses as $M_h$, $M_H$, and $M_A$ 
we stress that they are not respectively $CP$ even, $CP$ even, and $CP$ odd, 
but have some combination of these properties depending on the values
of $\th_{CP}$ and $\phi$.

The conversion between the parameters of Eq.\  \ref{e:pot} and 
these masses and mixings is carried out in the charged sector 
through\ben
\label{e:charged}
\la_4=\frac{2 M_{H^{\pm}}^2}{\upsilon^2},
\een
\noindent and in the neutral sector by writing
\ben
\label{e:X_pot}
\upsilon^2 X \equiv D^{-1}(\psi,\th_{CP},\phi)
 ~ M_P (M_h, M_H, M_A) ~ D(\psi,\th_{CP},\phi),
\een
where $M_P$ is a diagonal mass matrix given by 
\ben
\label{e:diag}
M_P  \equiv  {\rm Diag}\left[M_H^2,M_h^2,M_A^2\right], 
\een
and $D$ is the orthogonal matrix which diagonalises $X$.  Defining rotation
matrices in the usual way,
\ben
\label{e:mixing}
R_z(\alpha) = \left[\ba{ccc}\cos\alpha & \sin\alpha & 0 \\
                           -\sin\alpha & \cos\alpha & 0 \\
                            0 & 0 & 1\ea \right],
~~~~~~R_y(\alpha)=\left[\ba{ccc}\cos\alpha &0 & -\sin\alpha \\
                           0 &1 & 0 \\
                            \sin\alpha & 0 & \cos\alpha\ea \right],
\een
we can arrange for the mixing angles $\psi,\th_{CP},\phi$ to be the usual Euler
angles, through 
\ben
\label{e:tot_mixing}
D(\psi,\th_{CP},\phi) \equiv R_z(\psi)R_y(\theta)R_z(\phi).
\een

The $X (\psi,\th_{CP},\phi, M_h, M_H, M_A)$ of Eq.\ \ref{e:X_pot} can be 
obtained as a function of the parameters of Eq.\  \ref{e:pot}, by expanding
about the vacuum state Eq.\  \ref{e:minimum}, to give 
\bea
X(1,1)&=&\frac{1}{2}\left[4(\lambda_1+\lambda_3)\cos^2\beta\label{e:m11}
	+(\lambda_++\chi_1)\sin^2\beta\right],\\
X(1,2)&=&X(2,1)=\frac{1}{2}(4\lambda_3+\lambda_++\chi_1)\cos\beta\sin\beta,\label{e:m12}\\
X(1,3)&=&X(3,1)=\frac{1}{2}\chi_2\sin\beta,\label{e:m13}\\
X(2,2)&=&\frac{1}{2}\left[4(\lambda_2+\lambda_3)\sin^2\beta
	+(\lambda_++\chi_1)\cos^2\beta\right],\label{e:m22}\\
X(2,3)&=&X(3,2)=\frac{1}{2}\chi_2\cos\beta,\label{e:m23}\\
X(3,3)&=&\frac{1}{2}(\lambda_+-\chi_1).\label{e:m33}
\eea
Inverting Eqs.\  \ref{e:m11}-\ref{e:m33}
gives\footnote{ We have corrected two typographical errors 
from \cite{Grant:1999ci}: a swapped cos and sin in Eq.\  \ref{e:lambda1} and Eq.\  \ref{e:lambda2}, 
and a sign error in Eq.\  \ref{e:chi1}.}
\bea
\chi_2 &=&2\sqrt{X(1,3)^2+X(2,3)^2},\label{e:chi2}\\
\beta &=&\arctan\left[{X(1,3)}/{X(2,3)}\right],\\
\lambda_1&=&\left[X(1,1)\cos\beta-X(1,2)\sin\beta
              -2\lambda_3\cos2\beta\cos\beta\right]\frac{1}{2\cos^3\beta},\label{e:lambda1}\\
\lambda_2&=&\left[X(2,2)\sin\beta-X(1,2)\cos\beta
              +2\lambda_3\cos2\beta\sin\beta\right]\frac{1}{2\sin^3\beta},\label{e:lambda2}\\
\lambda_+&=&-2\lambda_3+X(1,2)\frac{1}{\sin\beta\cos\beta}+X(3,3),\label{e:lambda+}\\
\chi_1&=&-2\lambda_3+X(1,2)\frac{1}{\sin\beta\cos\beta}-X(3,3),\label{e:chi1}
\eea
\noindent where the $X$ above are the $X (\psi,\th_{CP},\phi, M_h, M_H, M_A)$ as given by Eq.\  \ref{e:X_pot}. And 
we have chosen $-\pi < 2\beta < \pi$ from which, depending on the sign of 
$X(1,2)$ and $X(1,3)$, we can set the sign of $\chi_2$.
Although it is unconventional to allow $\beta$ to take negative values, 
it is a natural consequence of allowing the mixing angles to vary over their
full range.
 
%%%%%%%%%%%%%%%%%%%%%%%%%%%%%%%%%%%%%%%%%%%%%%%%%%%%%%%%%%%%%%%%%%%%%%%%%%%%%%%%%%%%%
%%                                                                                 %%
%%                                                                                 %%
%%  	BOUNDEDNESS AND STABILITY OF HIGGS POTENTIAL			           %%
%%                                                                                 %%
%%%%%%%%%%%%%%%%%%%%%%%%%%%%%%%%%%%%%%%%%%%%%%%%%%%%%%%%%%%%%%%%%%%%%%%%%%%%%%%%%%%%%

\subsection{Boundedness and stability of the Higgs potential}
\label{s:boundedness_stab}
Before proceeding, we re-examine the conditions on our potential which derive  
from its boundedness and the stability of the vacuum state. For boundedness we
need 
consider only the quartic terms of Eq.\  \ref{e:pot} to find the large field 
behaviour of the potential.  
We write our doublets as 
\ben
\label{e:rewrite_doublets}
\phi_{\alpha}=\left|\sqrt{\rm{Q}_{\al}}\right|\left[\ba{c}\cos\rho_{\al}e^{i\ka_{\al}}
\\\sin\rho_{\al}e^{i\om\al}\ea\right],
\een
this will allow us to express the potential in 
terms of independent quantities. The quartic terms of 
Eq.\ \ref{e:pot} can then be written as
\ben
\label{e:pot_quartic}
V = a Q_1^2 + b Q_2^2 + c(\eta_1,\eta_2) Q_1 Q_2,
\een
where
\bea
\label{e:eta_1}
\eta_1 &=& \cos\rho_{1}\cos\rho_{2}\cos(\kappa_2-\kappa_1)
+ \sin\rho_{1}\sin\rho_{2}\cos(\om_2-\om_1),\\
\label{e:eta_2} 
\eta_2 &=& \cos\rho_{1}\cos\rho_{2}\sin(\kappa_2-\kappa_1)
+ \sin\rho_{1}\sin\rho_{2}\sin(\om_2-\om_1),
\eea
and
\ben
\label{e:a} 
a = \la_1 + \la_3,
\een
\ben
\label{e:b} 
b = \la_2 + \la_3,
\een 
\ben
\label{e:c} 
c(\eta_1,\eta_2) = 2\la_3 + \la_4 +( \la_+ - \la_4 + \chi_1) \eta_1^2 
+ ( \la_+ - \la_4 - \chi_1) \eta_2^2 +2 \chi_2\eta_1 \eta_2.
\een
The variables 
$Q_1$, $Q_2$, $\eta_1$, and $\eta_2$ are then independent. 
Furthermore, $Q_1$ and $Q_2$ are by definition non-negative, and 
$\eta_1$ and $\eta_2$ are constrained to lie in the unit disc
\ben
\label{e:eta_constriant} 
0\leq\eta_1^2+\eta_2^2\leq1.
\een

The potential can now be viewed
as a quadratic form in $Q_1$, $Q_2$, in which case the form must be positive
for all values of $\eta_1$, $\eta_2$ in the unit disc.   
If $c_{min}(\eta_1,\eta_2)$ is the minimum value of $c(\eta_1,\eta_2)$ 
for all $\eta_1$ and $\eta_2$, 
the condition for the form to be positive and the potential bounded are  
\bea\label{e:con_eig_1} 
a + b &\geq& 0, \\
\label{e:con_eig_2}
a b - \frac{c^2_{min}}{4} &\geq& 0.
\eea
On substituting the values of $a$, $b$, and $c_{min}$ into 
Eqs.\ \ref{e:con_eig_1} and \ref{e:con_eig_2}  we obtain
\bea\label{e:con_eig_param_1} 
\la_1 + \la_2 + 2\la_3 &\geq& 0,\\
\label{e:con_eig_param_2} 
4\lambda_1\lambda_2+4(\lambda_1+\lambda_2)\lambda_3 - 
(4\lambda_3+\lambda_C)\lambda_C &\geq& 0,
\eea
where 
\bea
\label{e:lambda_C} 
\lambda_C =  \left\{ \begin{array}{ll}
\la_+ - \left|\sqrt{\chi_1^2 -\chi_2^2}\right| 
& \mbox{if } \la_+ - \left|\sqrt{\chi_1^2 -\chi_2^2}\right| \geq \la_4 \\
\la_4   & \mbox{otherwise.} \end{array}
\right.
\eea
Eqs.\  \ref{e:con_eig_param_1} and \ref{e:con_eig_param_2} 
are the necessary and
sufficient conditions for a bounded quartic potential. 
In \cite{Grant:1999ci} we
considered only Eqs.\  \ref{e:con_eig_param_1} and 
\ref{e:con_eig_param_2} for the second case of Eq.\  \ref{e:lambda_C}.

The condition for the vacuum of Eq.\ \ref{e:minimum} to be a minimum is simply 
\ben
\label{e:vac_locmin}
m_h^2>0,~~~~~m_H^2>0,~~~~~m_A^2>0,~~~~~m_{H^{\pm}}^2>0.
\een
On substituting masses and mixings from Eqs.\  \ref{e:charged} 
and \ref{e:X_pot}, and Eqs.\  \ref{e:chi2}-\ref{e:chi1} 
into the inequalities 
Eqs.\  \ref{e:con_eig_param_1} and \ref{e:con_eig_param_2}
we could derive six conditions directly on masses and mixing angles. 
Vice versa, by substituting the expressions for the masses in to the
parameters of the potential, six
conditions could be obtained directly on the parameters of
Eq.\  \ref{e:pot}. In practice, 
we picked masses and mixings, calculated the parameters 
of Eq.\  \ref{e:pot}, and then verified 
that Eqs.\  \ref{e:con_eig_param_1} and \ref{e:con_eig_param_2} held. 

\subsection{Global minimisation}

While the constraints of Eq.\ \ref{e:vac_locmin} guarantee 
that Eq.\ \ref{e:minimum} is a minimum of the potential, 
they do not guarantee that it is a global minimum. We are 
dealing with a large number of parameters, and 
before we proceed we need to be aware that for some 
regions of this parameter space the minimum of Eq.\ \ref{e:minimum} 
is not a global minimum. We were unable to find all 
but the simplest analytic conditions on the parameters of our potential 
that constrained Eq.\ \ref{e:minimum} to be a global minimum. 

Our approach was perforce numerical: 
we ran the Maple
extremisation routine {\tt extrema}
which took as input parameters the masses and mixings mentioned above. 
However, we found this extremisation routine was not fully reliable and did
not find all the extrema. 
We instead adapted the code written to find sphaleron solutions to find
extrema with constant fields, and looked
for configurations with negative energy. 
In Appendix \ref{a:extrema} we give more details of our 
numerical method of finding global minima.

%%%%%%%%%%%%%%%%%%%%%%%%%%%%%%%%%%%%%%%%%%%%%%%%%%%%%%%%%%%%%%%%%%%%%%%%%%%%%%%%%%%%%
%%                                                                                 %%
%%                                                                                 %%
%%  	SPHALERON ANSATZ AND SPHERICAL SYMMETRY				           %%
%%                                                                                 %%
%%%%%%%%%%%%%%%%%%%%%%%%%%%%%%%%%%%%%%%%%%%%%%%%%%%%%%%%%%%%%%%%%%%%%%%%%%%%%%%%%%%%%

\section{Sphaleron ansatz and spherical symmetry}
\label{s:ansatz}

A sphaleron is a static, unstable solution to the field equations representing
the highest energy field configuration in a path connecting one vacuum to
another.  It is easiest to look for spherically symmetric solutions, and 
so we use the spherically symmetric ansatz of \cite{Ratra:1988dp}, 
extended to allow $P$, $C$ \cite{Kastening:1991nw}, and
$CP$ violation \cite{Grant:1999ci}:

\ben\label{e:ansatz_h}
\phi_{\alpha}=\frac{1}{2}\frac{\upsilon}{\sqrt{2}}(F_{\alpha}
+{i}G_{\alpha}\hat{x}^a\sigma^{a})
\left[\ba{c}0 \\ 1\ea\right]
\een
\ben\label{e:ansatz_g0}
W_0=\frac{1}{\sqrt{2}}\frac{1}{g}A_0\hat{x}^a\frac{\sigma^a}{2i}
\een
\ben\label{e:ansatz_gi}
W_i=\frac{1}{\sqrt{2}}\frac{1}{g}\left[\frac{(\sqrt{2}
+\beta)}{r}\varepsilon_{aij}\hat{x}_j
      +\frac{\alpha}{r}(\delta_{ai}-\hat{x}_a\hat{x}_i)
      +A_1\hat{x}_a\hat{x}_i\right]\frac{\sigma^a}{2i}
\een
where $F_{\alpha}=a_{\alpha}+ib_{\alpha}$ and 
$G_{\alpha}=c_{\alpha}+id_{\alpha}$, and $F_{\al}$, $G_{\al}$, $\al$, $\be$, $A_0$, and $A_1$ are functions of the radial co-ordinate $r$.

We work in the radial gauge where $A_1$ is zero, and as we are looking for 
static solutions we set $A_0$ to zero.
We have scaled separately the Higgs 
and gauge parts of this ansatz so that the kinetic contribution to the energy is of the form 
$\frac{1}{2} f_{A}'^{2} $, where
$f_A$ generically denotes the fields $a_{\al},b_{\al},c_{\al},d_{\al},\al,\be$.

Under the $P$, $C$, and $CP$ transformations of Eqs.\ \ref{e:Pxm}-\ref{e:CPxmg}, where we have set $\theta_\al=0$, 
the fields $f_A$, $A_0$, and $A_1$ transform as shown in Table \ref{t:cp_xms}.

\begin{table}[!htb]
\begin{center}
\vspace{0.5cm}
\begin{tabular}{|c|c|c|}\hline
$P$ & $C$ & $CP$ \\ \hline
$a_{\alpha}\rightarrow + a_{\alpha}$  & $a_{\alpha}\rightarrow +a_{\alpha}$ & $a_{\alpha}\rightarrow +a_{\alpha}$ \\ 
$b_{\alpha}\rightarrow +b_{\alpha}$  & $b_{\alpha}\rightarrow -b_{\alpha}$ & $b_{\alpha}\rightarrow -b_{\alpha}$ \\ 
$c_{\alpha}\rightarrow -c_{\alpha}$  & $c_{\alpha}\rightarrow +c_{\alpha}$ & $c_{\alpha}\rightarrow -c_{\alpha}$ \\ 
$d_{\alpha}\rightarrow -d_{\alpha}$  & $d_{\alpha}\rightarrow -d_{\alpha}$ & $d_{\alpha}\rightarrow +d_{\alpha}$ \\ 
$\alpha\rightarrow -\alpha$  & $\alpha\rightarrow +\alpha$ & $\alpha\rightarrow -\alpha$ \\ 
$\beta\rightarrow +\beta $ & $\beta\rightarrow +\beta$ & $\beta\rightarrow +\beta$  \\ 
$A_0\rightarrow -A_0$ & $A_0\rightarrow +A_0$ & $A_0 \rightarrow -A_0$ \\ 
$A_1\rightarrow -A_1$ & $A_1\rightarrow +A_1$ & $A_1\rightarrow -A_1$  \\ \hline
\end{tabular}
\caption{\label{t:cp_xms} $P$, $C$, and $CP$ transformations for the fields of ansatz \ref{e:ansatz_h}-\ref{e:ansatz_gi}.}
\vspace{0.5cm}
\end{center}
\end{table}
On substuting ansatz Eqs.\  \ref{e:ansatz_h}-\ref{e:ansatz_gi} into 
the Lagrangian \ref{e:lag} we find the static energy functional
\bea
\label{e:en_H}
E[f_A]& = & \frac{M_{W}}{g^2}\int dr\,d\theta\,d\phi\,\,r^{2}\,\sin\theta\,\,\, \left[K +V_{H} \right]
\eea 
where $r$ is in units of $M_{W}^{-1}$, and
\bea
K& = & K_{0}+K_{1}\hat{x}_{3},\label {e:kin_K}  \\
V_{H}& = & V_{0}+V_{1}\hat{x}_{3}+V_{2}\hat{x}_{3}\hat{x}_{3}.\label {e:pot_H}
\eea 
$K_0$, $K_1$, $V_0$, $V_1$, and $V_2$ are given in Appendix \ref{a:SEF}, and
$\hat{x}_3 = 2{\phi^v_1}^\dagger\sigma^a\phi^v_2 x^a/\up_1\up_2$ is 
the third component of a unit radial vector.  Hence this ansatz is potentially
inconsistent if $K_1$, $V_1$, and $V_2$ are non-zero.

If the field configuration conserves $C$: 
$F_{\alpha}=a_{\alpha}$ and $G_{\alpha}=c_{\alpha}$, 
and we have the usual ansatz of \cite{Ratra:1988dp}. This gives 
$K_{1}$=0 and $V_{1}$=0, although $V_{2}$ may be non-zero if 
$M_A \ne M_{H^\pm}$, and the field configuration has $c_\alpha\ne0$. 
If the field configuration conserves $P$: $G_\al = 0$, and
again all three of the dangerous terms $K_1$, $V_1$, and $V_2$ vanish.
In the presence of two Higgs doublets Bachas, Tinyakov, and Tomaras \cite{Bachas:1996ap} (for RWS) 
and Kleihaus \cite{Kleihaus:1999bh} (for bisphalerons)
used a $C$ conserving ansatz and worked with parameters
for which $M_A = M_{H^\pm}=0$ and thereby conserved spherical
symmetry. On introducing $C$ violating terms Kastening, Peccei, and Zhang \cite{Kastening:1991nw} 
used a $P$ conserving ansatz to find the ordinary sphaleron,
while in extending to the MSSM Moreno, Oaknin, and Quiros \cite{Moreno:1997zm} used a $C$ and $P$ conserving ansatz for the sphaleron, and so again   
neither \cite{Kastening:1991nw} nor \cite{Moreno:1997zm} would have noticed any departure from spherical symmetry.

The functions $K_{0}$, $V_{0}$, and
$V_{2}$ for the $C$ conserving ansatz, 
and the conditions on parameters and solutions which conserve exact 
spherical symmetry are given in Appendix \ref{a:SEF}. 
If we allow an ansatz which does not conserve $P$, $C$, or $CP$ 
$F_{\alpha}=a_{\alpha}+ib_{\alpha}$ and $G_{\alpha}=c_{\alpha}+id_{\alpha}$,
and $K_{1}$, $V_{1}$, and $V_{2}$ can all
be non zero. $K_0$, $K_1$, $V_0$, $V_1$, and $V_2$ for this case are also given in Appendix \ref{a:SEF}.

Our strategy is to assume $f_A \equiv f_A(r)$ and integrate over
$\hat{x}_{3}=\cos\theta$ 
of Eqs.\ \ref{e:en_H}--\ref{e:pot_H} to give
\bea
\label{e:en_symm}
E [f_A]& = & \frac{M_{W}}{\al_W}\int dr\,\,r^{2}\,\,\, 
\left[K_{0}+V_{0}+\frac{1}{3} V_{2} \right].
\eea 
If solutions, corresponding
to extrema of Eq.\  \ref{e:en_symm}, have field profiles for which  
$K_{1}=0$, $V_{1}=0$, and $V_{2}=0$, 
then the solutions are exactly spherically symmetric, and the ansatz has 
succeeded.  Otherwise,
the solutions are not exactly spherically symmetric, with $K_{1}$, $V_{1}$, and
$V_{2}$ measuring the departure from spherical symmetry. We can then regard 
Eq.\ \ref{e:en_symm} as the first term in an expansion in spherical harmonics,
and our procedure finds a good approximation to the $l=0$ modes provided that 
$K_{1}$, $V_{1}$, and $V_{2}$ are all small in comparison to $K_0$ and $V_0$.

In our previous paper \cite{Grant:1999ci} we assumed spherical symmetry at the level of the
static energy functional by imposing 
\ben
\label{e:oldsphsymm}
F_{\al}=\la (r)
G_{\al},
\een 
which is too restrictive when it comes to finding $C$ and $P$ violating solutions in
$C$ violating theories.
 
%%%%%%%%%%%%%%%%%%%%%%%%%%%%%%%%%%%%%%%%%%%%%%%%%%%%%%%%%%%%%%%%%%%%%%%%%%%%%%%%%%%%%
%%                                                                                 %%
%%                                                                                 %%
%%  	PROPERTIES OF SOLUTIONS						           %%
%%                                                                                 %%
%%%%%%%%%%%%%%%%%%%%%%%%%%%%%%%%%%%%%%%%%%%%%%%%%%%%%%%%%%%%%%%%%%%%%%%%%%%%%%%%%%%%%

\subsection{Properties of solutions}
\label{s:properties}
We can classify solutions according to which of the symmetries $C$, $P$, and $CP$ they
preserve.  The ordinary (Klinkhamer--Manton \cite{Klinkhamer:1984di}) SU(2) sphaleron
preserves both $C$, and $P$, and its extension to a $C$ conserving two
Higgs doublet theory therefore has 
$\alpha =0$, $b_{\al} =0$, $c_{\al} =0$, and $d_{\al} =0$.  Kunz and Brihaye \cite{Kunz:1989sx} and Yaffe \cite{Yaffe:1989ms} showed
that, with one Higgs doublet, there exist $P$ violating solutions at large Higgs
mass with lower energy than the ordinary sphaleron, this solution is named
the bisphaleron as it occurs in $P$ conjugate pairs.  The appearance of a
bisphaleron solution is signalled by the ordinary sphaleron developing an
extra negative eigenvalue as the Higgs mass increases.  In a $C$ conserving 
theory these solutions are $C$ invariant 
and have $b_{\al} =0$ and $d_{\al} =0$, 
and are distinguished from the ordinary sphaleron by non-zero
$c_\al$ and $\al$.  To date they have been investigated with only $M_h$, $M_H$, and $\tan\beta$ non zero,
which corresponds to $M_{H^{\pm}}=M_A=0$ in a $C$ conserving theory, where they maintain spherical symmetry. However with 
$M_{H^{\pm}} \neq M_A$ or a non-zero $\theta_{CP}$; $V_2$, or $K_1$, $V_1$, and $V_2$ respectively can all be non-zero.
Hence, departure from spherical symmetry is generic, even in the pure SU(2) two
doublet model.

Bachas, Tinyakov, and Tomaras \cite{Bachas:1996ap} investigated two Higgs doublets models and found
more $P$ violating solutions at lower Higgs masses than the bisphaleron.
Although again occurring in $P$ conjugate pairs, they are
distinguished from the bisphaleron in that their boundary conditions require
more than one Higgs doublet: the
two Higgs fields have a relative winding around
the 3-sphere of
gauge-inequivalent field values of constant $|\phi_1|$ and $|\phi_2|$.  Thus we
refer to them as relative winding or RW sphalerons or RWS.  
If we refer just to a sphaleron, we shall henceforth generally mean the
ordinary $P$ and $C$ conserving sphaleron.
Note that RW sphalerons are spherically symmetric in $C$ conserving theories
only when $M_A = M_{H^\pm}$. 

The defining characteristic of a sphaleron is that
it represents the highest point of a minimum energy path
starting and ending in the vacuum, along which the Chern-Simons number changes
by $\pm1$. 
The Chern-Simons number is defined as
\bea
{n}_{\mathrm{CS}} & =& \frac{g^2}{16\pi^2}\varepsilon_{ijk}\int d^3x\left[W^a_i\partial_{j}W^a_k
+\frac{1}{3}g\varepsilon^{abc}W^a_iW^b_jW^c_k\right] \\
&= &\frac{g^2}{32\pi^2}\int d^3xK^0,
\eea 
where $\partial_{\mu}K^{\mu}=F^a_{\mu\nu}\tilde{F}^{a\mu\nu}.$
Under a gauge transformation, ${n}_{\mathrm{CS}}$ changes by an integer: hence,
field configurations with integer ${n}_{\mathrm{CS}}$ are gauge equivalent to
the vacuum $W^a_i=0$. One should also note that ${n}_{\mathrm{CS}}$ is odd
under $CP$.

Ordinary sphalerons have half-integer Chern-Simons number $n_{CS}$, which by
choice of a suitable gauge can be taken to be precisely
$1/2$. However, Yaffe found
that the bisphalerons pairs had $n_{CS} = 1/2 \pm \nu$, where $\nu$ was
typically fairly small, and depended on the parameters in the Higgs potential.
Bachas, Tinyakov, and Tomaras did not calculate the Chern-Simons number of their
relative winding sphalerons pairs, but we also find them to come in pairs with 
$n_{CS} = 1/2 \pm \nu$.  That solutions which spontaneously violate $CP$ in
this way should come in such pairs is clear, as field configurations
with $n_{CS} = 1/2 - \nu$ can be obtained from one with $n_{CS} = 1/2 + \nu$ by
a combination of a $CP$ and a gauge transformation.

%%%%%%%%%%%%%%%%%%%%%%%%%%%%%%%%%%%%%%%%%%%%%%%%%%%%%%%%%%%%%%%%%%%%%%%%%%%%%%%%%%%%%
%%                                                                                 %%
%%                                                                                 %%
%%  	FINDING SOLUTIONS						           %%
%%                                                                                 %%
%%%%%%%%%%%%%%%%%%%%%%%%%%%%%%%%%%%%%%%%%%%%%%%%%%%%%%%%%%%%%%%%%%%%%%%%%%%%%%%%%%%%%

\section{Finding solutions}
\label{s:finding_solns}
\subsection{Method}\label{s:method}
%%%%%%%%%%%%%%%%%%%%%%%%%%%%%%%%%%%%%%%%%%%%%%%%%%%%%%%%%%%%%%%%%%%%%%%%%%%%%%%%%%%%%
%%                                                                                 %%
%%                                                                                 %%
%%  	METHOD								           %%
%%                                                                                 %%
%%%%%%%%%%%%%%%%%%%%%%%%%%%%%%%%%%%%%%%%%%%%%%%%%%%%%%%%%%%%%%%%%%%%%%%%%%%%%%%%%%%%%

We will be finding solutions to a static energy functional of the form
\ben
\label{e:E_static_gen}
E[f_A] = \frac{M_W}{\al_W} \int dr\, \,{\cal E}(f_A),
\een
where 
\ben
\label{e:E_den_static_gen}
{\cal E}(f_A) = \frac{1}{2}f_{G}'^{2} +\frac{1}{2}r^2f_{H}'^{2}
+ P(f_A).
\een
Here, $P(f_A)$ is a polynomial in the 10 fields $f_A$, which 
we divide into gauge fields $f_{G}= \al,\be$ and Higgs fields $f_{H}=a_{\al},
b_{\al}, 
c_{\al},
d_{\al}. $ 

We use a Newton method, following \cite{Yaffe:1989ms}, which is an efficient way of finding
extrema (and not just minima).  The method can be briefly characterised as
updating the fields $f_A$ by an amount $\delta f_A$, given by the solution of 
\ben
\label{e:Newt_raph}
\frac{\delta^{2} {\cal E} }{\delta f_B \delta f_A}  ~~ \delta f_B = -\frac{\delta {\cal E} }{\delta f_A},
\een
which we can abbreviate
as ${\cal E}^{''} \delta f = - {\cal E}^{'}$. Provided ${\cal E}^{''}$ has no
zero eigenvalues, the equation has a unique solution, subject to boundary
conditions which we detail below. 
We sometimes added a fraction of $\delta f$ which, although slower,
occasionaly produced a more stable convergence. 
The procedure is started from an initial guess for $f_A$, and then
repeated with each 
improved configuration, until ${\cal E}^{'} $ is small enough so that $\delta f \simeq 0$.

A particular advantage to using this method is that because we are calculating ${\cal E}^{''}$, it is straight forward to
get the negative curvature eigenvalues, $\om^2$, from the diagonalistion of ${\cal E}^{''}$ at each solution. To achieve this 
we use
\ben
\label{e:eignevals}
 \frac{1}{2}{\cal E}^{''}
\left[\ba{c}\de f_{G}\\  r\, \de f_{H} \ea\right] = \om^2 
\left[\ba{c}\de f_{G}\\  r\, f_{H} \ea\right],
\een
from 
\ben
\label{e:2nd_der}
\de^2 E[f_A] =  \frac{M_W}{\al_W}\int dr\,\,\left[\ba{c}\de f_{G}\\  
r\, \de f_{H} \ea\right]^T  \frac{1}{2} {\cal E}^{''}
\left[\ba{c}\de f_{G}\\  r\, \de f_{H} \ea\right],
\een
where it is understood that the $\cal E^{''}$ of 
Eqs.\ \ref{e:eignevals} and \ref{e:2nd_der} 
has been differentiated with respect to $f_G$ and $r f_H$, and not as 
in the Newton method of Eq.\ \ref{e:Newt_raph} with respect to $f_G$ and $f_H$.

\subsection{Boundary conditions}
\label{s:bcs}
%%%%%%%%%%%%%%%%%%%%%%%%%%%%%%%%%%%%%%%%%%%%%%%%%%%%%%%%%%%%%%%%%%%%%%%%%%%%%%%%%%%%%
%%                                                                                 %%
%%                                                                                 %%
%%  	BOUNDARY CONDITIONS						           %%
%%                                                                                 %%
%%%%%%%%%%%%%%%%%%%%%%%%%%%%%%%%%%%%%%%%%%%%%%%%%%%%%%%%%%%%%%%%%%%%%%%%%%%%%%%%%%%%%

Next we turn our attention to boundary conditions. Before we 
look at specific conditions for different solutions, 
we consider the terms of Eq.\ \ref{e:K^G_0_cp} of Appendix \ref{a:SEF}, (up
to numerical factors)
\bea
\label{e:boundaryterms} 
K^G_0 & \propto & \frac{1}{r^2}(\al^{2}+\be^{2} -2)^2
+(a_{\al}^{2}+b_{\al}^{2}+c_{\al}^{2}+d_{\al}^{2}) (\al^{2}+\be^{2} +2) \nonumber\\ 
& & +\sqrt{2}\be (a_{\al}^{2}+b_{\al}^{2}-c_{\al}^{2}-d_{\al}^{2})
- 2\sqrt{2}\al (a_{\al}c_{\al}+b_{\al}d_{\al}).  
\eea
We introduce new fields $\chi$, $\K_\al$, $\L_\al$, $\Psi$, and $\Th_\al$
defined by 
\bea
-\be + i \al & = & \sqrt{2} \chi \exp (i\Psi), \\
a_{\al} + i c_{\al} & = & 2 \frac{\upsilon_{\al}}{\upsilon} 
\K_{\al} \exp (i \Th_{\al}),\\
b_{\al} + i d_{\al} & = & 2 \frac{\upsilon_{\al}}{\upsilon} 
\L_{\al} \exp (i \Th_{\al}),\\
\eea
\noindent and rewrite (\ref{e:boundaryterms}) as 
\bea
\label{e:bcs_en_gauge}
K_0 & \propto & \frac{1}{4 r^2}(\chi^2-1)^2  \nonumber \\
& &+ (2\chi^2 + 2) \left[ \cos^2\beta(\K_1^2+\L_1^2) 
+ \sin^2\beta (\K_2^2+\L_2^2)\right]  \nonumber \\
& &- 4\chi \cos^2\beta (\K_1^2+\L_1^2)\Re[\exp(-i\Psi+i2\Theta_1)]  \nonumber \\
& &- 4\chi\sin^2\beta (\K_2^2+\L_2^2)\Re[\exp(-i\Psi+i2\Theta_2)] .
\eea
We have a boundary condition from the finiteness of the energy density, 
due to the first term in Eq.\ \ref{e:bcs_en_gauge} which can be expressed as 
\bea
\label{e:gaugebc}
\chi^2 \rightarrow 1  & \rm{as}  & r \rightarrow 0.
\eea 
From the finiteness of the gauge current density (which is proportional to the 
second, third, and fourth terms in Eq.\ \ref{e:bcs_en_gauge}) and using
Eq.\ \ref{e:gaugebc}, we also have
\bea
\label{e:higgsbc}
\left. \begin{array}{ll}
(\K_1^2+\L_1^2)\Re[\exp(-i\Psi+i2\Theta_1)] & \rightarrow \K_1^2+\L_1^2 \\
(\K_2^2+\L_2^2)\Re[\exp(-i\Psi+i2\Theta_2)] & \rightarrow \K_2^2+\L_2^2 \end{array}
\right\} {\mathrm as } \ r \rightarrow 0.
\eea 
To satisfy Eq.\ \ref{e:higgsbc} we require
\bea
\label{e:Theta}
 \rm{either}\,\, \left. \begin{array}{ll}
\K_1^2+\L_1^2 & \rightarrow 0 \\
\K_2^2+\L_2^2 & \rightarrow 0 \end{array}
\right\}  & \rm{or} &
\left. \begin{array}{ll}
\Theta_1 & \rightarrow \Psi/2 + n_1 \pi \\
\Theta_2 & \rightarrow \Psi/2 + n_2 \pi \end{array}
\right\} \rm{as } \ r \rightarrow 0,
\eea 
where $n_1,n_2 \in {\mathbb Z}$. Eq.\ \ref{e:Theta} can be rewritten as
\bea
\label{e:Theta12}
\left. \begin{array}{rcl}
\rm{either}& \K_\al^2+\L_\al^2 & \rightarrow 0 \\
\rm{or}& \Theta_1 - \Theta_2 & \rightarrow (n_1-n_2) \pi \end{array}
\right\} \rm{as } \ r \rightarrow 0.
\eea
Eqs.\ \ref{e:gaugebc} and \ref{e:Theta12} are then our boundary conditions as 
$r \rightarrow 0$.
The boundary conditions as $r \rightarrow \infty$ can be obtained
from finiteness of $K_0$, (Eq. \ref{e:K_0})  and of $V_0$ (Eq. \ref{e:pot_0}).

The ordinary sphaleron 
satisfies Eq.\ \ref{e:Theta12} by having 
\ben (\K_\al^2+\L_\al^2)|_{r=0} = 0.
\een
The full set of boundary conditions for the sphaleron 
are given in Table \ref{t:BCsphal}.

Bisphaleron pairs have different boundary conditions. 
To satisfy Eq.\ \ref{e:Theta12}, where $\delta$ is a small positive angle, they have 
\ben 2\Theta_{1}|_{r=0} =
2\Theta_{2}|_{r=0} = \Psi|_{r=0} \equiv 2\Theta = -\pi \pm \delta.
\een 
The boundary conditions on the $f_A$ of these solutions are given
in Table \ref{t:BCbisphal}.

Relative winding sphalerons pairs 
satisfy Eq.\ \ref{e:Theta12} through 
\ben
2(\Theta_{1}-\pi ) |_{r=0} = 2\Theta_{2}|_{r=0} = \Psi|_{r=0} = -\pi \pm \delta.
\een 
From Eq.\ \ref{e:Theta12} we see that since $n_1 =n_2$ for bisphalerons while 
$n_1 = n_2+1$ for RWS, RWS unlike bisphalerons can only occur in multi-doublet theories. 
The integers $n_1$ and
$n_2$ represent the winding numbers of the Higgs fields around the 3-spheres of
constant $|\phi_1|$ and $|\phi_2|$, with only their difference having any
gauge-invariant meaning. 
The RWS boundary conditions are given in Table \ref{t:BCRWsphal}.

\begin{table}[ht]
\begin{center}
\begin{tabular}{|c | c c c c c c|}\hline 
$r \rightarrow 0$  & $ \al \rightarrow 0$ & $ \be \rightarrow \sqrt{2} $
 & $a_{\al} \rightarrow 0$ & $ b_{\al} \rightarrow 0$ & $c_{\al} \rightarrow 0$ & $d_{\al} \rightarrow 0$  \\ \hline\hline
$r \rightarrow \infty $ &  $ \al \rightarrow 0 $ & $ \be \rightarrow -\sqrt{2}$ 
& $a_{1} \rightarrow 2\cos\beta$ & $b_{1} \rightarrow 0$ & $c_{1} \rightarrow 0$ & $d_{1} \rightarrow 0$ \\
 &   &  & $a_{2} \rightarrow 2\sin\beta$ & $b_{2} \rightarrow 0$ & $c_{2} \rightarrow 0$ & $d_{2} \rightarrow 0$ \\  \hline
\end{tabular}
\caption{\label{t:BCsphal} Boundary
conditions for the ordinary $C$, and $P$ conserving sphaleron.}
\end{center}
%\end{table}
%\begin{table}[ht]
\begin{center}
\begin{tabular}{|c |c c c| }\hline 
$r \rightarrow 0$  & $ \al \rightarrow \sqrt{2}\sin2\Theta$ & $a_{1} \rightarrow 2 \K_1\cos\beta\cos\Theta$ & $ a_{2} \rightarrow 2 \K_2\sin\beta\cos\Theta$   \\ 
& $ \be \rightarrow -\sqrt{2}\cos2\Theta$ & $b_{1} \rightarrow 2 \L_1\cos\beta\cos\Theta$ & $ b_{2} \rightarrow 2 \L_2\sin\beta\cos\Theta$  \\
 & & $c_{1} \rightarrow 2 \K_1\cos\beta\sin\Theta$ & $ c_{2} \rightarrow 2 \K_2\sin\beta\sin\Theta$ \\ 
 & & $d_{1} \rightarrow 2 \L_1\cos\beta\sin\Theta$ & $ d_{2} \rightarrow 2 \L_2\sin\beta\sin\Theta$\\ \hline 
\end{tabular} 
\caption{\label{t:BCbisphal}Boundary conditions at the origin for
the ($P$ violating) bisphaleron.  The boundary conditions at infinity are the
same as for the sphaleron, Table \ref{t:BCsphal}.}
\end{center}
%\end{table}
%\begin{table}[ht]
\begin{center}
\begin{tabular}{|c | c c c| }\hline 
$r \rightarrow 0$  & $ \al \rightarrow \sqrt{2}\sin\Psi$ & $a_{1} \rightarrow 2 \K_1\cos\beta\cos\Theta_1$ & $ a_{2} \rightarrow 2 \K_2\sin\beta\cos\Theta_2$   \\  
& $ \be \rightarrow -\sqrt{2}\cos\Psi$ & $b_{1} \rightarrow 2 \L_1\cos\beta\cos\Theta_1$ & $ b_{2} \rightarrow 2 \L_2\sin\beta\cos\Theta_2$  \\ 
 & & $c_{1} \rightarrow 2 \K_1\cos\beta\sin\Theta_1$ & $ c_{2} \rightarrow 2 \K_2\sin\beta\sin\Theta_2$ \\ 
 & & $d_{1} \rightarrow 2 \L_1\cos\beta\sin\Theta_1$ & $ d_{2} \rightarrow 2 \L_2\sin\beta\sin\Theta_2$\\ \hline 
\end{tabular}
\caption{\label{t:BCRWsphal}Boundary conditions at the origin for
the ($P$ violating) RWS.
 The boundary conditions at infinity are the
same as for the sphaleron, Table \ref{t:BCsphal}.}
\end{center}
\end{table}

\subsection{Numerical performance}
\label{ss:numchecks}

The details of the implementation of the algorithm and the boundary counditions
are relegated to Appendix \ref{a:numerics}.  We checked the 
accuracy of our code by evaluating the 
energy, negative curvature eigenvalues and Chern-Simons number for some of the same parameters as 
Yaffe \cite{Yaffe:1989ms} and Bachas, Tinyakov, and Tomaras \cite{Bachas:1996ap}, and found good agreement. These can be seen in Table\ \ref{t:yaffeBTT}.

\begin{table}[ht]
\begin{center}
\begin{tabular}{|c|c c c c|c c c c|c c|}\hline
$m$  & $E_{sph} $ & $ -\om^2_1$ & $-\om^2_2 $& $-\om^2_3 $& $E_{bi} $
& $-\om^2_1 $& $-\om^2_2 $ & $n_{CS}$& $E_{RWS} $& $-\om^2_1 $   \\ \hline
$5$ & $4.435$ & $5.391$   & $\cdots$ & $\cdots$ & $\cdots$  
& $\cdots$ & $\cdots$ & $\cdots$ & $\cdots$ & $\cdots$   \\ 
$6$ & $4.531$ & $6.217$   & $0.279$ & $\cdots$ & $\cdots$  
& $\cdots$ & $\cdots$ & $\cdots$ & $4.528$ & $5.171$   \\ 
$7$ & $4.609$ & $7.171$   & $1.225$ & $\cdots$ & $\cdots$  
& $\cdots$ & $\cdots$ & $\cdots$ & $4.587$ & $4.147$   \\
$10$ & $4.778$ & $11.22$  & $5.962$ & $\cdots$ & $\cdots$  
& $\cdots$ & $\cdots$ & $\cdots$ & $4.668$ & $3.090$   \\ 
$13$ & $4.888$ & $17.70$  & $13.27$ & $0.316$ & $4.886$  
& $11.86$ & $6.546$ & $0.454$ & $4.700$ & $2.773$   \\ 
$15$ & $4.942$ & $23.49$  & $19.49$ & $0.926$ & $4.930$  
& $8.447$ & $2.349$ & $0.428$ & $4.711$ & $2.670$   \\ 
$30$ & $5.147$ & $101.4$  & $98.55$ & $3.212$ & $5.031$  
& $5.207$ & $\cdots$ & $0.387$ & $4.734$ & $2.451$   \\
$50$ & $5.243$ & $292.7$  & $290.1$ & $4.734$ & $5.052$
& $4.874$ & $\cdots$ & $0.380$ & $4.738$ & $2.403$   \\ \hline
\end{tabular}
\caption{\label{t:yaffeBTT} Energy ($M_W/\alpha_W$), negative
eigenvalues ($M_W^2$), and Chern-Simons number
for $m =M_H/M_W=M_h/M_w$ and $\tan\beta=1$, for some of the same
parameters as \cite{Yaffe:1989ms} and \cite{Bachas:1996ap}.
The solution with energy $E_{bi}$ was reached by
perturbing the
ordinary sphaleron in the direction of the eigenvector
with eigenvalue $-\om^2_3$,
and the solution with energy $E_{RWS}$
was reached by a perturbation with
eigenvalue $-\om^2_2 $. If we refer
to Fig.\ 2 of \cite{Bachas:1996ap} we see that the
bisphaleron branch itself bifurcates at the point where it no longer has
two negative eigenvalues, and we note as a point of
interest that
the eigenvector with eigenvalue $-\om^2_2 $ takes us to the solution
with lowest energy and not the $S_1$ of \cite{Bachas:1996ap}.
The $n_{CS}$ of the RWS for equal $CP$ even Higgses,
and $\tan\beta=1$ with all other parameters zero is 1/2, this is not the case generally.
The agreement with \cite{Yaffe:1989ms} and \cite{Bachas:1996ap} is excellent.
}
\end{center}
\end{table}
The numerical scheme worked excellently, with typical convergence after five
to fifteen iterations of $1 \times 10^{-13}$ in 
the sum of absolute change in all fields at all points.
 The few problems we did encounter were: 
(1) sometimes the initial configuration for a RW sphaleron  
was so close to 
the sphaleron that the Newton extremisation found the 
original sphaleron, particularly 
at points in parameter space near the bifurcation point, and (2) the Newton extremisation sometimes found
the vacuum 
from the inital configuration for a RW sphaleron . The first was solved by 
using a higher mass RW sphaleron as initial conditions for minimisation, 
and the second problem by updating each minimisation 
not with $\delta f_{\alpha}$ but with a fraction of it.

We ran simultaneously two codes. One with the $C$ conserving 
ansatz, and the other with the $C$ and $P$ violating ansatz. 
In the absence of $C$ violation the two codes were identical. 
With 101 point instead of 51, the difference in energy, $n_{CS}$, and eigenvalues was 
at most of order 0.5 \% of the value with 51 points.

\section{Results}
\label{s:results}
%%%%%%%%%%%%%%%%%%%%%%%%%%%%%%%%%%%%%%%%%%%%%%%%%%%%%%%%%%%%%%%%%%%%%%%%%%%%%%%%%%%%%
%%                                                                                 %%
%%                                                                                 %%
%%  	SPECIFIC RESULTS						           %%
%%                                                                                 %%
%%%%%%%%%%%%%%%%%%%%%%%%%%%%%%%%%%%%%%%%%%%%%%%%%%%%%%%%%%%%%%%%%%%%%%%%%%%%%%%%%%%%%

\subsection{No $CP$ violation, $M_A$ = $M_{H^\pm} = 0$}

In order to compare with previous work, we firstly examine the unrealistic
limit of $M_A$ = $M_{H^\pm} = 0$, with no explicit $CP$ violation in the
potential.  We set the parameters $\la_3=0$ and $\tan\beta = 6$, and scanned
through $M_h$ and $M_H$ between 0 and 800 GeV.

%%%%%%%%%%%%%%%%%%%%%%%%%%%%%%%%%%%%%%%%%%%%%%%%%%%%%%%%%%%%%%%%%%%%%%%%%%%%%%%%%%%%%
%%                                                                                 %%
%%   Results I: 4 figures. Contour in M_h M_H space. All other parameters zero     %%
%%                                                                                 %%
%%%%%%%%%%%%%%%%%%%%%%%%%%%%%%%%%%%%%%%%%%%%%%%%%%%%%%%%%%%%%%%%%%%%%%%%%%%%%%%%%%%%%
\begin{figure}[!ht]
\centering{
\includegraphics[height=8cm,width=9cm]{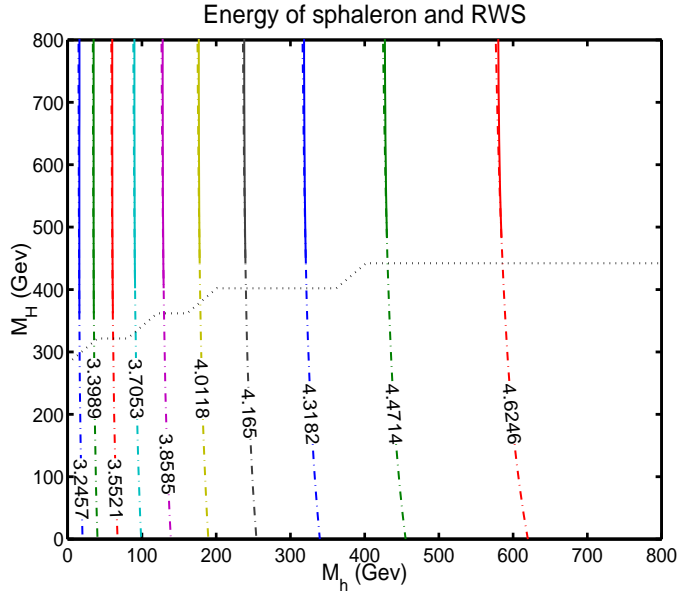}\
\caption{\label{f:energy1} Contours in $M_h$, $M_H$ space of the energy
of the sphaleron 
(dashes),and of the RWS (solid), in units of $M_W/\alpha_W$.
Below the dotted line the sphaleron is the only solution. Above the dotted line, both 
solutions exist. The input parameters are $\tan\beta=6$ with all other parameters zero.}
}
\end{figure}

\begin{figure}[!ht]
\centering{
\includegraphics[height=8.0cm,width=9cm]{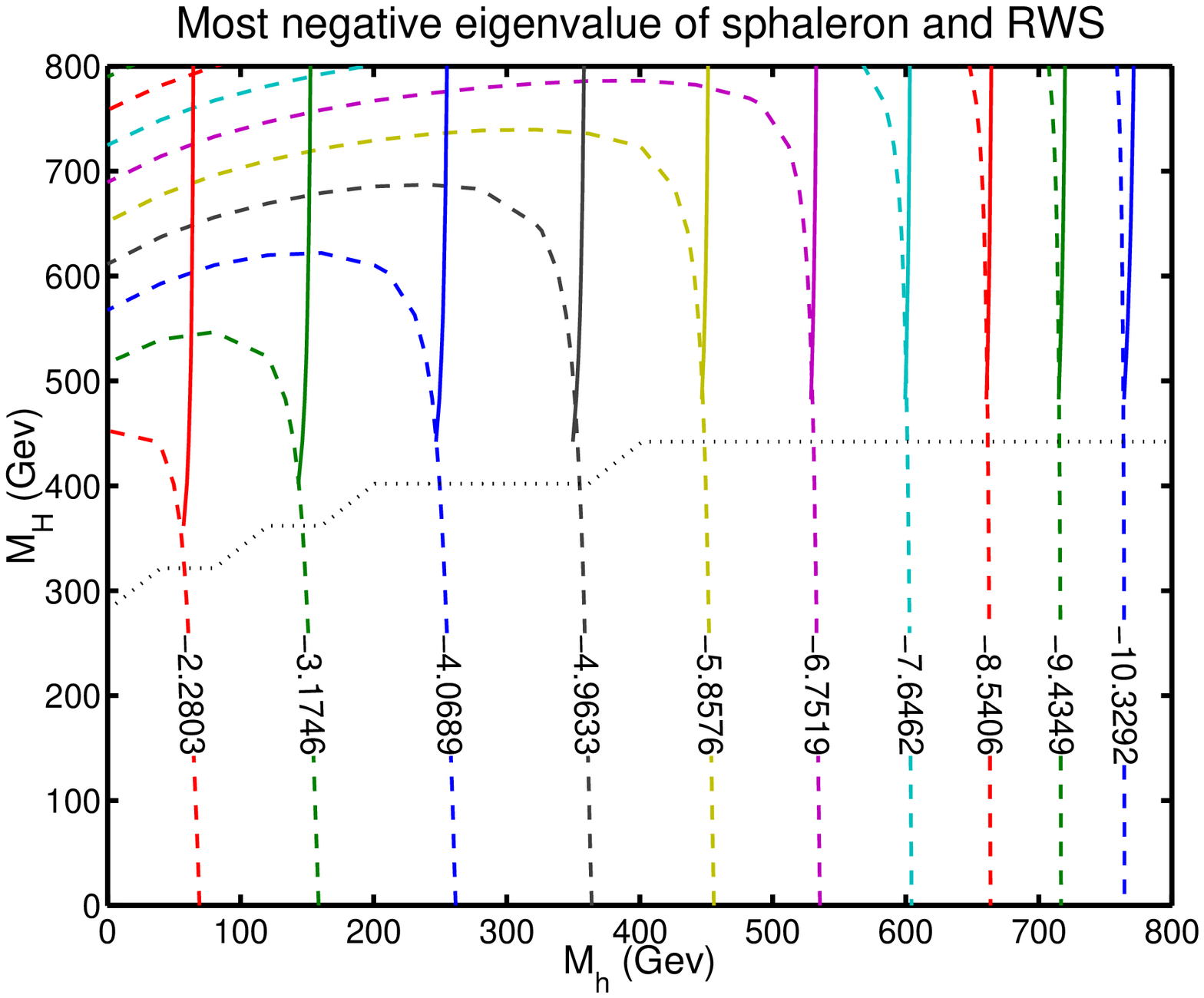} 
\vspace{1.0cm}\\
\includegraphics[height=8.0cm,width=9cm]{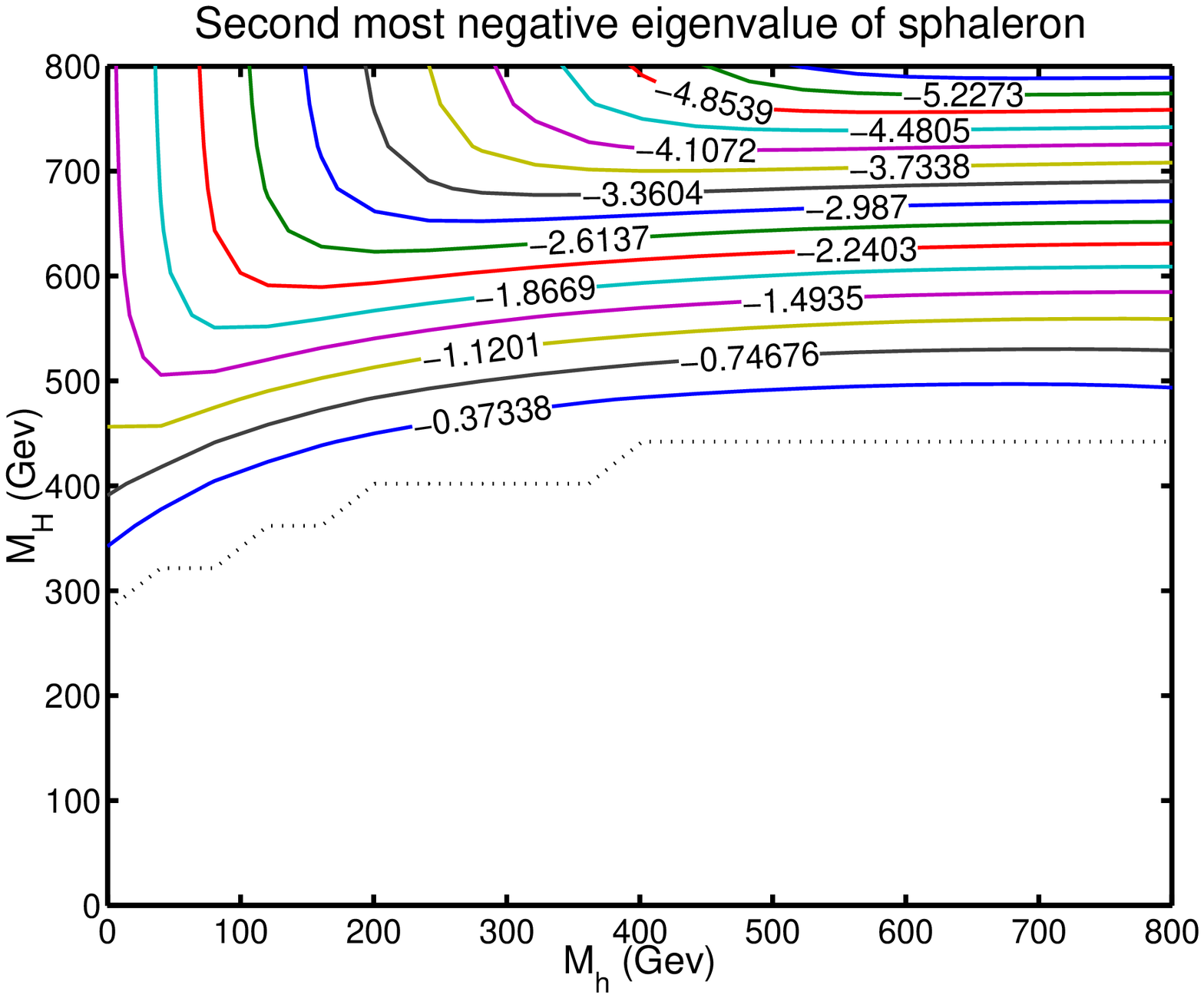} 
\vspace{0.5cm}\\
\caption{\label{f:first_eig1}\label{f:second_eig1}Contours in $M_h$, $M_H$ space of the eigenvalue in units of $M_W^2$. 
The top figure shows the most negative eigenvalule of the sphaleron (dashes), and of the RWS (solid). 
The bottom figure shows the second most negative 
eigenvalue  of the sphaleron.
Below the dotted line the sphaleron is the only solution. 
Above the dotted line, both solutions exist. 
The input parameters are $\tan\beta=6$ with all other parameters zero.} }
\end{figure}

\begin{figure}[!ht]
\centering{
\includegraphics[height=8cm,width=9cm]{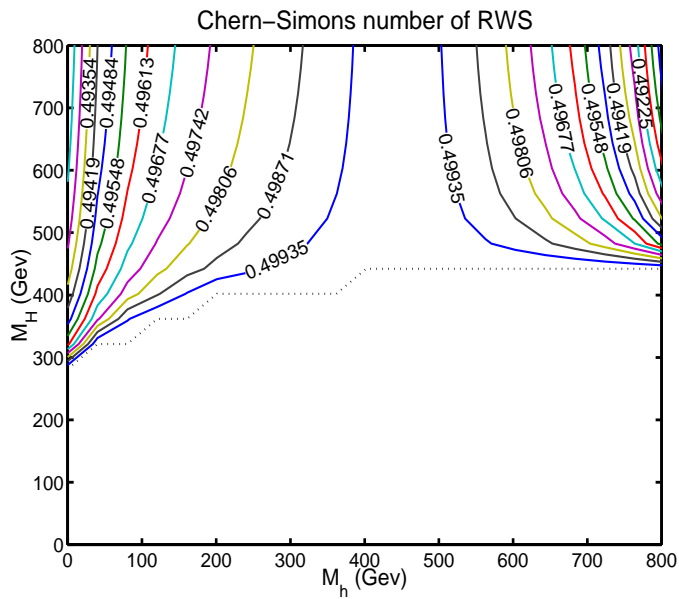}
\caption{\label{f:chern1} Contours in $M_h$, $M_H$ space of the Chern-Simons
number of the RWS. 
Below the dotted line only the sphaleron solution exists, with $n_{CS}=1/2$.
The input parameters are $\tan\beta=6$ with all other parameters zero.}
}
\end{figure}

Figs.\ \ref{f:energy1}--\ref{f:chern1} show contours in the
$M_h$ and $M_H$ plane. 
The contours are respectively of 
energy (Fig. \ref{f:energy1}), most negative eigenvalue and second most negative eigenvalue 
(Fig. \ref{f:second_eig1}), and $n_{CS}$ (Fig. \ref{f:chern1}) of the sphaleron and relative winding sphaleron.
When we show equal contours of both solutions the sphalerons are shown as dashes, and the RWS as solid.
Below the black horizontal dotted line, shown on all four contour plots, only the sphaleron solution exists, 
above the black dotted line both solutions
exist. The sphaleron never develops a third negative eigenvalue, nor the RWS a second negative eigenvalue. 
The solutions maintained exact spherical symmetry: $V_2$ was zero throughout;
this was expected as both $\theta_{CP}=0$, and $M_A=M_{H^{\pm}}=0$.
These contours are from the same potential as used by BTT \cite{Bachas:1996ap} and contain some of the parameter space they scanned. 
Where we overlap we agree with their results, 
and we confirm their observation that the second negative eigenvalue appears
when one of the Higgs has a largeish mass, ($M_H\sim 5M_W$). 
For low values of this heavier
mass the lighter Higgs needs to be as light as possible; 
i.e. for the existence of relative winding sphalerons it is preferable to have 
the two Higgs masses, $M_h$ and $M_H$, well separated.

Fig.\
\ref{f:energy1} shows both the energy of the sphaleron and the energy of the RWS, there is almost no difference between their
energies, and the energy depends mainly on the mass of the lighter Higgs.
Figure \ref{f:first_eig1} shows the most negative eigenvalue of both the sphaleron and RWS, and we see that there is 
a large difference between the values of negative eigenvalues for the different solutions; the negative eigenvalue of
the sphaleron can be double that for the relative winding sphaleron for the same point in parameter space. 
Fig.\ \ref{f:second_eig1} also shows the second negative eigenvalue of the sphaleron. The second most negative
eigenvalue belongs to the perturbation which leads 
to the RW sphaleron in configuration space.

Looking at Fig.\ \ref{f:chern1} we see that the Chern-Simons number 
of the RW sphaleron is generally not a half.
There is a line in the contour space where $n_{CS}=1/2$. This occurs, for $\tan\beta=1$, along the line of $M_h=M_H$,
 and shifts in the contour plane for different values of $\tan\beta$. 
We have only shown here solutions with $n_{CS}\leq 1/2$. Each of these solutions with $n_{CS}\leq 1/2$ has a $P$ conjugate partner, 
with Chern-Simons $n_{CS}^{con} \geq1/2$,
such that  $n_{CS}+n_{CS}^{con} =1$.

\subsection{No $CP$ violation, $M_A = 3M_W$, $M_{H^\pm} = 2M_W$}
\begin{figure}[!htb]
\centering{
\includegraphics[height=8cm,width=9cm]{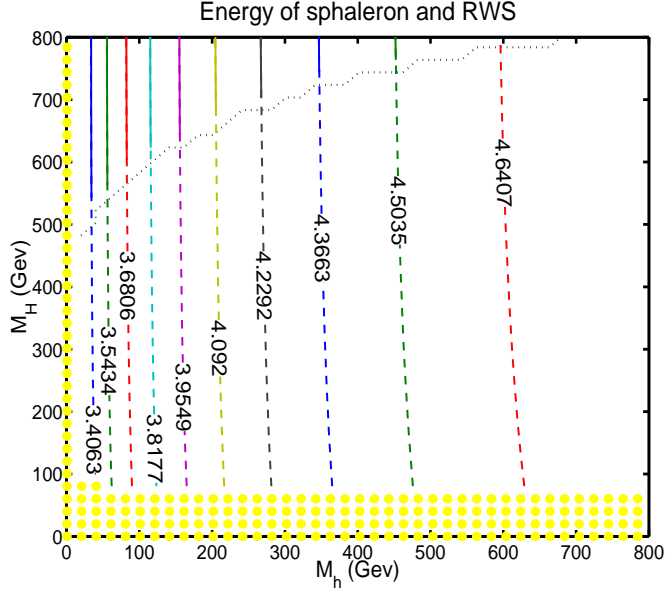}\
\caption{\label{f:energy3} Contours in $M_h$, $M_H$ space of the energy
of the sphaleron
(dashes),and of the RWS (solid), in units of $M_W/\alpha_W$.
Below the dotted line the sphaleron is the only solution, while above, both 
solutions exist. For the dotted area the potential is unbounded. 
The input parameters are $\tan\beta=6$, $M_A=241$ GeV, $M_{H^{\pm}}=161$ GeV, and $\lambda_3=-0.05$.}
}
\end{figure}

\begin{figure}[!htb]
\centering{
\includegraphics[height=8.0cm,width=9cm]{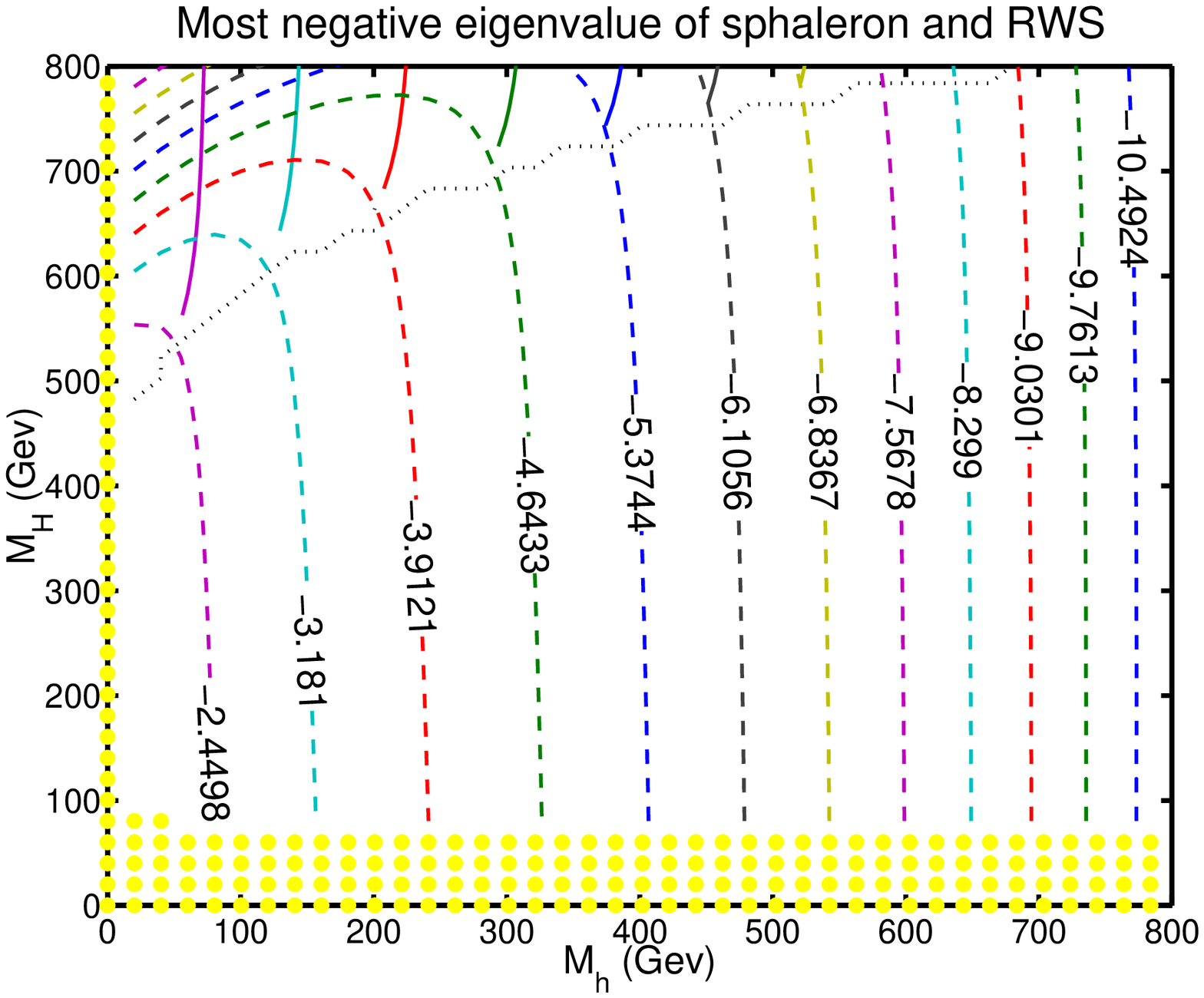}
\vspace{0.5cm}\\
\includegraphics[height=8.0cm,width=9cm]{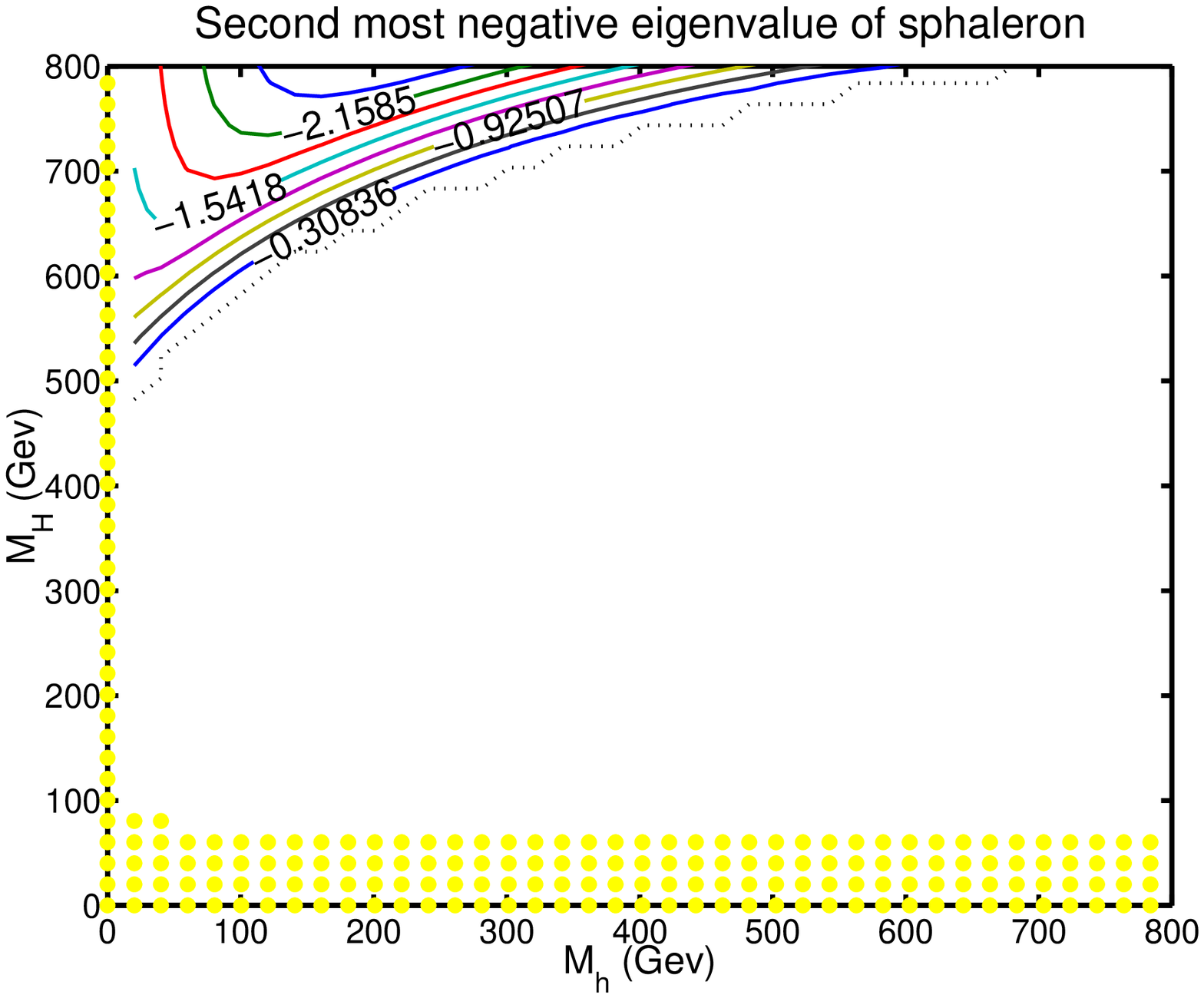}
\caption{\label{f:first_eig3}\label{f:second_eig3}Contours in $M_h$, $M_H$ 
space of eigenvalues in units of $M_W^2$. 
The top figure shows the most negative eigenvalue of the sphaleron (dashes), and of the RW
sphaleron (solid). 
The bottom figure shows the second most negative
eigenvalue  of the sphaleron.
Below the dotted line the sphaleron is the only solution. 
Above the dotted line, both solutions exist. For the dotted region the potential is unbounded.  
The input parameters are $\tan\beta=6$, $M_A=241$ GeV, $M_{H^{\pm}}=161$ GeV, and $\lambda_3=-0.05$.
}
}
\end{figure}

\begin{figure}[!htb]
\centering{
\includegraphics[height=8cm,width=9cm]{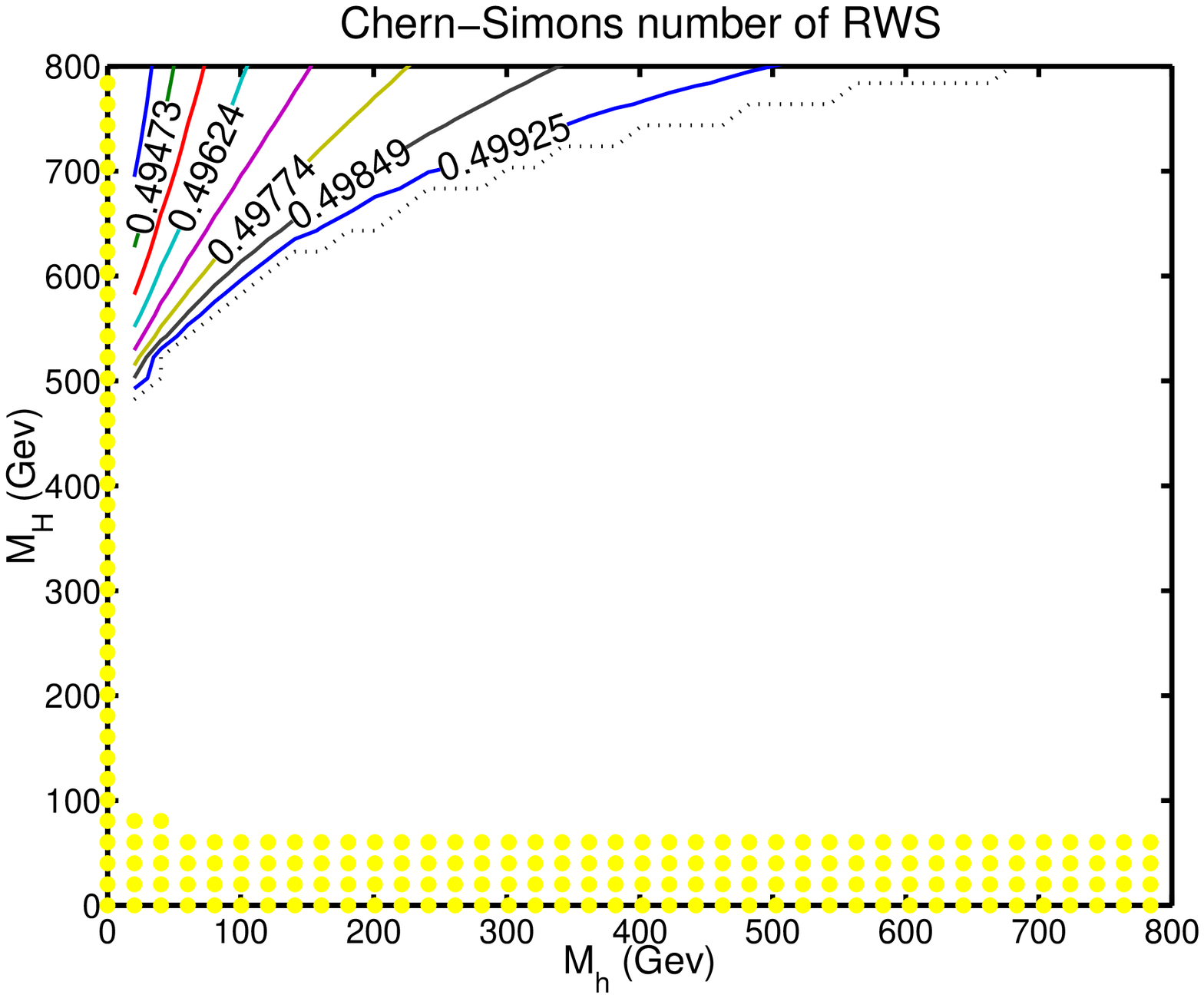}
\caption{\label{f:chern3} Contours in $M_h$, $M_H$ space of the Chern-Simons number of the RW
sphaleron.
Below the dotted
line only the sphaleron solution exists, with $n_{CS}=0.5$. For the dotted region the potential is unbounded. 
The input parameters are $\tan\beta=6$, $M_A=241$ GeV, $M_{H^{\pm}}=161$ GeV, and $\lambda_3=-0.05$.}}
\end{figure}

Figs.\ \ref{f:energy3}--\ref{f:chern3}
show contours in $M_h$, $M_H$ space of
energy (Fig.\ \ref{f:energy3}), 
most negative eigenvalue of the sphaleron and RWS, (Fig.\ \ref{f:first_eig3} top), 
and second most negative eigenvalue of the sphaleron (Fig. \ref{f:second_eig3} bottom),
and $n_{CS}$ (Fig. \ref{f:chern3}) of the sphaleron and the relative winding sphaleron.
Again when both solutions are shown the sphaleron is dashes, and the RWS solid.

%%%%%%%%%%%%%%%%%%%%%%%%%%%%%%%%%%%%%%%%%%%%%%%%%%%%%%%%%%%%%%%%%%%%%%%%%%%%%%%%%%%%%%%
%%                                                                                   %%
%%   Results II: 4 figures.Contour in M_h M_H space. M_A,M_H^{\pm},\lam_3 non-zero   %%
%%                                                                                   %%
%%%%%%%%%%%%%%%%%%%%%%%%%%%%%%%%%%%%%%%%%%%%%%%%%%%%%%%%%%%%%%%%%%%%%%%%%%%%%%%%%%%%%%%

%end third plots

For these figures we took
$M_A=241$ GeV, $M_{H^{\pm}}=161$ GeV, again with no explicit 
$CP$ violation. We set the parameters $\la_3=-0.05$,
and $\tan\beta=6$, and scanned through $M_h$ and $M_H$ between 0 and 800 GeV, 
with 20 GeV increments.
Again below the black dotted line, shown on all four contour plots,
only the sphaleron solution exits, while above both solutions exist. 
We see that the RW sphaleron 
solutions still persist for a large region of the parameter space. The dotted region 
at low $M_H$ was unbounded according to Eqs.\ \ref{e:con_eig_param_1} and \ref{e:con_eig_param_2}. These solutions did not 
maintain exact spherical symmetry corresponding to $V_2=0$, 
but the maximum value of energy due to the  $V_2$ term 
was 0.6\% of the energy due
to $V_0$.

The solutions have the same general features as those at zero $M_A$ and
$M_{H^\pm}$: the RW sphaleron appears at widely separated $M_H$ and $M_h$. 
While the energies of the two solutions 
in Fig.\ \ref{f:energy3} are almost indistinguishable,
the most negative eigenvalue (Fig. \ref{f:first_eig3} top), of the sphaleron can be double that of the 
RW sphaleron.
We show the value of the second most negative eigenvalue of the sphaleron in
Fig.\ \ref{f:second_eig3} (bottom). 
The sphaleron never developed a third negative eigenvalue, nor the RW sphaleron
a second negative eigenvalue. 
In Fig.\
\ref{f:chern3} we show the Chern-Simons number of the RW sphaleron,
and again for every solution shown with 
$n_{CS}=1/2-\nu$ there 
is a $P$ conjugate solution with $n^{con}_{CS}=1/2+\nu$.

\subsection{$CP$ violation, $M_A=8M_W$, $M_{H^\pm}=2M_W$}
\label{ss:resCPvioln}

\begin{figure}[!hbt]
\centering{
\includegraphics[height=8cm,width=9cm]{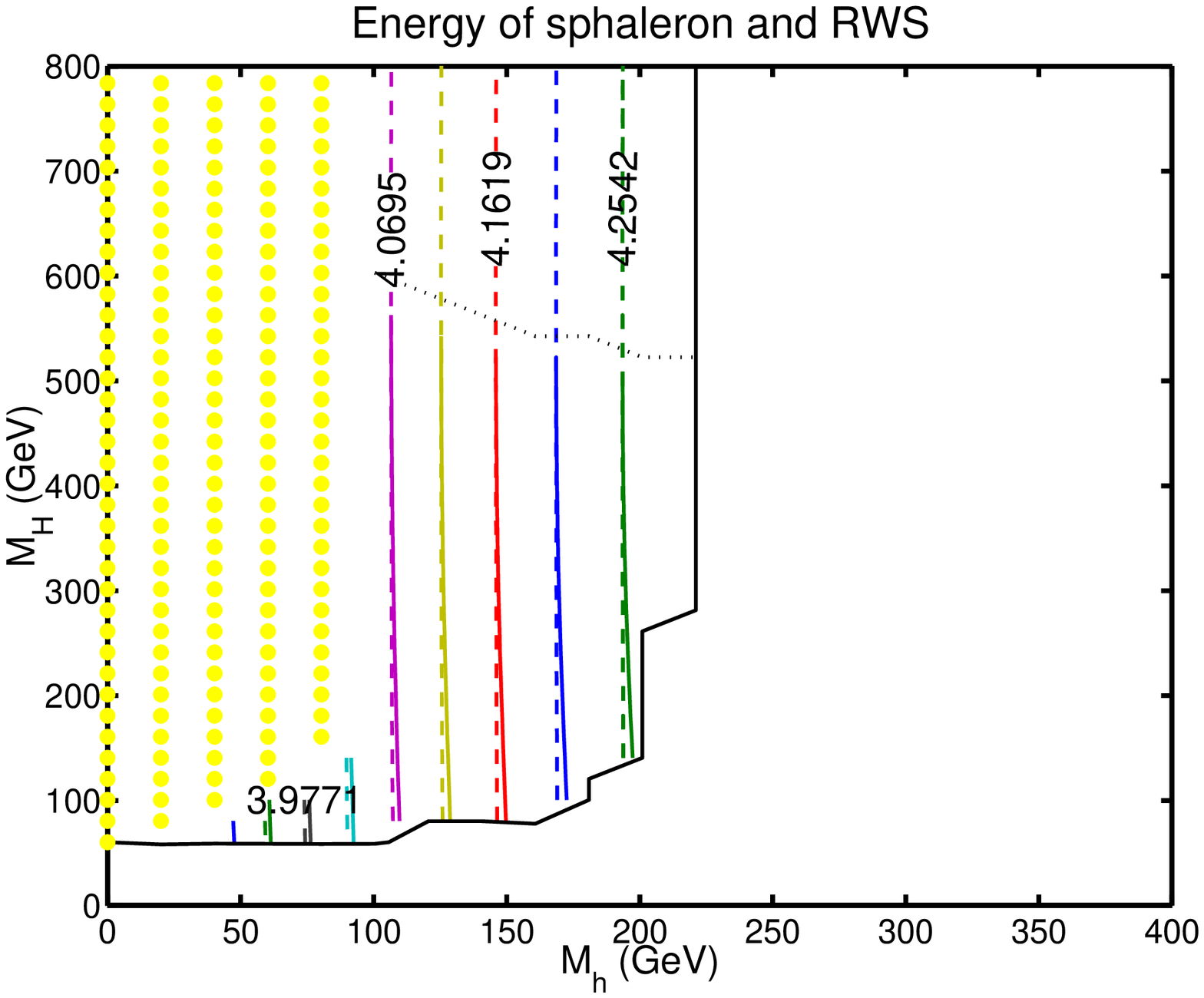}
\vspace{0.5cm}\\
\includegraphics[height=8.0cm,width=9cm]{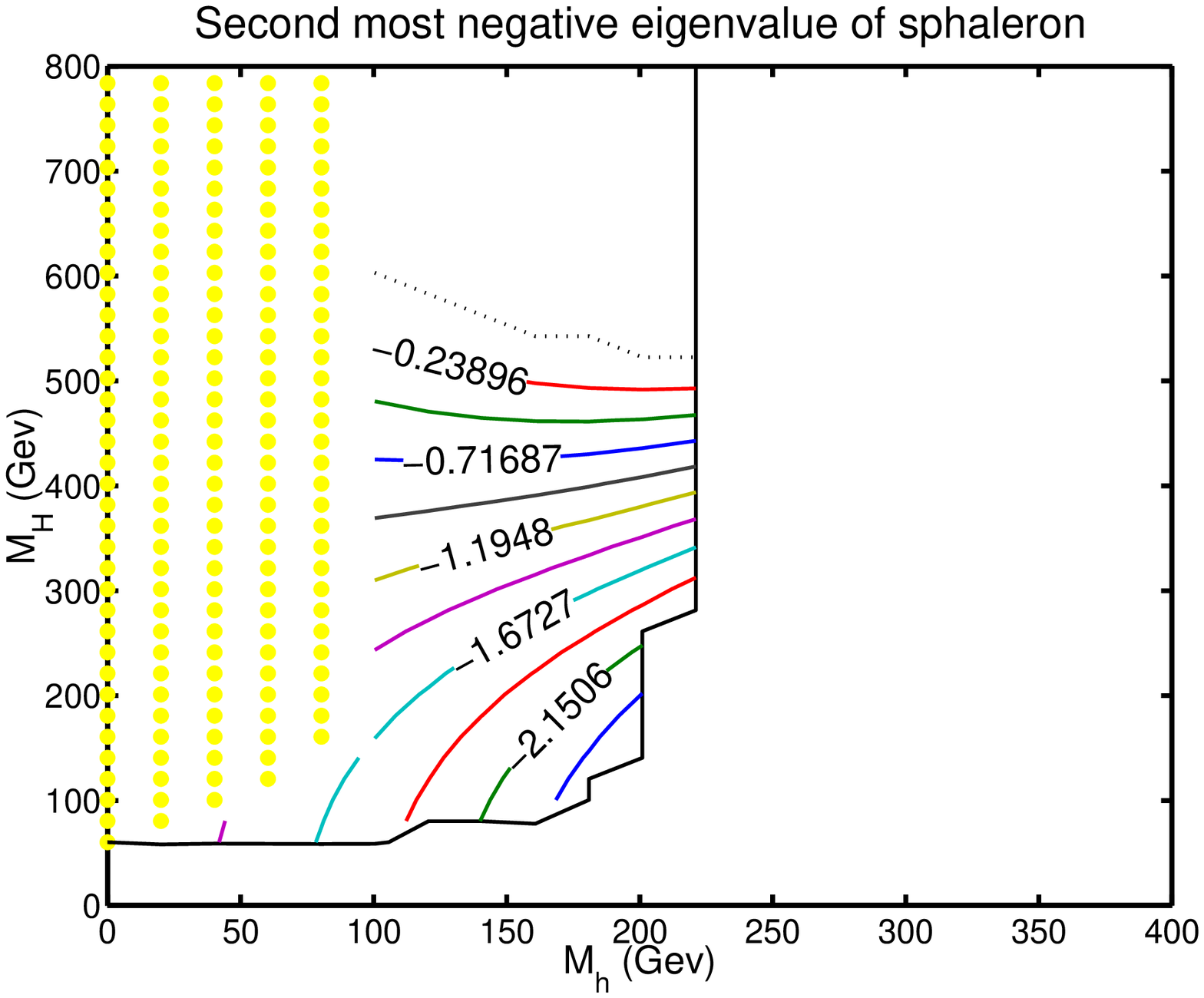}
\caption{\label{f:en4}\label{f:second_eig4} Top: contours in $M_h$, $M_H$ space of energy in units of $M_W/\alpha_W$
of the sphaleron
(dashes),and of the RW sphaleron (solid).
Bottom: contours in $M_h$, $M_H$ space of second negative eigenvalue ($M_W^2$) of the sphaleron.
Above the dotted line the sphaleron is the only solution, while below both
solutions exist. For the blank area Eq.\ \ref{e:minimum} is not the global minimum.
For the dotted area the potential is unbounded. The input parameters are $\theta_{CP}=0.49\pi$, $\phi=0.1\pi$, $\psi=0.0$,
$M_A=643$ GeV, $M_{H^\pm}=161$ GeV, and $\lambda_3$=3.0. $\tan\beta$=3.1.}
}
\end{figure}

\begin{figure}[!hbt]
\centering{
\includegraphics[height=8.0cm,width=9cm]{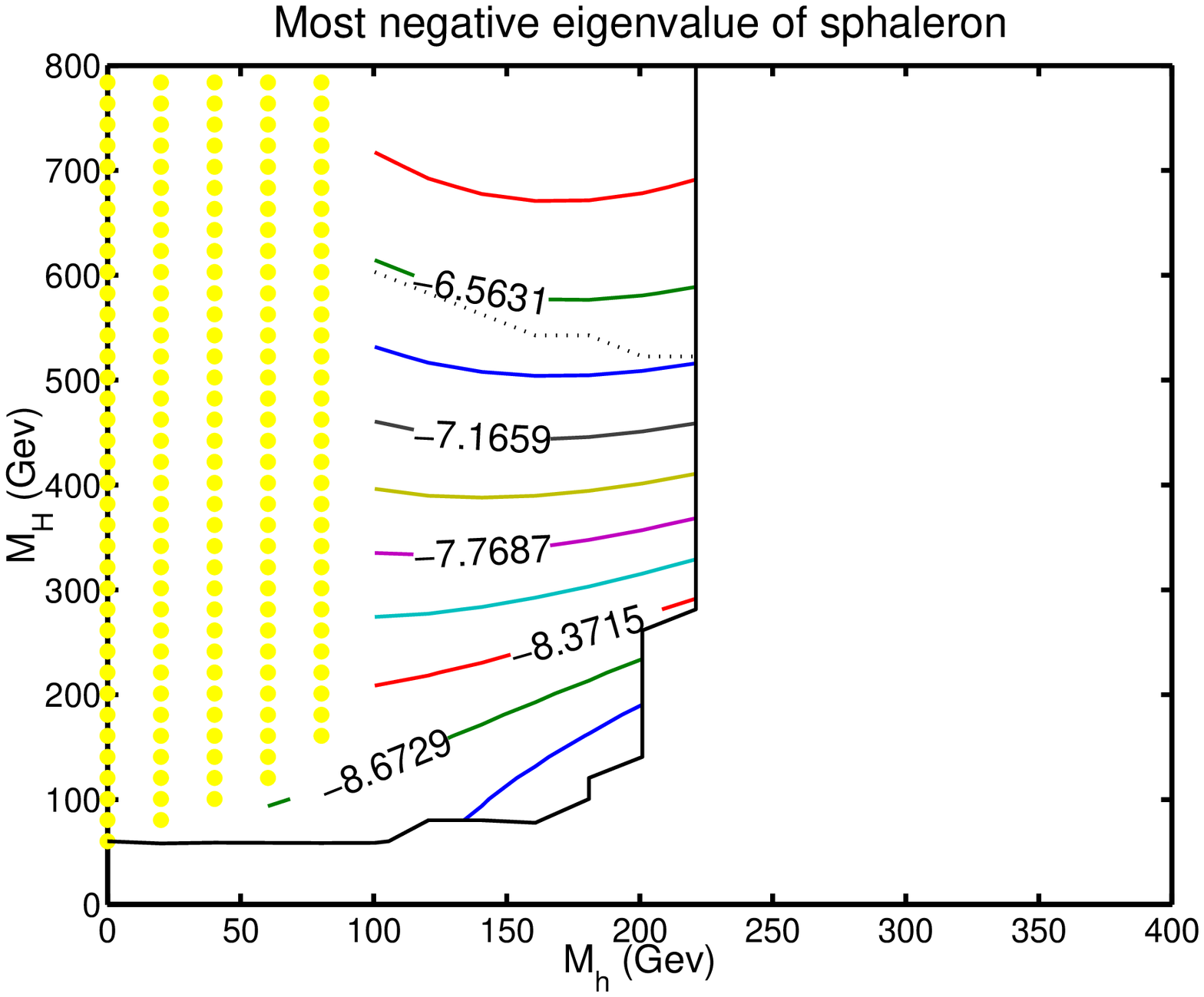}
\vspace{1.0cm}\\
\includegraphics[height=8.0cm,width=9cm]{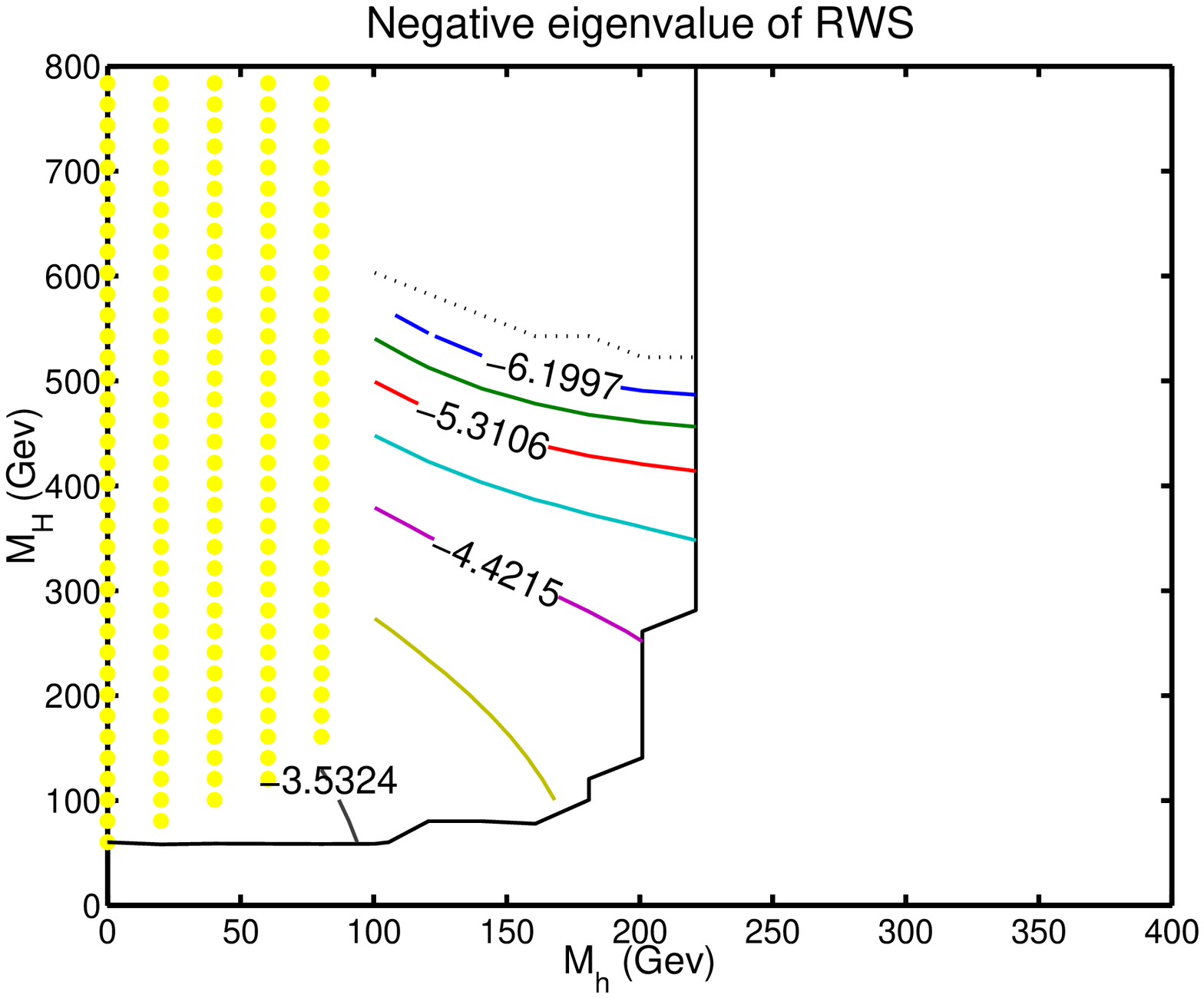}
\caption{\label{f:first_eig4}\label{f:eig_RWS4}Contours in $M_h$, $M_H$ space of the most negative eigenvalue ($M_W^2$)
of the sphaleron (top) and of the relative winding sphaleron (bottom).
Above the dotted line the sphaleron is the only solution, while below both solutions exist.
For the blank area Eq.\ \ref{e:minimum} is not the global minimum.
For the dotted area the potential is unbounded.
The input parameters are $\theta_{CP}$=0.49$\pi$, $\phi$=0.1$\pi$, $\psi$=0.0,
$M_A=643$ GeV, $M_{H^\pm}=161$ GeV, and $\lambda_3$=3.0. $\tan\beta$=3.1.}
}
\end{figure}

\begin{figure}[!hbt]
\centering{
\includegraphics[height=8cm,width=9cm]{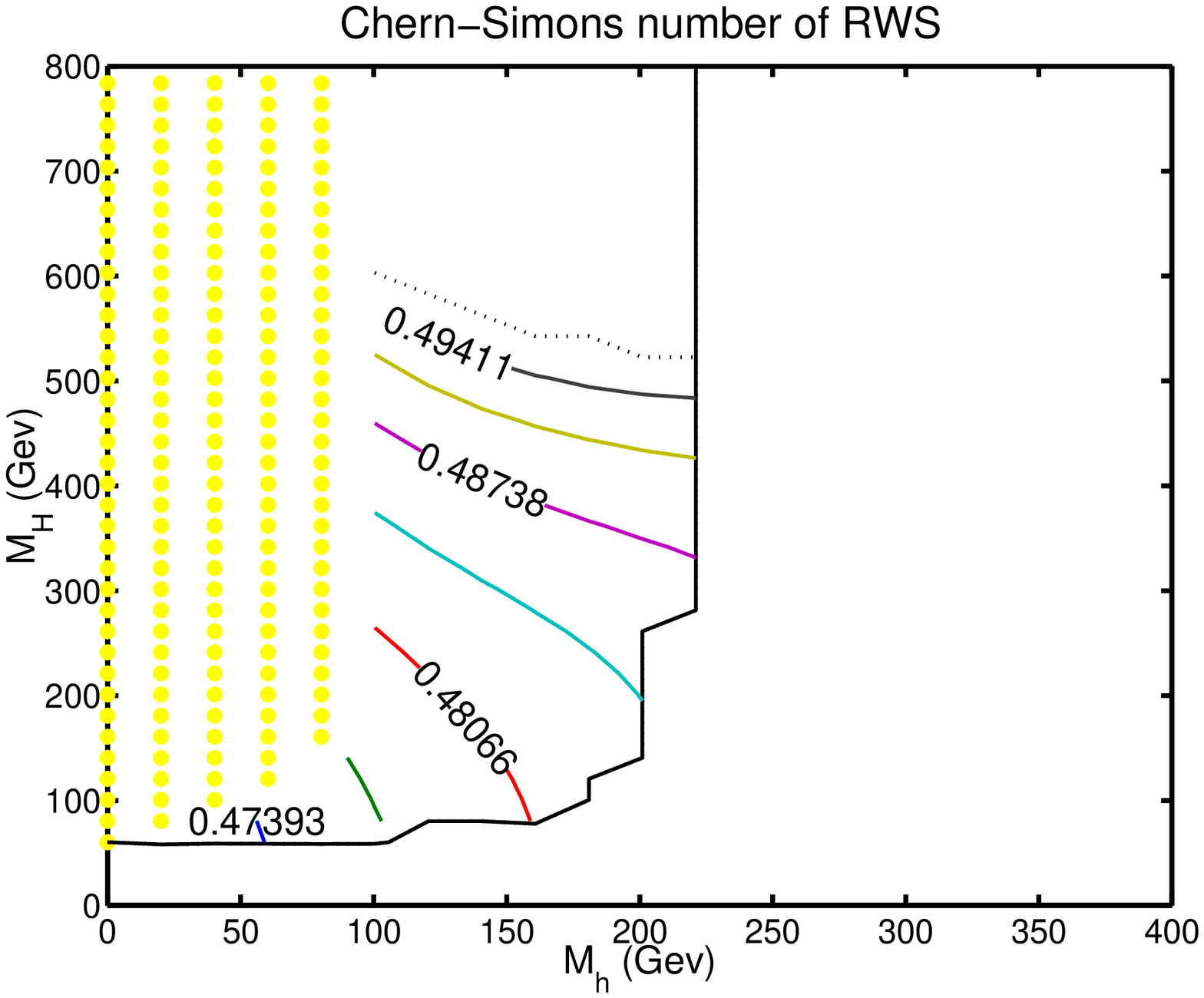}
\caption{\label{f:chern4} Contours in $M_h$, $M_H$ space of the Chern-Simons number of the RWS.
Above the dotted line only the sphaleron solution exists, with $n_{CS}=1/2$.
For the blank area Eq.\ \ref{e:minimum} is not the global minimum.
For the dotted area the potential is unbounded.
The input parameters are $\theta_{CP}$=0.49$\pi$, $\phi$=0.1$\pi$, $\psi$=0.0,
$M_A=643$ GeV, $M_{H^\pm}=161$ GeV, and $\lambda_3$=3.0.  $\tan\beta$=3.1.}}
\end{figure}

Figs.\ \ref{f:en4}--\ref{f:chern4}
show contours in $M_h$, $M_H$ space of
energy and 
second negative eigenvalue (Fig.\ \ref{f:second_eig4}),
most negative eigenvalue (Figs.\
\ref{f:first_eig4}) and Chern-Simons number (Fig.\
\ref{f:chern4}) of the sphaleron 
and relative winding sphaleron. 
Sphaleron contours are shown as dashed lines 
and RW sphaleron contours as solid when present on the same graph.

%%%%%%%%%%%%%%%%%%%%%%%%%%%%%%%%%%%%%%%%%%%%%%%%%%%%%%%%%%%%%%%%%%%%%%%%%%%%%%%%%%%%%%%
%%                                                                                   %%
%%   Results III: 5 figures.Contour in M_h M_H space.With CP violation               %%
%%                                                                                   %%
%%%%%%%%%%%%%%%%%%%%%%%%%%%%%%%%%%%%%%%%%%%%%%%%%%%%%%%%%%%%%%%%%%%%%%%%%%%%%%%%%%%%%%%

For these figures we took
$M_A=643$ GeV, $M_{H^{\pm}}=161$ GeV, this time with
$CP$ violation: $\theta_{CP}=0.49\pi$.
The remaining parameters were $\phi=0.1\pi$, $\psi$=0.0,  and $\lambda_3$=3.0,
giving $\tan\beta$=3.1.
We scanned through $M_h$ between 0 and 400 GeV, and $M_H$ between 0 and 800
GeV, with 20 GeV increments.
The dotted region at low $M_h$ was unbounded
according to Eqs.\ \ref{e:con_eig_param_1} and \ref{e:con_eig_param_2}, and for the white out area,
surrounded by the solid black line, the minimum of Eq.\ \ref{e:minimum}
was not the global minimum.

As with the previous contour plots,
a large region of parameter space contained relative winding sphalerons. For these input parameters, though, due to the large
$CP$ violating mixing angle, the role of the
large Higgs mass $M_H$ is taken on by $M_A$.
Since, from previous contour plots, the relative winding sphaleron solution prefers regions of parameter space
where there is a large separation in values of the heaviest (in this case the $M_A$) and the lightest
(in this case $M_h$, and $M_H$) Higgs masses,
the relative winding sphaleron solultions exist for the lower part of the contour plot, and not the upper part.
Referring to
Figs.\ \ref{f:en4}--\ref{f:chern4}: above the black dotted line the
sphaleron is the only solution, while below the black dotted line both the sphaleron and the relative winding sphaleron exist,
this is opposite
to the behaviour in the absence of $CP$ violation.

From Fig.\
\ref{f:en4} (top) the energy of the two solutions is as before almost the same.
The second negative eigenvalue of the sphaleron is
shown in the lower half of Fig.\ \ref{f:second_eig4}. The sphaleron does not develop a third
negative eigenvalue, nor the RW sphaleron
a second negative eigenvalue. We show the most negative eigenvalue of
the sphaleron and the RW sphaleron
(Fig.\ \ref{f:eig_RWS4}) on separate graphs,
and again their respective negative
eigenvalues can be very different at the same point in the contour plane.
We then show the Chern-Simons numbers for the RW sphaleron
in Fig.\ \ref{f:chern4}. Note that
we only show solutions with $n_{CS} \leq 1/2$: again, there
are parity conjugate partners to each of these RW sphalerons, and
the $n_{CS}$ of the RW sphaleron and of its parity partner add up to one.

There is no breaking in the degeneracy of the relative winding sphaleron pairs in energy, eigenvalues, or absolute
difference from 1/2 of Chern-Simons number, due to the presence of $CP$ violation. The solutions are not exactly spherically symmetric,
and have non zero values for all three of $K_1$, $V_1$, and $V_2$. The values of $K_1$, $V_1$, and $V_2$ as a percentage of
the Higgs potential energy are each never more than 0.5 \%.

\subsection{MSSM parameter space}
\begin{figure}[!hbt]
\centering{
\includegraphics[height=8cm,width=9cm]{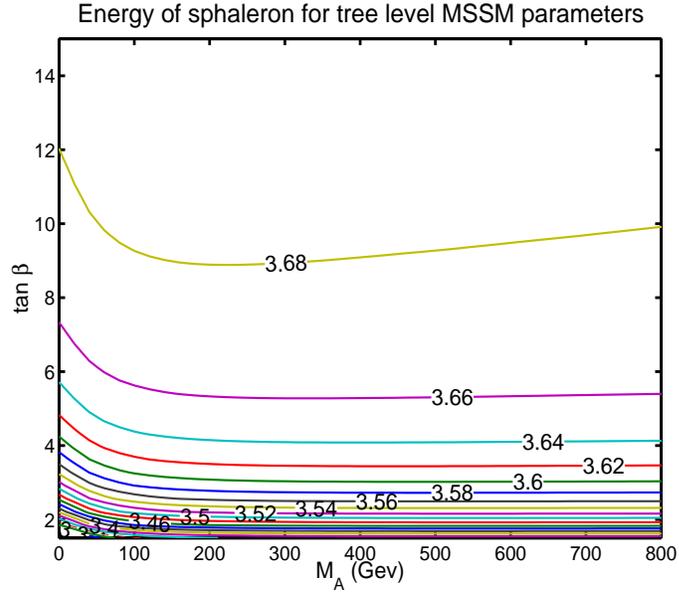}\
\vspace{2.0cm}\\
\includegraphics[height=8cm,width=9cm]{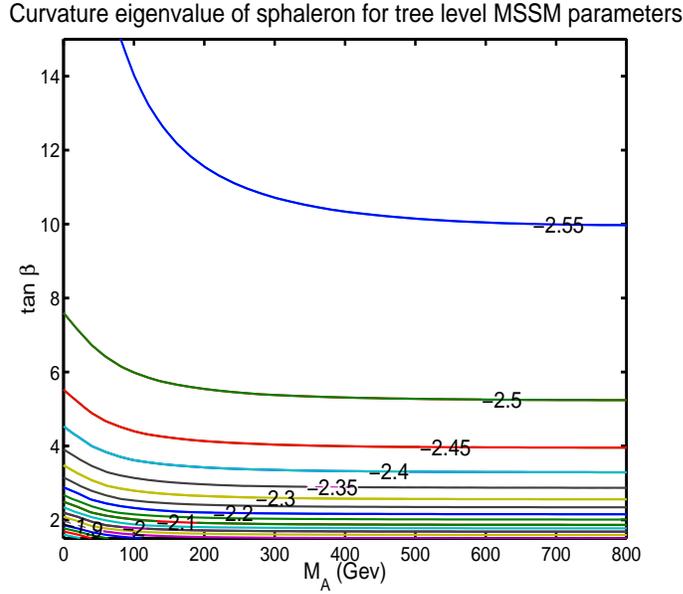}
\vspace{1.0cm}\\
\caption{\label{f:en_MSSM}\label{f:first_eig_MSSM} Contours in $M_A$, 
$\tan \beta$ space of the sphaleron for tree level MSSM parameters. The top figure shows 
energy ($M_W/\alpha_W$) of the sphaleron. The bottom figure shows 
negative curvature
eignevalue ($M_W^2$) of the sphaleron.}}
\end{figure}

\begin{figure}[!hbt]
\centering{
\includegraphics[height=8cm,width=9cm]{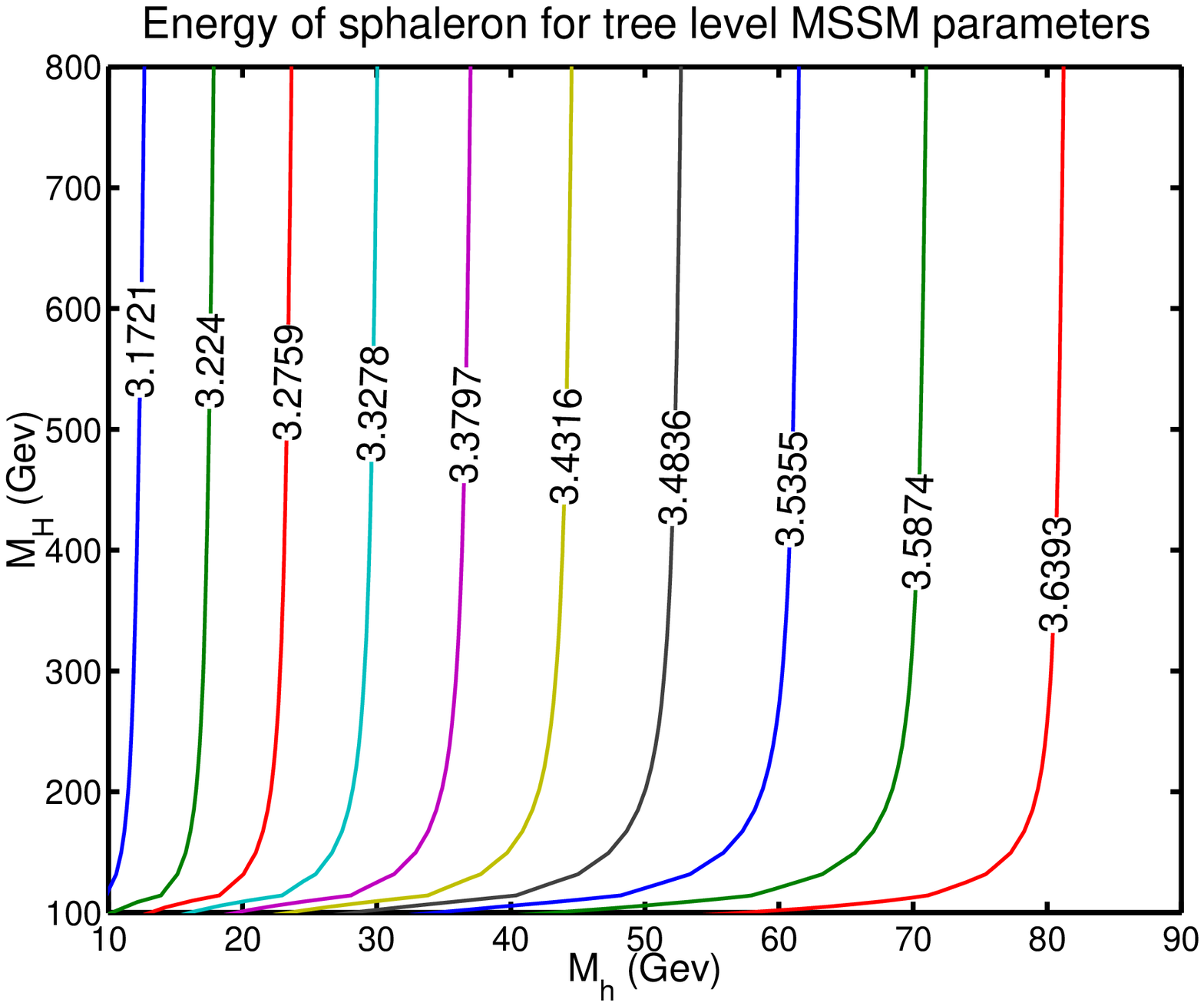}\
\vspace{2.0cm}\\
\includegraphics[height=8cm,width=9cm]{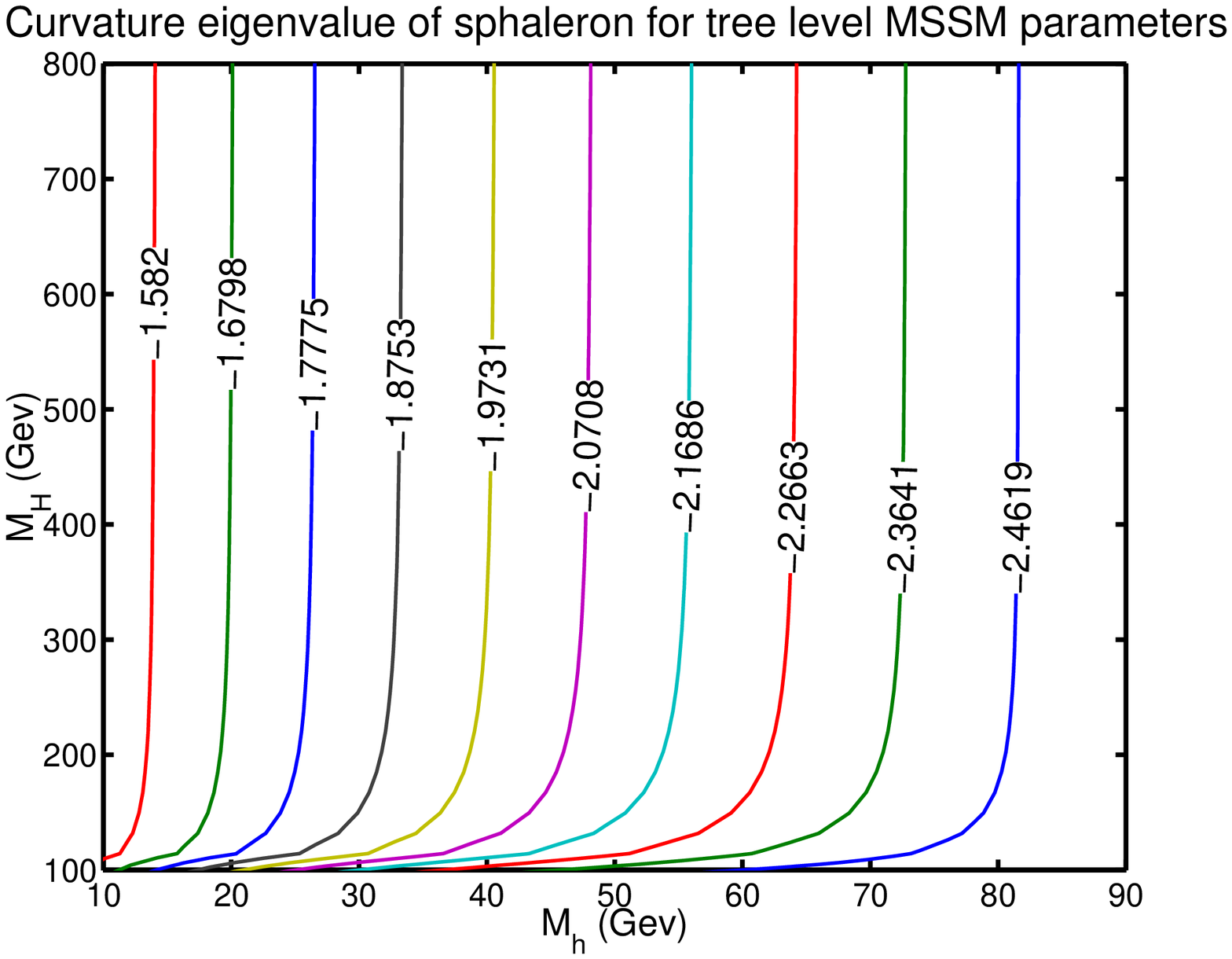}
\vspace{0.5cm}\\
\caption{\label{f:en_MSSM_2}\label{f:first_eig_MSSM_2}Contours in $M_h$, $M_H$ space of the energy ($M_W/\alpha_W$)
of the sphaleron for tree level MSSM parameters. The top figure shows 
energy ($M_W/\alpha_W$) of the sphaleron. The bottom figure shows 
negative curvature
eignevalue ($M_W^2$) of the sphaleron.}}
\end{figure}

Next
we scan through tree level MSSM parameter space. Fig.\
\ref{f:en_MSSM} shows
the scan in $M_A$, $\tan\beta$ space. Fig.\
\ref{f:first_eig_MSSM_2} shows
the scan in $M_h$, $M_H$ space. We plot contours
of energy (top) and negative eigenvalue (bottom) for each of these scans.

%%%%%%%%%%%%%%%%%%%%%%%%%%%%%%%%%%%%%%%%%%%%%%%%%%%%%%%%%%%%%%%%%%%%%%%%%%%%%%%%%%%%%%%%%%%%%
%%                                                                                         %%
%%   Results VI: 2 figures.Contours M_h,M_H, and tan beta, M_A for treelevel MSSM          %%
%%                                                                                         %%
%%%%%%%%%%%%%%%%%%%%%%%%%%%%%%%%%%%%%%%%%%%%%%%%%%%%%%%%%%%%%%%%%%%%%%%%%%%%%%%%%%%%%%%%%%%%%

For the range of parameters we show the sphaleron 
did not develop a second negative eigenvalue. There was no departure from spherical symmetry, 
as only the $a_{\alpha}$ field 
of the Higgs ansatz and the $\beta$ field of the gauge ansatz were ever
non-zero. From these four contours 
(Figs. \ref{f:en_MSSM} and \ref{f:first_eig_MSSM_2}) we agree with 
the general result of \cite{Moreno:1997zm} that the energy of 
the sphaleron is sensitive
to mainly $M_h$ and $\tan\beta$, although their results should be 
more accurate as they included 1-loop radiative corrections.
There were no relative winding sphalerons for the 
range of parameters explored.

\subsection{Sphaleron energy and $CP$ violation}
\begin{figure}[!hbt]
\centering{
\includegraphics[height=8.0cm,width=9cm]{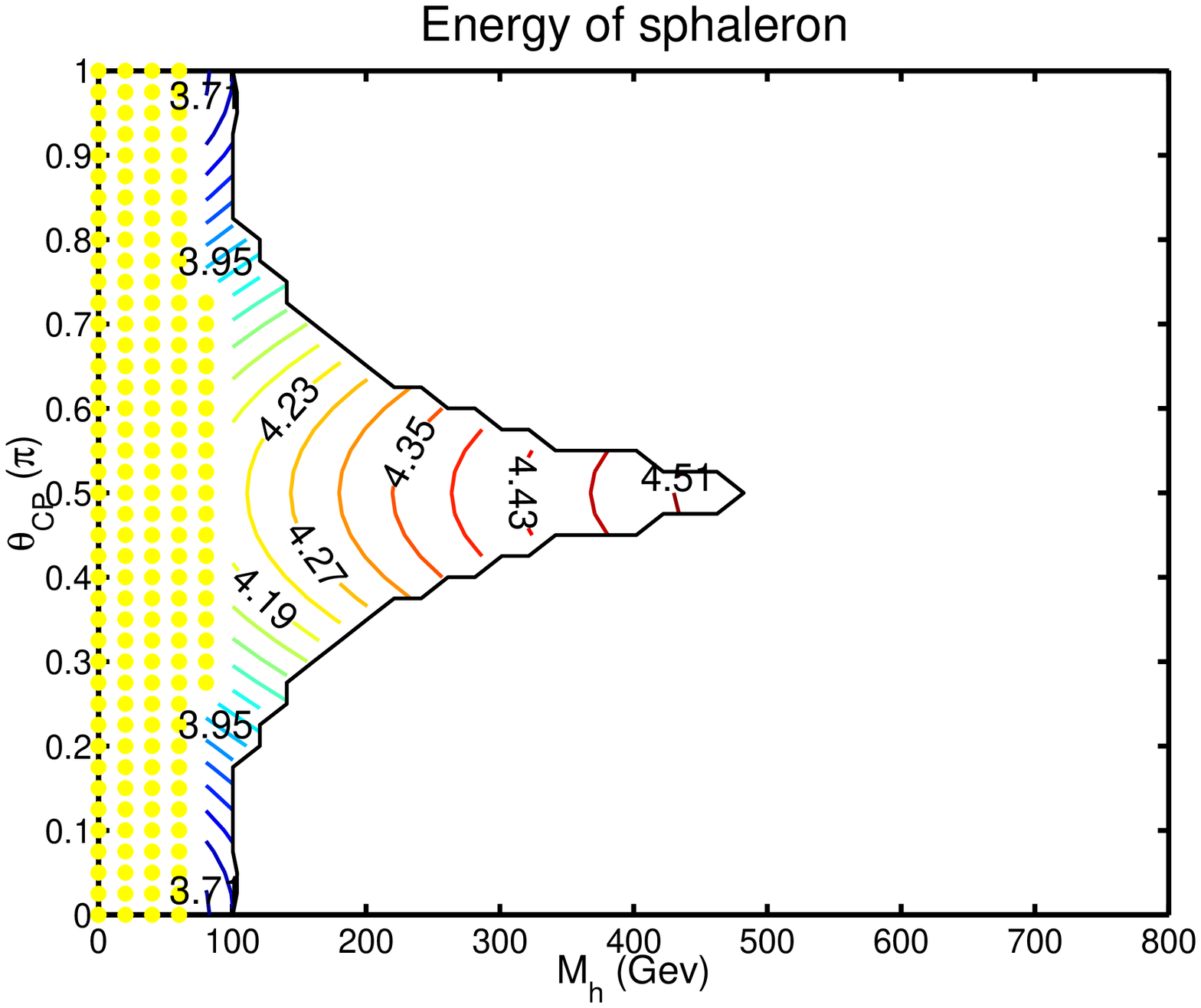}
\vspace{1.0cm}\\
\includegraphics[height=8cm,width=9cm]{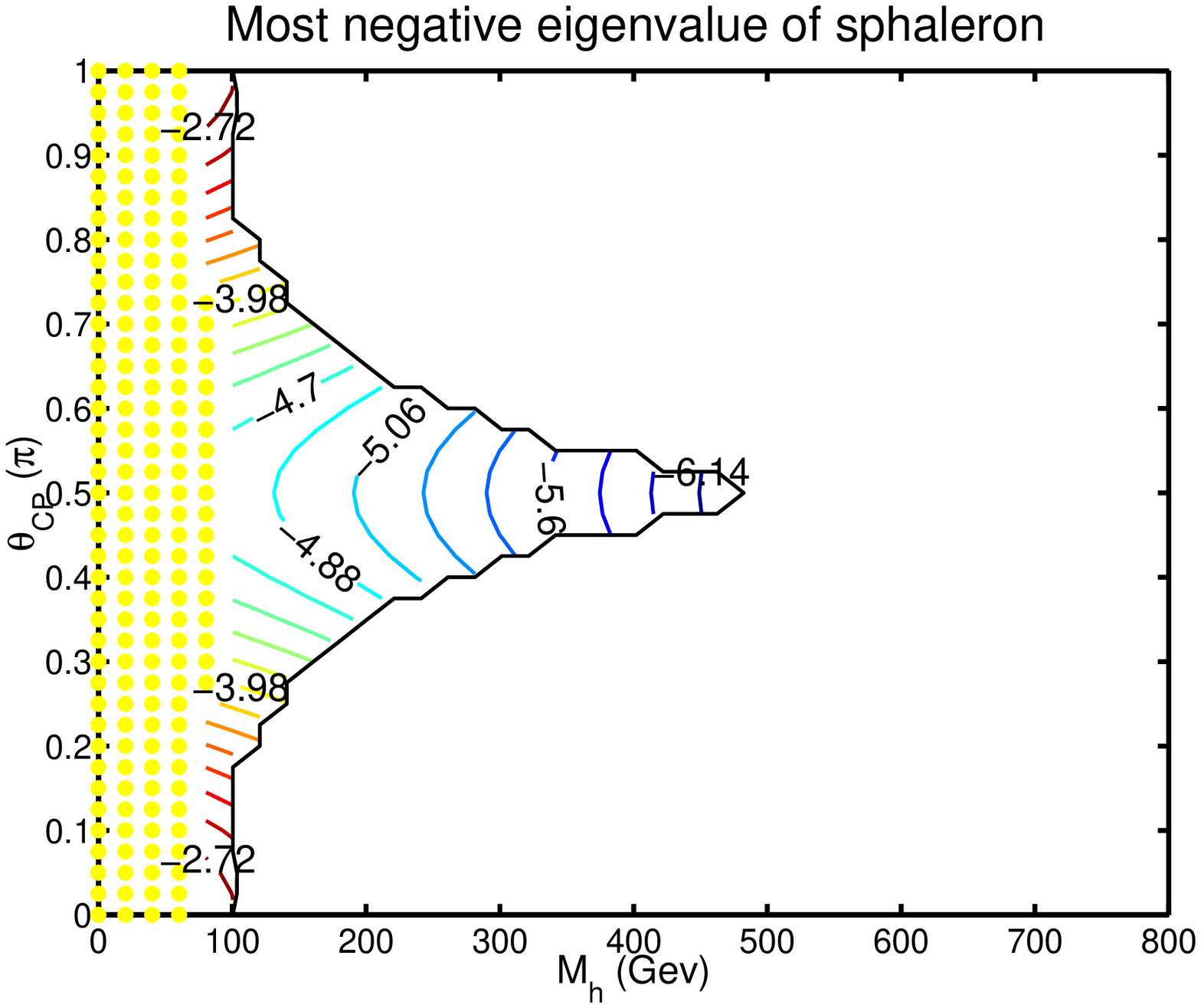}
\caption{\label{f:en5}\label{f:first_eig5} 
Top: contours in $M_h$, $\theta_{CP}$ space of the energy ($M_w/\alpha_w$) of sphaleron,
and bottom: of the negative eigenvalue of the sphaleron ($M_W^2$).
For this region of parameter space the sphaleron is the only solution. 
For the blank region Eq.\ \ref{e:minimum} is not the global minimum. 
For the dotted region the potential is unbounded. 
The input parameters are  $\phi$=0.125$\pi$, $\psi$=0.0, $M_H=110$ GeV, 
$M_A=500$ GeV, $M_{H^\pm}=500$ GeV, and $\lambda_3$=0.0.  $\tan\beta$=2.4.
}}
\end{figure}

We
recall that a $CP$ violating mixing angle can have a large effect on the properties of the 
sphaleron. Here (Fig. \ref{f:en5}) we scan through $M_h$, $\theta_{CP}$ space and show 
the energy of the sphaleron and the negative eigenvalue of the sphaleron for input parmaters 
$\phi$=0.125$\pi$, $\psi$=0.0, $M_H=110$ GeV, 
$M_A=500$ GeV, $M_{H^\pm}=500$ GeV, and $\lambda_3$=0.0, these give $\tan\beta$=2.4. For the dotted region 
at low $M_h$
the potential was unbounded, and for the blank region, bordered by the solid black line, the minimum of Eq.\ \ref{e:minimum}
was not the global minimum of the static energy functional.
For this region of parameter space the sphaleron never developed a second negative curvature eigenvalue. 

%%%%%%%%%%%%%%%%%%%%%%%%%%%%%%%%%%%%%%%%%%%%%%%%%%%%%%%%%%%%%%%%%%%%%%%%%%%%%%%%%%%%%%%%%%%%%
%%                                                                                         %%
%%   Results VII: 2 figures.Contours M_h,theta_CP, of energy and first eig of sphaleron    %%
%%                                                                                         %%
%%%%%%%%%%%%%%%%%%%%%%%%%%%%%%%%%%%%%%%%%%%%%%%%%%%%%%%%%%%%%%%%%%%%%%%%%%%%%%%%%%%%%%%%%%%%%

%fifth plots

The energy of the sphaleron (Fig. \ref{f:en5}: top) is dependent upon the value of the $CP$ violating mixing angle, 
and changes by about fourteen percent as the mixing angle varies between its minimum and its maximum. The energy is,
in the presence of $CP$ violation, still sensitive to the lightest Higgs mass. 

The negative eigenvalue (Fig. \ref{f:en5}: bottom) also has this strong dependence 
on the $CP$ violating mixing angle, with an increase of over fifty percent
as the mixing angle varies. Also the dependence on $M_h$, although not as dramatic as the effect of $CP$ violation, is still present.

\subsection{Field profiles}

\subsubsection{Sphaleron and RW sphaleron}

Next we show the field profiles for the sphaleron, relative winding sphaleron,
and conjugate relative winding sphaleron 
for a point in the contour
plot of Section \ref{ss:resCPvioln} corresponding to a $CP$ violating theory 
with 
$M_A=8M_W$, $M_{H^\pm}=2M_W$, $M_h=1.25M_W$,
and $M_H=1.5M_W$.  We recall that the mixing angles were 
$\theta_{CP}$=0.49$\pi$, $\phi$=0.1$\pi$, $\psi$=0.0, and the 
coupling $\la_3=3.0$.

\begin{figure}[!hbt]
\centering{
\includegraphics[height=8cm,width=9cm]{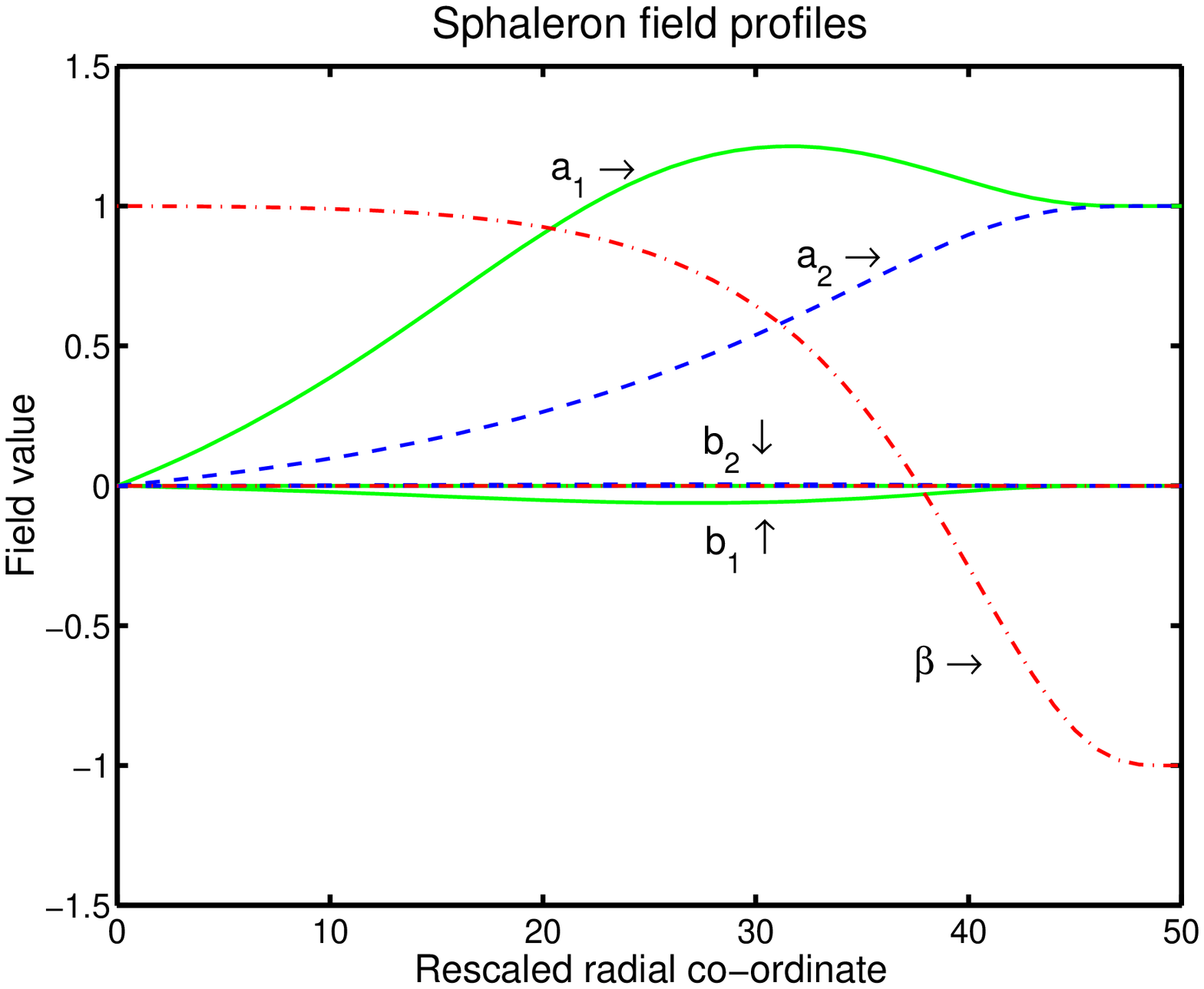}
\vspace{2.0cm}\\
\includegraphics[height=5.0cm,width=9cm]{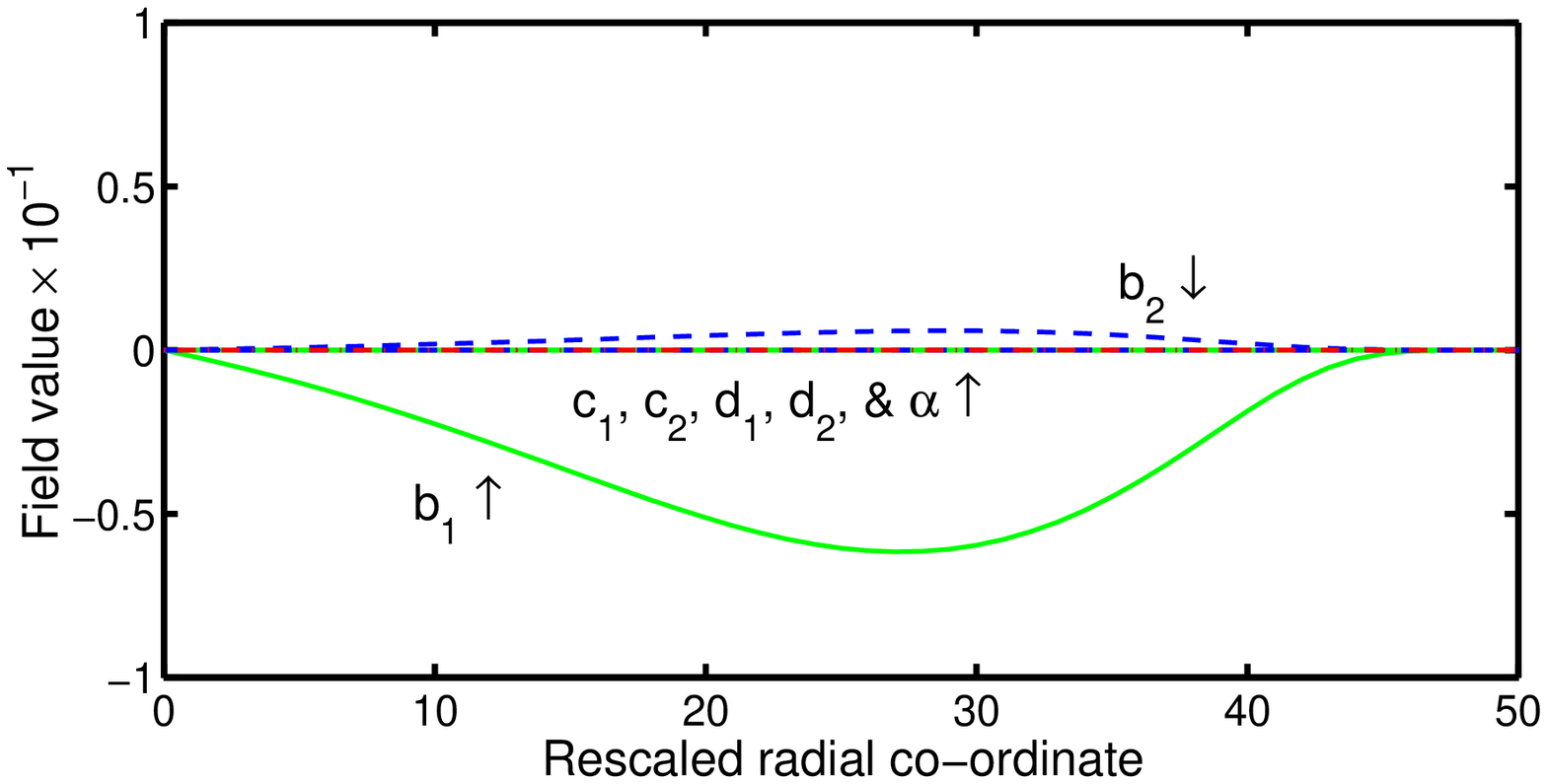}
\vspace{2.0cm}\\
\caption{\label{f:sph1}\label{f:sph2}The sphaleron field profiles (top), and 
the profiles for $b_1$ and $b_2$ in more detail (bottom). $c_{\al}=d_{\al}=\alpha=0$.
This configuration has energy=$4.053$ $M_W /\al_W$, $n_{CS}$=1/2, 
and two negative curvature eigenvalues $-8.696 M_W^2$, and $-1.754 M_W^2$. 
Input parameters are: $\theta_{CP}$=0.49$\pi$, $\phi$=0.1$\pi$, $\psi$=0.0, $M_h=101$ GeV, $M_H=121$ GeV, 
$M_A=643$ GeV, $M_{H^\pm}=161$ GeV, and $\lambda_3$=3.0.
These give $\tan\beta$=3.1, $\lambda_1=26.29$, $\lambda_2=-2.59$, $\lambda_+=0.91$, $\lambda_4=0.85$, 
$\chi_1=0.42$, and $\chi_2=0.41$.}
}
\end{figure}

\begin{figure}[!hbt]
\centering{
\includegraphics[height=8cm,width=9cm]{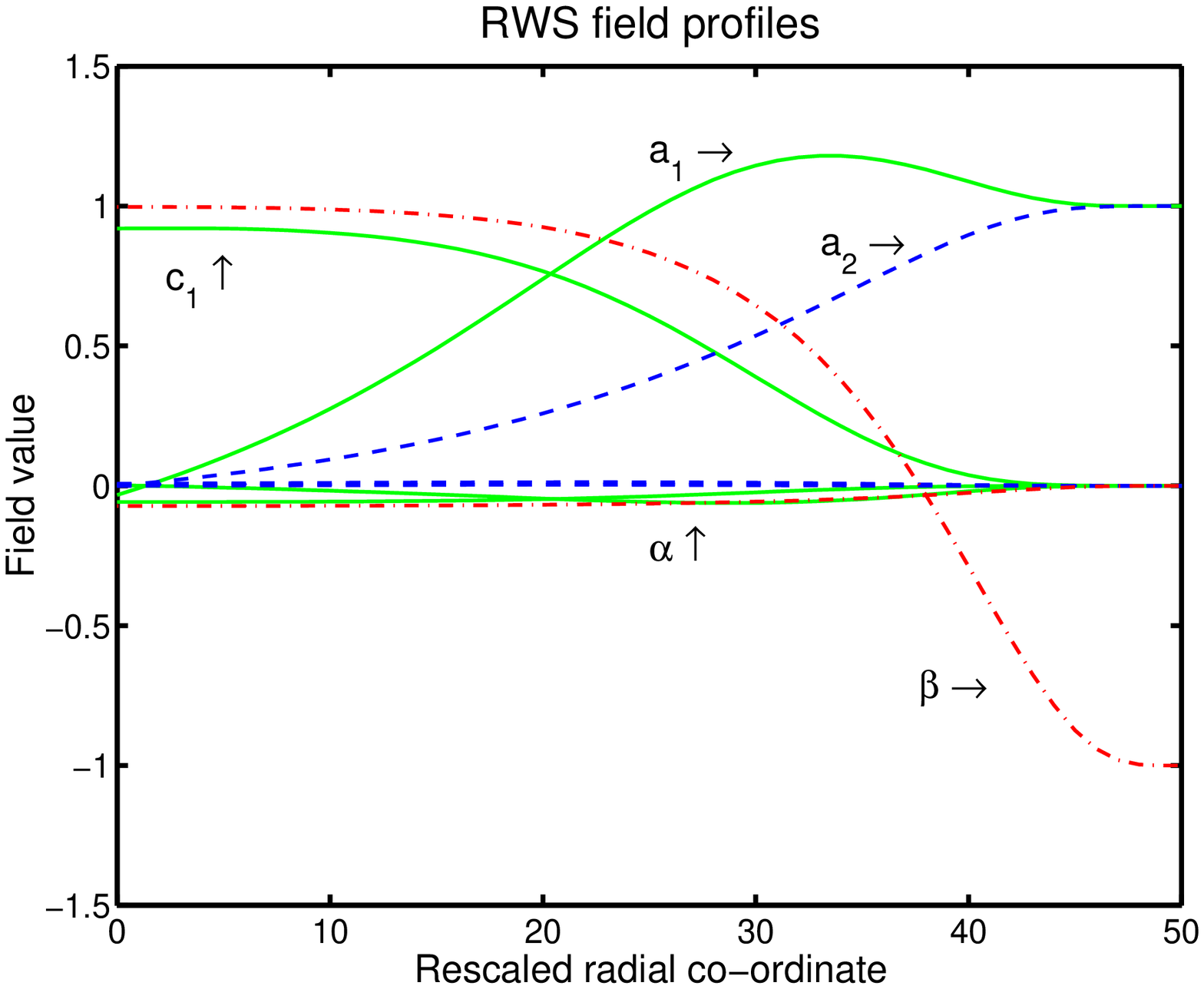}
\vspace{2.0cm}\\
\includegraphics[height=5.0cm,width=9.0cm]{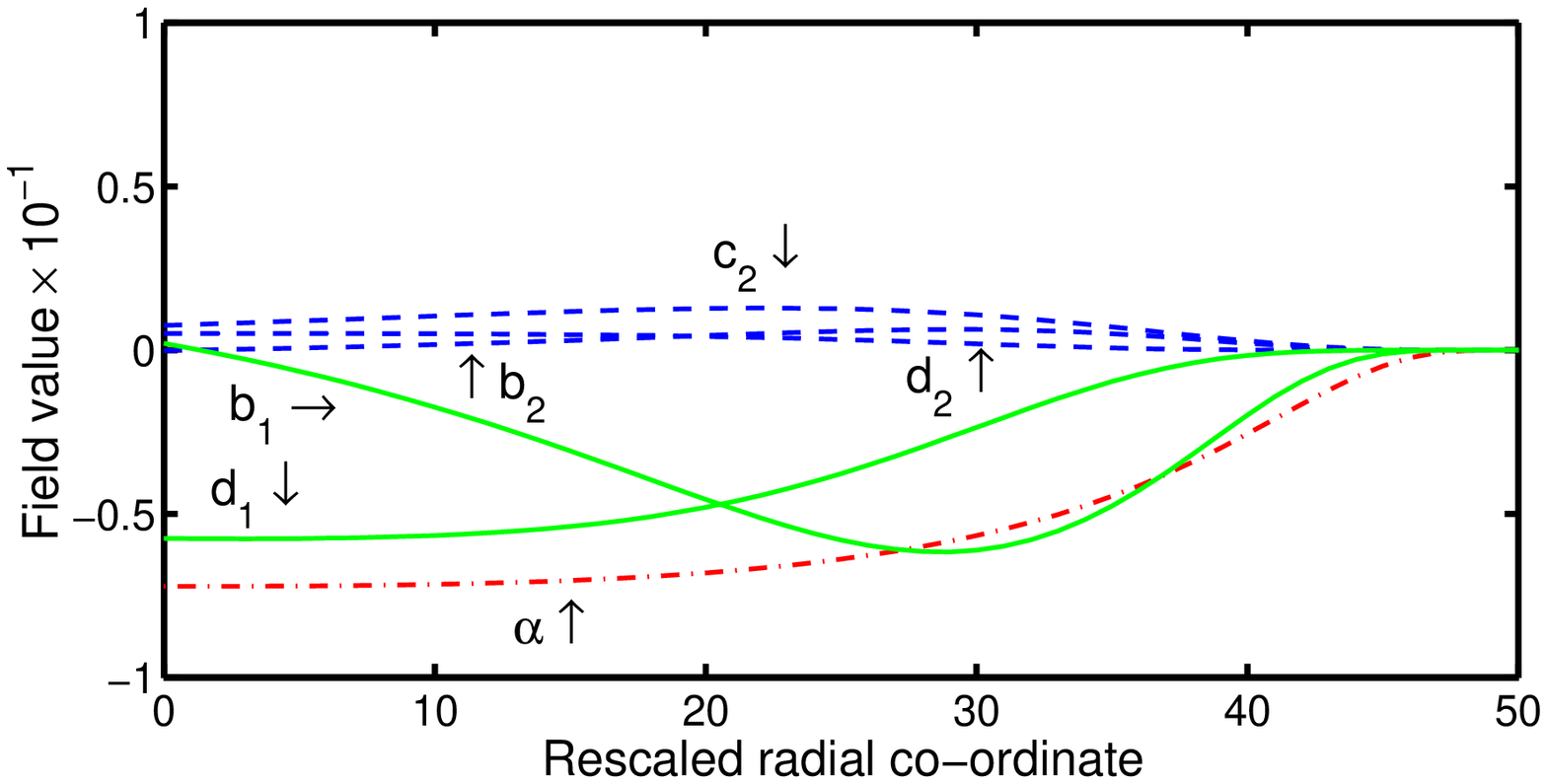}
\vspace{2.0cm}\\
\caption{\label{f:RWS1}\label{f:RWS2} 
The RW sphaleron field profiles (top) and the profiles for $b_1$, $b_2$, $c_2$, $d_1$, $d_2$, and $\alpha$ in more detail (bottom).
This configuration has energy=$4.047$ $M_W /\al_W$, $n_{CS}$=0.478, and one negative curvature 
eigenvalue $-3.637 M_W^2$.
Input parameters are: $\theta_{CP}$=0.49$\pi$, $\phi$=0.1$\pi$, $\psi$=0.0, $M_h=101$ GeV, $M_H=121$ GeV, 
$M_A=643$ GeV, $M_{H^\pm}=161$ GeV, and $\lambda_3$=3.0.
These give $\tan\beta$=3.1, $\lambda_1=26.29$, $\lambda_2=-2.59$, $\lambda_+=0.91$, $\lambda_4=0.85$, 
$\chi_1=0.42$, and $\chi_2=0.41$.}}
\end{figure}

\begin{figure}[!hbt]
\centering{
\includegraphics[height=8cm,width=9cm]{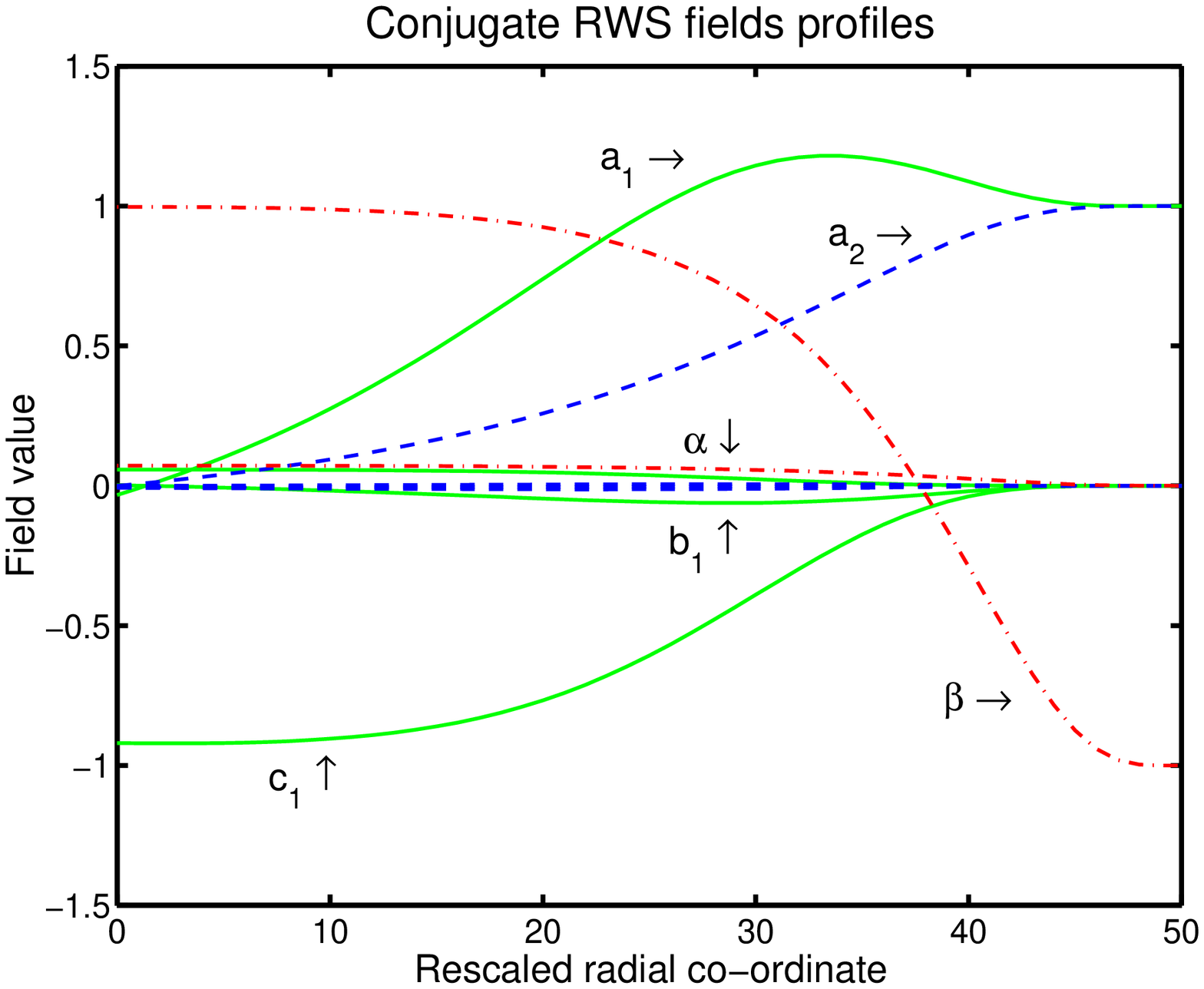}
\vspace{2.0cm}\\
\includegraphics[height=5.0cm,width=9.0cm]{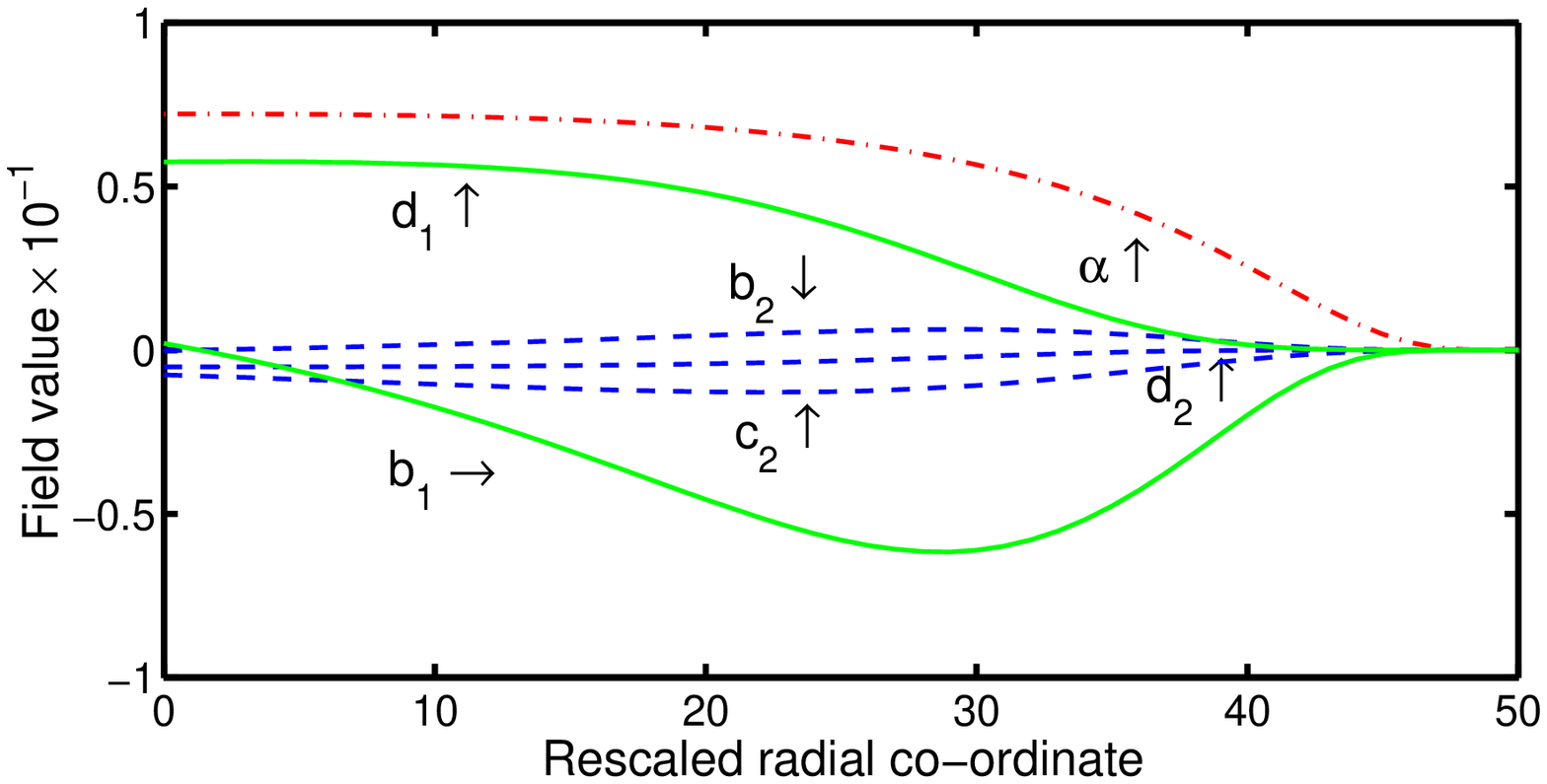}
\vspace{2.0cm}\\
\caption{\label{f:conRWS2}\label{f:conRWS1} 
The conjugate RW sphaleron 
field profiles (top), and the profiles for $b_1$, $b_2$, $c_2$, $d_1$, $d_2$, and $\alpha$ in more detail (bottom).
This configuration has energy=$4.047$ $M_W /\al_W$, $n_{CS}$=0.522, and 
one negative curvature eigenvalue $-3.637 M_W^2$.
Input parameters are: $\theta_{CP}$=0.49$\pi$, $\phi$=0.1$\pi$, $\psi$=0.0, $M_h=101$ GeV, $M_H=121$ GeV, 
$M_A=643$ GeV, $M_{H^\pm}=161$ GeV, and $\lambda_3$=3.0.
These give $\tan\beta$=3.1, $\lambda_1=26.29$, $\lambda_2=-2.59$, $\lambda_+=0.91$, $\lambda_4=0.85$, 
$\chi_1=0.42$, and $\chi_2=0.41$.}
}
\end{figure}

\begin{figure}[!hbt]
\centering{
\includegraphics[height=8cm,width=9cm]{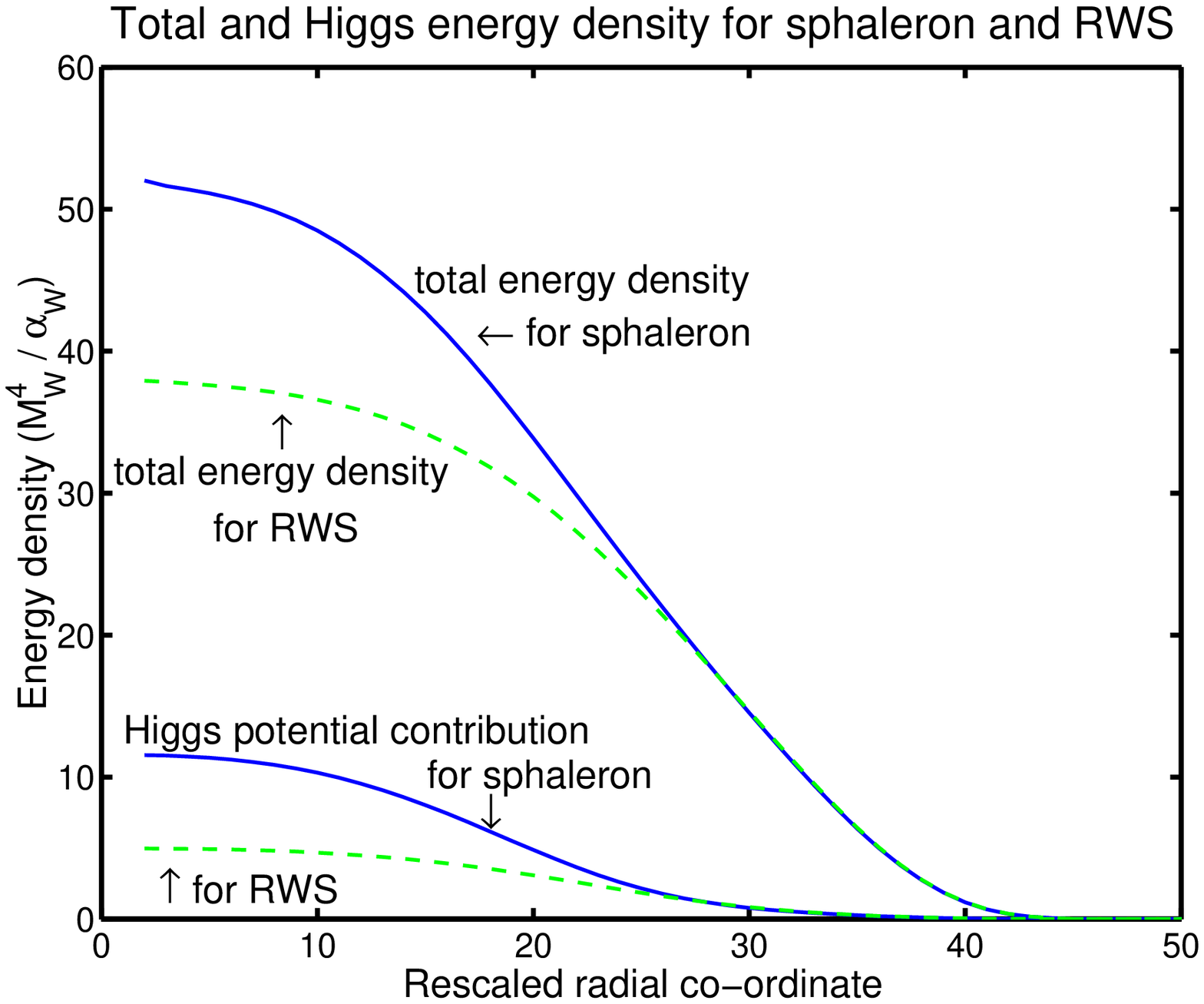}
\vspace{2.0cm}\\
\includegraphics[height=5.0cm,width=9.0cm]{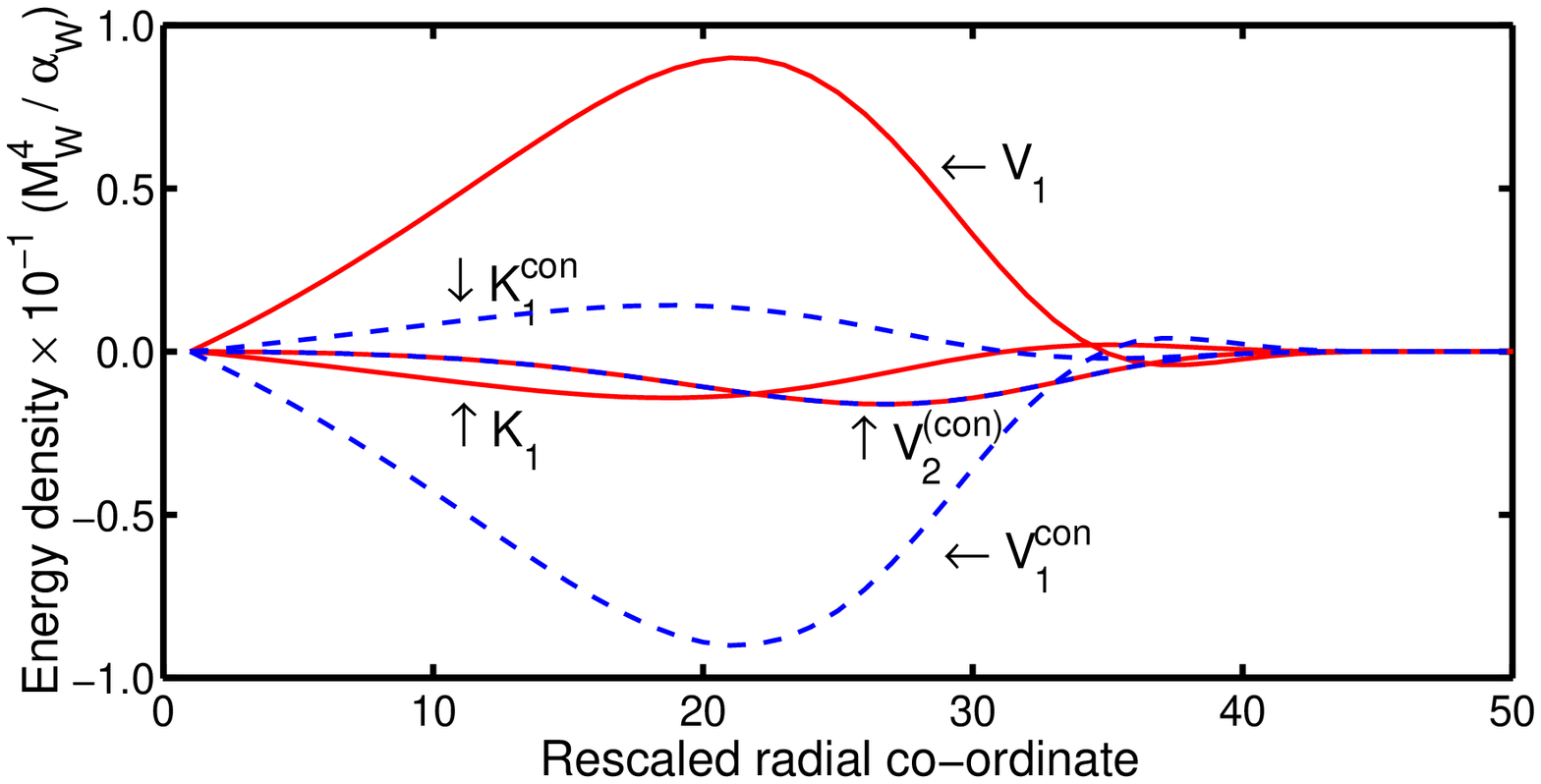}
\vspace{1.0cm}\\
\caption{\label{f:en_ss_rws}\label{f:en_den_rws} The top of the figure shows the total and the Higgs potential contribution 
to energy density in units of $M_W^4/\al_W$ for the sphaleron (solid) and the RWS (dashes). The bottom figure shows
$K_1$, $V_1$, and $V_2$ for the RWS (solid) and its conjugate (dashes) in the same units.
Both $K_1$ and $V_1$ are equal to their values for conjugate solutions, but have opposite sign. 
$V_2$ is equal to its value for the conjugate solution.
Input parameters are: $\theta_{CP}$=0.49$\pi$, $\phi$=0.1$\pi$, $\psi$=0.0, $M_h=101$ GeV, $M_H=121$ GeV, 
$M_A=643$ GeV, $M_{H^\pm}=161$ GeV, and $\lambda_3$=3.0. 
These give $\tan\beta$=3.1, $\lambda_1=26.29$, $\lambda_2=-2.59$, $\lambda_+=0.91$, $\lambda_4=0.85$, 
$\chi_1=0.42$, and $\chi_2=0.41$.}
}
\end{figure}

Before we proceed we check whether this point in parameter space is
phenomenologically viable at zero temperature, as $M_h=1.25M_W$ is ruled out if
the $hZZ$ coupling is too large. 
We calculate the couplings $g_{hZZ}$, $g_{HZZ}$, and $g_{AZZ}$ according to 
\cite{Carena:2000yi}
%http://delphiwww.cern.ch/~pubxx/www/delsec/papers/public/papers.html#0274
using the values of input parameters used in 
Figs.\ \ref{f:sph1}--\ref{f:en_den_rws}, 
and compare them with the latest particle data \cite{Bock:2000gk}.

Using
\bea
g_{hZZ}&=& D[1,1]\cos\beta  +D[2,1]\sin\beta \\
g_{HZZ}&=& D[1,2]\cos\beta  +D[2,2]\sin\beta \\
g_{AZZ}&=& D[1,3]\cos\beta  +D[2,3]\sin\beta 
\eea
where $D$ is given by Eq.\ \ref{e:tot_mixing}, we obtain, for the paramaters of figures \ref{f:sph1}-\ref{f:en_den_rws}
\bea
g^2_{hZZ}&=& 0.081 \\
g^2_{HZZ}&=& 0.824 \\
g^2_{AZZ}&=& 0.095 
\eea
which for masses $M_h=101$ GeV, $M_H=121$ GeV, and $M_A=643$ GeV are with in experimental bounds. 
Although we have 
labelled the Higgses with subscripts $h$, $H$, and $A$; because of the values of the mixings 
$\phi=0.1\pi$, $\theta_{CP}=0.49\pi$, $\psi=0.0$, while the particle with subscript $h$ is $CP$ even, 
those with subscript $H$, and $A$ are a mix of $CP$ even and $CP$ odd.

%%%%%%%%%%%%%%%%%%%%%%%%%%%%%%%%%%%%%%%%%%%%%%%%%%%%%%%%%%%%%%%%%%%%%%%%%%%%%%%%%%%%%%%%%%%%%
%%                                                                                         %%
%%   Results IV: 4 figures.Field profiles,energy density. For point in previous conours    %%
%%                                                                                         %%
%%%%%%%%%%%%%%%%%%%%%%%%%%%%%%%%%%%%%%%%%%%%%%%%%%%%%%%%%%%%%%%%%%%%%%%%%%%%%%%%%%%%%%%%%%%%%

%fields profile RWS plots 

%end fields profile RWS plots 
We then plot the energy density of the two types of solution, and
the values of $K_1$, $V_1$, and $V_2$ as a function of the rescaled radial co-ordinate for the sphaleron, RW
sphaleron, and conjugate RW sphaleron.
We recall that the departure of $K_1$, $V_1$, and $V_2$ from zero signals the
breakdown of the spherically symmetric ansatz, and their size relative to the
total energy density indicates the seriousness of the breakdown.

It is convenient to plot the field values rescaled according to 
\bea
f_G=\frac{f_G}{\sqrt{2}}, 
~~~~ f_H=\frac{\upsilon}{\upsilon_{\alpha} } \frac{f_H }{2},
\eea
as then the asymptotic values are either 0 or $\pm1$.

The ordinary sphaleron field profiles are plotted in Fig.\ \ref{f:sph1}
as a function of the rescaled radial points. 
The solution has non zero values of $a_{\alpha}$, $b_{\alpha}$, and $\beta$ as expected for a field configuration that
preserves $P$ but violates $C$, due to the presence of a $C$ violating parameter in the potential.
 The sphaleron has Chern-Simons number 1/2, two negative eigenvalues (-8.696
$M^2_W$, and -1.754 $M^2_W$),
and has energy $4.053M_W /\al_W$. 

The relative winding sphaleron field configurations, shown in Fig.\
\ref{f:RWS1}, have non zero values for all fields. 
The solution violates $P$ spontaneously and $C$ explicitly, and violates the combination $CP$. 
It has one negative eigenvalue (-3.637 $M^2_W$), energy less than its 
ordinary sphaleron ($4.047M_W
/\al_W$), and Chern-Simons number 0.478. Its parity congugate partner, shown in Figure \ref{f:conRWS1},
has field profiles identical to a $P$ transformation of the RWS: that is 
$c_{\alpha}\rightarrow -c_{\alpha}$, $d_{\alpha}\rightarrow -d_{\alpha}$, and $\alpha\rightarrow -\alpha$, with all other 
fields remaining unchanged. The solution has identical energy, and eigenvalue to its $P$ conjugate solution,
and its Chern-Simons 
number is 0.522.

Next we show (Fig. \ref{f:en_den_rws}: top) the energy density of the sphaleron, and the RW
sphaleron, 
and in detail (Fig. \ref{f:en_ss_rws}: bottom) 
the values of $K_1$, $V_1$, and $V_2$ for the RWS in units of energy density. 
$K_1$ and $V_1$  are equal in value, but opposite in sign 
for the conjugate pair, $V_2$ is equal in value and equal in sign. These deviations from spherical symmetry are 
of order one part in $10^3$ for these values of parameters.

\subsubsection{Bisphaleron}

%%%%%%%%%%%%%%%%%%%%%%%%%%%%%%%%%%%%%%%%%%%%%%%%%%%%%%%%%%%%%%%%%%%%%%%%%%%%%%%%%%%%%%%%%%%%%
%%                                                                                         %%
%%   Results V: 2 figures.Field profiles,energy density for bisphaleron                    %%
%%                                                                                         %%
%%%%%%%%%%%%%%%%%%%%%%%%%%%%%%%%%%%%%%%%%%%%%%%%%%%%%%%%%%%%%%%%%%%%%%%%%%%%%%%%%%%%%%%%%%%%%
%fields profile bisphaleron plots 

\begin{figure}[!hbt]
\centering{
\includegraphics[height=8cm,width=9cm]{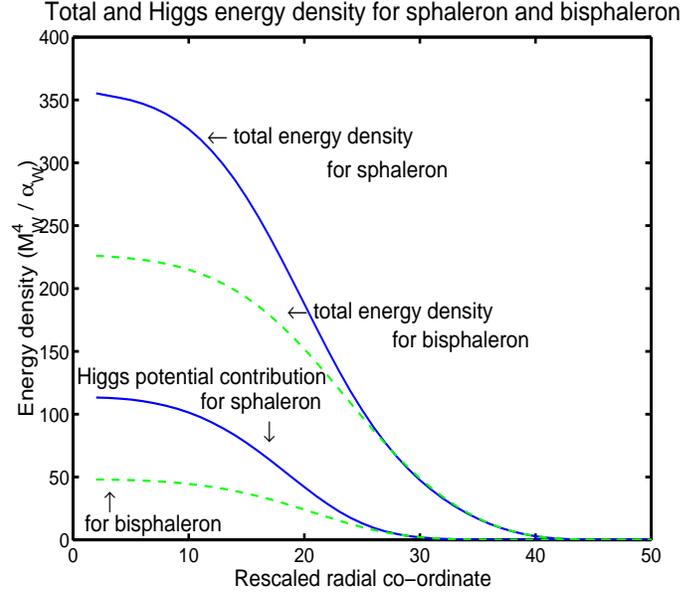}
\vspace{2.0cm}\\
 \includegraphics[height=5.0cm,width=9.0cm]{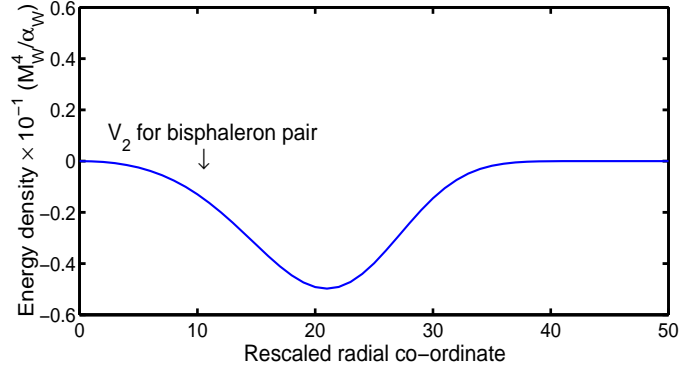}
\vspace{2.0cm}\\
\caption{\label{f:bisphal_ss}\label{f:bisphal_en} The top figure shows total and Higgs potential contribution to energy density 
($M_W^4/\al_W$)
for the sphaleron (solid) and the bisphaleron (dashes).
The bottom figure shows $V_2$ for the bisphaleron solution and its conjugate.
$V_2$ for both the bisphaleron and conjugate solution are equal.
Input parameters are: $\tan\beta$=6.0, $\theta_{CP}$=0.0, $\phi$=0.0, $\psi$=0.0, $M_h$=15.0$M_W$, $M_H$=17.0$M_W$, 
$M_A$=2.0$M_W$, $M_{H^\pm}$=3.0$M_W$, and $\lambda_3$=-0.1.}
}
\end{figure}

\begin{figure}[!hbt]
\centering{
\includegraphics[height=8cm,width=9cm]{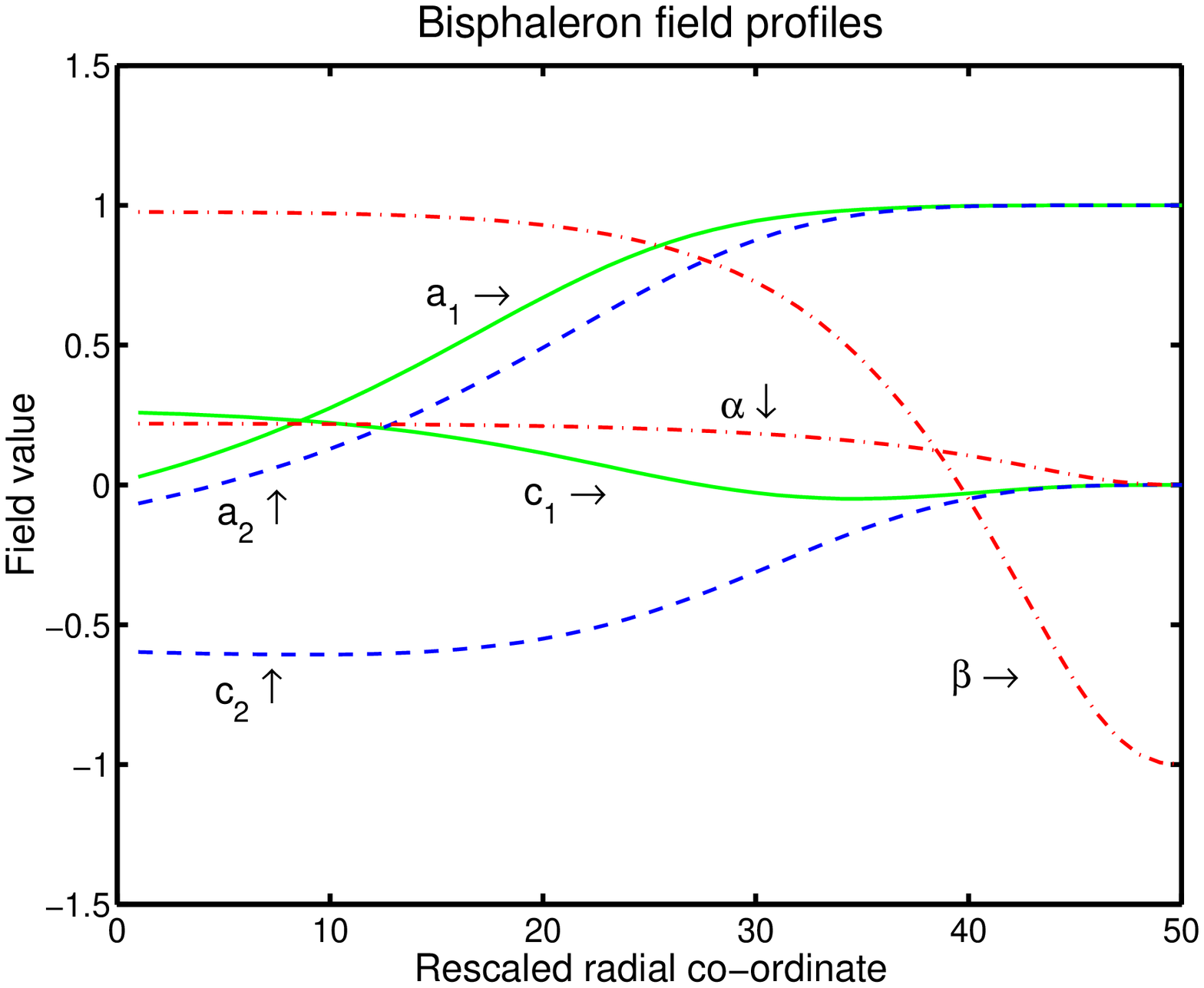}
\caption{\label{f:bisphal}The bisphaleron field profiles for 
$\tan\beta$=6.0, $M_h$=15.0$M_W$, $M_H$=17.0$M_W$, $M_A$=2.0$M_W$,
$M_{H^{\pm}}=3.0M_W $ and $\lambda_3$=-0.1. It has energy=$4.932$ $M_W /\al_W$, $n_{CS}$=0.569,
two negative curvature eigenvalues $-11.915M_W^2$, and $-6.788 M_W^2$. 
Its conjugate partner is the identical solution under $P$ conjugation
($\alpha\rightarrow - \al$), and has $n_{CS}$=0.431.
}}
\end{figure}

%fields profile bisphaleron plots

For completeness we detail the bisphaleron fields profiles for non zero $M_A$ and $M_{H^{\pm}}$, and show their departure from 
spherical symmetry. Figs. \ref{f:bisphal_ss} and \ref{f:bisphal} concern this bisphaleron. 
We have chosen masses which are perhaps unrealistically 
large, in order to reach the part of parameter space where the bisphaleron
exists: 
$\tan\beta$=6.0, $M_h$=15.0$M_W$, $M_H$=17.0$M_W$, $M_A$=2.0$M_W$,
$M_{H^{\pm}}$=3.0$M_W$ and $\lambda_3$=-0.1, with no CP violation. For these input parameters 
$\lambda_1=567.6$, $\lambda_2=12.4$, $\lambda_+=0.627$, $\lambda_4=1.923$, 
$\chi_1=-0.227$, and $\chi_2=0.0$.

The energy density and departure from spherical symmetry are shown 
in Figure \ref{f:bisphal_en}.
The $CP$ invariance means that $b_{\alpha}=d_{\alpha}=0$,
and hence $K_1$ and $V_1$ vanish. The departure from spherical symmetry is entirely in the $V_2$ term 
shown in units of energy density
($M_W^4/\al_W$)
in the lower half of Fig. \ref{f:bisphal_en}. The departure from spherical symmetry is of order 1 part in $10^4$.

The configuration in Fig.\ \ref{f:bisphal} has energy=$4.932$ $M_W /\al_W$, $n_{CS}$=0.569, 
it has two negative curvature eigenvalues -11.915 $M_W^2$, and -6.788 $M_W^2$.
Its associated sphaleron has energy=$4.943$ $M_W /\al_W$ with $n_{CS}$=1/2, 
and three negative curvature eigenvalues $-23.823 M_W^2$, $-13.249 M_W^2$, and $-0.933 M_W^2$. 
Its conjugate bisphaleron has identical energy, and negative curvature eigenvalues, but $n_{CS}$=0.431;
so again
the $n_{CS}$ of the bisphaleron and its conjugate add to one.

\section{Conclusions}
\label{s:conclusions}
%%%%%%%%%%%%%%%%%%%%%%%%%%%%%%%%%%%%%%%%%%%%%%%%%%%%%%%%%%%%%%%%%%%%%%%%%%%%%%%%%%%%%
%%                                                                                 %%
%%                                                                                 %%
%%  	CONCLUSIONS							           %%
%%                                                                                 %%
%%%%%%%%%%%%%%%%%%%%%%%%%%%%%%%%%%%%%%%%%%%%%%%%%%%%%%%%%%%%%%%%%%%%%%%%%%%%%%%%%%%%%

In this paper we have made a thorough study of the properties of sphalerons in 
two Higgs doublet SU(2) gauge theories.  Using a spherically symmetric 
approximation, we have performed scans in the 
physical parameter space defined by the masses and mixing angles of the Higgs 
particles, recording the energy, lowest eigenvalues, and the Chern-Simons 
number, with results recorded in Figs.\ \ref{f:energy1}--\ref{f:first_eig5}. 
We have also shown the profiles of the fields of our ansatz for selected 
solutions in Figs.\ \ref{f:sph1}--\ref{f:bisphal}.

We can draw a number of broad conclusions from these results.  Firstly, for 
a wide range of parameters, the minimum energy sphaleron is not the natural 
generalisation of the Klinkhamer-Manton sphaleron \cite{Klinkhamer:1984di} 
with vanishing Higgs fields 
at the origin, but a parity violating pair of relative winding (RW) 
sphalerons, first identified by Bachas, Tinyakov, and Tomaras 
\cite{Bachas:1996ap}.  These are 
related to the bisphalerons or deformed sphalerons found in one doublet 
models by Yaffe 
\cite{Yaffe:1989ms} and Kunz and Brihaye \cite{Kunz:1989sx}, but are
specific to two Higgs doublet models. 
This pair was always degenerate in energy, as is to be expected from a parity
conserving Lagarangian. This degeneracy
is lifted when Standard Model fermions are included \cite{Nolte:1995pz}.

The favoured regions of parameter space for RW sphalerons to exist are 
those where there is a large difference in the masses of the neutral Higgses. 
The mass of the heavier Higgs can be as low as $5M_W$. Bisphalerons appear at 
yet higher heavy Higgs masses, but were always more massive than the RW 
sphalerons in the parameter space we explored.

The appearance of extra sphaleron solutions is signalled by the ordinary 
sphaleron developing another negative eigenvalue: thus where the RW sphaleron 
exists the ordinary sphaleron has two negative eigenvalues, and three where 
the bisphaleron exists also.  The lowest energy sphaleron must have exactly 
one negative eigenvalue. The numerically calculated eigenvalues 
of a solution not only aid its identification, but are important for 
accurate calculation of the baryon number violation rate: if the negative
eigenvalue of the lowest energy sphaleron solution is $\omega_-^2$, then the 
rate is proportional to $|\omega_-|$ 
\cite{Arnold:1987mh}.  The difference 
between the most negative eigenvalue of the sphaleron and the negative 
eigenvalue of the RW sphaleron could be well over a factor of two.

The most important quantity for the calculation of the $B$ violation rate is 
normally the sphaleron energy. There is however very little difference in the 
energies of the ordinary and RW sphaleron: typically less than 1\% in the 
range of parameters we surveyed.  Thus the main contribution to the error 
in the rate from using the ordinary sphaleron comes from the negative 
eigenvalue.  One must not only use the correct eigenvalue but also 
include a factor of two in the 
RW sphaleron rate, one for each of the two degenerate parity conjugate 
solutions.  However, this leads only to logarithmic corrections to the 
sphaleron energy bound (\ref{e:SphEneBou}).

The most important parameter for the sphaleron energy was found to be the 
mass of the lightest Higgs, in accordance with previous studies.  However, 
we were able to extend our work on the dependence of the energy on the 
$CP$ violating mixing angle $\theta_{CP}$ \cite{Grant:1999ci} to show that 
there was an strong dependence on this quantity as well,
with the sphaleron energy varying by $\sim$ 15 \% as $\theta_{CP}$ was 
adjusted through its allowed range. We note as well 
that we were unable to find a region of parameter space 
for which RW sphalerons existed over
a wide range of $\th_{CP}$, for which the potential was bounded, 
and for which Eq.\ \ref{e:minimum} was the global minimum.

Although we used a spherically symmetric ansatz, we found that two Higgs 
doublet sphalerons are generically not spherically symmetric. This means 
that our results are approximate: however, the departure from spherical 
symmetry, as measured by the relative size of the symmetry violating terms 
in the static energy functional, was less than 0.2\%, and so this is not 
a serious problem for the accuracy of our results.  A larger correction is 
to be expected when one considers the full SU(2)$\times$U(1) theory at 
non-zero $\theta_W$, for which one also has to abandon the spherically 
symmetric ansatz and resort to an axially symmetric one instead \cite{Brihaye:1994ib}.

Another source of error is the neglect of radiative and thermal corrections.
Ideally one should work out the determinants 
of fluctuation matrices
\cite{Carson:1990rf,Carson:1990jm,Baacke:1994aj,Diakonov:1996xz}.
One can also find solutions using the 1-loop 
finite temperature effective potential \cite{Moreno:1997zm}. 
This is an implicit gradient expansion, neglecting 
finite temperature corrections to gradient terms, which turn out to be 
small \cite{Moore:1996jv}.
Such computations are model-dependent: 
one first computes radiatively corrected couplings in the static energy 
functional, and then the sphaleron energy.
Our approach decouples the computation of the radiative corrections, 
for we can take masses and angles to 
be their 1-loop corrected values. Although this neglects cubic terms
and terms of dimension
higher than 4 in the potential, it is an easy way of improving on the
tree-level calculation, without sacrificing too much accuracy, as the
contribution to the energy from the Higgs potential can be seen from Figs.\
\ref{f:en_den_rws} and \ref{f:bisphal_en} to be small.

Despite these sources of error, we can conclude 
the calculations of the sphaleron 
energy in $CP$ conserving models cannot safely be 
applied to 
$CP$ violating electroweak theories, and that the 
sphaleron bound on the mass of the lightest Higgs in $CP$
violating theories requires further investigation.

%\acknowledgements
We wish to thank Mikko Laine and Neil McNair for helpful discussions.
This work was conducted on the SGI Origin platform using COSMOS
Consortium facilities, funded by HEFCE, PPARC and SGI.
We acknowledge computing support from
the Sussex High Performance Computing Initiative.

\appendix

\section{Parametrization of two-doublet potentials}
\label{a:param_pot}
In Section
\ref{s:higgs_pot} we wrote the two Higgs doublet potential as Eq.\ \ref{e:pot}. Here we write two common 
forms of the most general two Higgs doublet potential. Firstly we write  
\bea
\label{eapp:potMSSM}
V(\phi_{1},\phi_{2}) & = & m_{1}^2\phi_{1}^{\dagger}\phi_{1}+m_{2}^2\phi_{2}^{\dagger}\phi_{2}
                         +m_{12}^2\phi_{1}^{\dagger}\phi_{2}+m_{12}^{2*}\phi_{2}^{\dagger}\phi_{1}\nonumber\\ 
                        & & \ell_{1}(\phi_{1}^{\dagger}\phi_{1})^2 +\ell_{2}(\phi_{2}^{\dagger}\phi_{2})^2
                            +\ell_{3}\phi_{1}^{\dagger}\phi_{1}\phi_{2}^{\dagger}\phi_{2}
                            +\ell_{4}\phi_{1}^{\dagger}\phi_{2}\phi_{2}^{\dagger}\phi_{1}     \nonumber\\
                        & &  +\ell_{5}\phi_{1}^{\dagger}\phi_{2}\phi_{1}^{\dagger}\phi_{2}
                        +\ell_{5}^{*}\phi_{2}^{\dagger}\phi_{1}\phi_{2}^{\dagger}\phi_{1}    \nonumber\\
                        & &  +\ell_{6}\phi_{1}^{\dagger}\phi_{1}\phi_{1}^{\dagger}\phi_{2}
                        +\ell_{6}^{*}\phi_{1}^{\dagger}\phi_{1}\phi_{2}^{\dagger}\phi_{1} \nonumber\\
                        & &  +\ell_{7}\phi_{2}^{\dagger}\phi_{2}\phi_{1}^{\dagger}\phi_{2}
                        +\ell_{7}^{*}\phi_{2}^{\dagger}\phi_{2} \phi_{2}^{\dagger}\phi_{1},   
\eea
\noindent where the only complex parameters are the $m_{12}^{2}$, $\ell_5$, $\ell_6$, and $\ell_7$. This potential has 14 independent 
parameters. Imposing the discrete symmetry $\phi_1 \rightarrow \phi_1$, $\phi_2 \rightarrow -\phi_2$  on dimension four terms
will force $\ell_6=\ell_7=0$, and we have a potential with ten independent parameters. 

Writing the same potential as
\bea
\label{eapp:pot2doub}
V(\phi_{1},\phi_{2}) & = & (\lambda_{1}+\lambda_3)(\phi_{1}^{\dagger}\phi_{1}-\frac{\upsilon_{1}^2}{2})^{2}
                        +(\lambda_{2}+\lambda_3)(\phi_{2}^{\dagger}\phi_{2}-\frac{\upsilon_{2}^2}{2})^{2} \nonumber\\ 
                        & & +2\lambda_{3}(\phi_{1}^{\dagger}\phi_{1}-\frac{\upsilon_{1}^2}{2})
                                   (\phi_{2}^{\dagger}\phi_{2}-\frac{\upsilon_{2}^2}{2}) \nonumber\\
                        & & +\lambda_{4}\left[\phi_{1}^{\dagger}\phi_{1}\phi_{2}^{\dagger}\phi_{2}
                                     -\Re^2( \phi_{1}^{\dagger}\phi_{2})      
                                     -\Im^2( \phi_{1}^{\dagger}\phi_{2})\right] \nonumber\\ 
& & +\lambda_{5}(\Re( \phi_{1}^{\dagger}\phi_{2})-\frac{\upsilon_{1}\upsilon_{2}}{2}\cos\xi)^{2}         
+\lambda_{6}(\Im( \phi_{1}^{\dagger}\phi_{2})-\frac{\upsilon_{1}\upsilon_{2}}{2}\sin\xi)^{2} \nonumber\\ 
& & +\lambda_{7}(\Re( \phi_{1}^{\dagger}\phi_{2})-\frac{\upsilon_{1}\upsilon_{2}}{2}\cos\xi)
(\Im( \phi_{1}^{\dagger}\phi_{2})-\frac{\upsilon_{1}\upsilon_{2}}{2}\sin\xi)       \nonumber\\ 
& & +\mu_{1}(\phi_{1}^{\dagger}\phi_{1}-\frac{\upsilon_{1}^2}{2})
 (\Re( \phi_{1}^{\dagger}\phi_{2})-\frac{\upsilon_{1}\upsilon_{2}}{2}\cos\xi)\nonumber\\ 
& &  +\mu_{2}(\phi_{1}^{\dagger}\phi_{1}-\frac{\upsilon_{1}^2}{2})
    (\Im( \phi_{1}^{\dagger}\phi_{2})-\frac{\upsilon_{1}\upsilon_{2}}{2}\sin\xi)\nonumber\\ 
& & +\mu_{3}(\phi_{2}^{\dagger}\phi_{2}-\frac{\upsilon_{2}^2}{2})
 (\Re( \phi_{1}^{\dagger}\phi_{2})-\frac{\upsilon_{1}\upsilon_{2}}{2}\cos\xi) \nonumber\\ 
& &+\mu_{4}(\phi_{2}^{\dagger}\phi_{2}-\frac{\upsilon_{2}^2}{2})
    (\Im( \phi_{1}^{\dagger}\phi_{2})-\frac{\upsilon_{1}\upsilon_{2}}{2}\sin\xi),
\eea
where all the parameters are real, we again have a potential with 14 independent parameters. Imposing 
$\phi_1 \rightarrow \phi_1$, $\phi_2 \rightarrow -\phi_2$ on dimension four terms
we force four of these parameters $\mu_1=\mu_2=\mu_3=\mu_4=0$, and we have a ten parameter potential.

The advantage of writing the potential as Eq.\ \ref{eapp:pot2doub} is that the three of the parameters of the potential are 
$\xi$, $\upsilon_1$, and $\upsilon_2$, and that the zero of the potential is 
\bea
\label{eapp:zero}
\phi_{\alpha}=\frac{\upsilon_{\alpha}}{\sqrt{2}}\left[\ba{c}0 \\ e^{i\varphi_\alpha}\ea\right],
\eea
\noindent where $\varphi_1 = 0$, and  $\varphi_2 = \xi$. 

The relations between the parameters of Eq.\ \ref{eapp:potMSSM} and those of Eq.\ \ref{eapp:pot2doub} are

\bea
m_{1}^{2} & = & -(\lambda_1 + \lambda_3)\upsilon_1^2-\la_3\upsilon_2^2-\frac{\mu_1}{2}\upsilon_1\upsilon_2\cos\xi-\frac{\mu_2}{2}\upsilon_1\upsilon_2\sin\xi, \\
m_{2}^{2} & = & -(\lambda_2 + \lambda_3)\upsilon_2^2-\la_3\upsilon_1^2 -\frac{\mu_3}{2}\upsilon_1\upsilon_2\cos\xi-\frac{\mu_4}{2}\upsilon_1\upsilon_2\sin\xi,\\
\Re (m_{12}^{2}) & = & -\frac{\lambda_5}{2}\upsilon_1\upsilon_2\cos\xi-\frac{\la_7}{4}\upsilon_1\upsilon_2\sin\xi-\frac{\mu_{1}}{2}\upsilon_1^2-\frac{\mu_{3}}{2}\upsilon_2^2, \\
\Im (m_{12}^{2}) & = & -\frac{\lambda_5}{2}\upsilon_1\upsilon_2\sin\xi-\frac{\la_7}{4}\upsilon_1\upsilon_2\cos\xi-\frac{\mu_{2}}{2}\upsilon_1^2-\frac{\mu_{4}}{2}\upsilon_2^2,  \\
\ell_1 & = & \la_1 +\la_3, \\
\ell_2 & = & \la_2 +\la_3, \\
\ell_3 & = & 2\la_3 +\la_4, \\
\ell_4 & = & \frac{\la_5+\la_6 }{2}-\la_4, \\
\ell_5 &  = & \frac{1}{4}(\la_5-\la_6  - i \la_7), \\
\ell_6 &  = & \frac{1}{2}(\mu_1 - i \mu_2), \\
\ell_7 &  = & \frac{1}{2}(\mu_3 - i \mu_4). \\
\eea

We are free to redefine the fields $\phi_{\al}$ of Eqs.\ \ref{eapp:potMSSM} and \ref{eapp:pot2doub}. 
Rewriting Eq.\ \ref{eapp:pot2doub} with 
$\phi_{\al} \rightarrow \phi_{\al} e^{i\varphi_\al}$ gives
\bea
\label{eapp:pot2doub_redef}
V(\phi_{1},\phi_{2}) & = & (\lambda_{1}+\lambda_3)(\phi_{1}^{\dagger}\phi_{1}-\frac{\upsilon_{1}^2}{2})^{2}
                        +(\lambda_{2}+\lambda_3)(\phi_{2}^{\dagger}\phi_{2}-\frac{\upsilon_{2}^2}{2})^{2} \nonumber\\ 
                        & & +2\lambda_{3}(\phi_{1}^{\dagger}\phi_{1}-\frac{\upsilon_{1}^2}{2})
                                   (\phi_{2}^{\dagger}\phi_{2}-\frac{\upsilon_{2}^2}{2}) \nonumber\\
                        & & +\lambda_{4}\left[\phi_{1}^{\dagger}\phi_{1}\phi_{2}^{\dagger}\phi_{2}
                                     -\Re^2( \phi_{1}^{\dagger}\phi_{2})      
                                     -\Im^2( \phi_{1}^{\dagger}\phi_{2})\right] \nonumber\\ 
& & +(\lambda_{+}+\chi_1)(\Re( \phi_{1}^{\dagger}\phi_{2})-\frac{\upsilon_{1}\upsilon_{2}}{2})^{2}         
+(\lambda_{+}-\chi_1)\Im( \phi_{1}^{\dagger}\phi_{2})^{2} \nonumber\\ 
& & +\chi_2(\Re( \phi_{1}^{\dagger}\phi_{2})-\frac{\upsilon_{1}\upsilon_{2}}{2})
\Im( \phi_{1}^{\dagger}\phi_{2})     \\ 
& & +\tilde{\mu_{1}}(\phi_{1}^{\dagger}\phi_{1}-\frac{\upsilon_{1}^2}{2})
 (\Re( \phi_{1}^{\dagger}\phi_{2})-\frac{\upsilon_{1}\upsilon_{2}}{2}) 
 +\tilde{\mu_{2}}(\phi_{1}^{\dagger}\phi_{1}-\frac{\upsilon_{1}^2}{2})
    \Im( \phi_{1}^{\dagger}\phi_{2})\nonumber\\ 
& & +\tilde{\mu_{3}}(\phi_{2}^{\dagger}\phi_{2}-\frac{\upsilon_{2}^2}{2})
 (\Re( \phi_{1}^{\dagger}\phi_{2})-\frac{\upsilon_{1}\upsilon_{2}}{2})  
 +\tilde{\mu_{4}}(\phi_{2}^{\dagger}\phi_{2}-\frac{\upsilon_{2}^2}{2})
    \Im( \phi_{1}^{\dagger}\phi_{2}),\nonumber
\eea
and we now have a potential which is a function of 13 parameters, 
one less than both Eqs.\ \ref{eapp:potMSSM} and \ref{eapp:pot2doub}. 
Where these new parmeters are in terms of those of 
Eq.\ \ref{eapp:pot2doub}
\bea
\lambda_{+}&=&\frac{1}{2}(\lambda_{5}+\lambda_{6}),\\
\lambda_{-}&=&\frac{1}{2}(\lambda_{5}-\lambda_{6}),\\
\chi_{1}&=&\frac{\lambda_{7}}{2}\sin2\xi+\lambda_{-}\cos2\xi,\\
\chi_{2}&=&\frac{\lambda_{7}}{2}\cos2\xi-\lambda_{-}\sin2\xi,\\
\tilde{\mu_{1}}&=&\mu_{1}\cos\xi+\mu_{2}\sin\xi,\\
\tilde{\mu_{2}}&=&-\mu_{1}\sin\xi+\mu_{2}\cos\xi,\\
\tilde{\mu_{3}}&=&\mu_{3}\cos\xi+\mu_{4}\sin\xi,\\
\tilde{\mu_{4}}&=&-\mu_{3}\sin\xi+\mu_{4}\cos\xi.\\
\eea  

On imposing the discrete symmetry $\phi_1 \rightarrow \phi_1$, $\phi_2 \rightarrow -\phi_2$ on dimension four terms 
$\tilde{\mu_{1}}=\tilde{\mu_{1}}=\tilde{\mu_{1}}=\tilde{\mu_{1}}=0$, 
and we have a potential which is a function of nine parameters, again one less than the potentials of 
Eqs.\ \ref{eapp:potMSSM} and \ref{eapp:pot2doub} with the same symmetry imposed. 
This nine parameter potential is Eq.\ \ref{e:pot} of section \ref{s:higgs_pot} and is the potential we use throughout.
 
\section{Extrema of the potential}
\label{a:extrema}

Extrema of the potential given in Eq.\ \ref{e:pot}
occur at solutions to the four independent equations 
\ben
\label{eapp:soln_extrema}
\frac{\delta  V(X_i) }{\delta X_i} = 0,
\een
where $i=1,2,3,4,$ and $X_i$ are the $x_1$, $x_2$, $y_2$, and $z_2$ of 
\ben
\label{eapp:ext_phi}
\phi_{1}=\frac{\upsilon_{1}}{\sqrt{2}}\left[\ba{c}0\\ x_1 \ea\right],
~~~~~\phi_{2}=\frac{\upsilon_{2}}{\sqrt{2}}\left[\ba{c}z_2\\ x_2 + i y_2 \ea\right].
\een
A general bounded function of four variables with quartic and quadratic 
terms only can have up to $2^4$ minima.
%However the potential \ref{e:pot} has symmetries that reduce this number. 

The trivially found solutions to Eq.\ \ref{eapp:soln_extrema} are 
$x_1=\pm1$, $x_2=\pm1$, $y_2=z_2=0$ (i.e. Eq.\ \ref{e:minimum}), 
and $x_1=x_2=y_2=z_2=0$.
The only other solution we were able to find analytically was 
\bea
x_1& = & 0, \nonumber\\
x_2^2+y_2^2 +z_2^2-1 & = & \frac{\lambda_3}{(\lambda_2 +\lambda_3)\tan^2 \beta}, \nonumber \\ 
x_2&  = & \frac{-\chi_2}{(\lambda_+ +\chi_1)}y_2,  \label{e:ring_extrema}  
\eea  
these describe a circle with one zero eigenvalue, and potential energy 
\ben
V= \frac{\upsilon_1^2\upsilon_2^2}{4}
\left[\frac{\lambda_1 \lambda_2 +(\lambda_1 + \lambda_2)
\lambda_3} {(\lambda_2 + \lambda_3)\tan^2 \beta} +\lambda_+ + \chi_1\right] ,
\een
which may be less than zero for a potential obeying Eqs.\ \ref{e:con_eig_param_1}, 
\ref{e:con_eig_param_2}, and \ref{e:vac_locmin}, 
and is a zero of the other terms of the static energy functional \ref{e:en_H}.

To find numerically the global minimum, 
we implemented two methods.
Firstly, using the Maple extremisation routine {\tt extrema}, we looked for 
an extremum  
of $V(X_i)$ with negative energy somewhere in the chosen region
of parameter space.  As the 
vacuum in our parametrisation has
zero energy, this meant it was not the global minimum. 
We used this solution as an initial configuration for a simple relaxation 
algorithm, which is equivalent to setting ${\cal E ^{''}}$ 
of the Newton method (Eq.\ \ref{e:Newt_raph}) to unity. 
We then scanned though parameter space
relaxing to the global minimum at every point.

Our second method was to
use an initial configuration of $X_i=0$, find the eigenvalues
of the configuration,  
and add a perturbation in the direction of the most eigenfunction with the
most negative eigenvalue. We
then used the relaxation routine on this configuration. 
We did this for each point in parameter space, 
reinitialising to $X_i=0$ at each point.

\section{Static energy functional}
\label{a:SEF}

On substituting the ansatz of Eqs.\ \ref{e:ansatz_h}--\ref{e:ansatz_gi} into the Lagrangian \ref{e:lag} 
we obtain the static energy functional of Eq.\ \ref{e:en_H}.
Here we give the form of $K_{0}$, $K_{1}$, $V_{0}$, $V_{1}$, and $V_{2}$ for the $C$ conserving ansatz and for the $C$ and $P$ violating
ansatz.

In the absence of $C$ violation $F_{\alpha}=a_{\alpha}$ and $G_{\alpha}=c_{\alpha}$, and we have the usual ansatz of Ratra and Yaffee \cite{Ratra:1988dp}
where $K_{1}=V_{1}=0$ and $K_{0}$, $V_{0}$, and $V_{2}$ are
\bea
K_{0} & = & K^D_{0}+ K^G_{0},  \label{e:K_0} \\
 & &K^D_{0}  = \frac{1}{2r^2} \bigg[ a_{\al}^{'2}r^2  + c_{\al}^{'2}r^2 +\al^{'2}+ \be^{'2}\bigg] \label{e:K^D_0}, \\
 & &K^G_{0} =\frac{1}{2r^2} \bigg[ \frac{1}{4r^2}(\al^{2}+\be^{2} -2)^2 \nonumber \\ 
 & & \hspace{1.0cm}  +\frac{1}{4}(a_{\al}^{2}+c_{\al}^{2}) (\al^{2}+\be^{2} +2)  \nonumber \\
& & \hspace{1.0cm}   +\frac{\sqrt{2}\be }{2}(a_{\al}^{2}-c_{\al}^{2})- \sqrt{2}\al a_{\al}c_{\al} \bigg], \label{e:K^G_0} \\
V_{0}& = \frac{\upsilon^{2}}{16M_{W}^{2}} & \bigg[ (\lambda_{1}+\lambda_{3})\left(a_{1}^{2}+c_{1}^{2}-4\cos^{2}\beta\right)^{2}\nonumber\\
  & &+(\lambda_{2}+\lambda_{3})\left(a_{2}^{2}+c_{2}^{2}-4\sin^{2}\beta\right)^{2} \nonumber\\
  & &+2\lambda_{3}\left(a_{1}^{2}+c_{1}^{2}-4\cos^{2}\beta\right)
                                   \left(a_{2}^{2}+c_{2}^{2}-4\sin^{2}\beta\right) \nonumber\\
  & &+\lambda_{4}\left(a_{1}c_{2}-a_{2}c_{1}\right)^{2}\nonumber\\
  & &+(\lambda_{+}+\chi_{1})\left(a_{1}a_{2}+c_{1}c_{2}-4\cos\beta\sin\beta\right)^{2} \bigg], \label{e:pot_0} \\
V_{2}& = \frac{\upsilon^{2}}{16M_{W}^{2}}&\bigg[(-\lambda_{4}+\lambda_{+}-\chi_{1})\left(a_{1}c_{2}-a_{2}c_{1}\right)^{2}  \bigg]. \label{e:pot_2} 
\eea
This ansatz will maintain spherical symmetry if $V_2=0$. The condition $V_2=0$ 
is met if $\lambda_{4}=\lambda_{+}-\chi_{1}$, or equivalently if 
$M_{H^{\pm}}=M_{A}$. In cases where $M_{H^{\pm}}\ne M_{A}$, the spherical 
symmetry of a field configuration will still be maintained if 
$a_{1}c_{2}=a_{2}c_{1}$, as the $V_2$ 
terms vanish from the energy density.  
The ordinary sphaleron comes into this class of 
configurations since $c_1=c_2=0$. However, it is still important to include
this term as it affects the form of $\cal E^{''}$ used in Eq.\ \ref{e:eignevals} to calculate the curvature eigenvalues.

In the presence of $C$ violation  $b_{\alpha}$ and $d_{\alpha}$ are no longer zero 
and $K_{0}$, $K_{1}$, $V_{0}$, $V_{1}$, and $V_{2}$ are

\bea
K_{0} & = & K^D_{0}+ K^G_{0}, \label{e:K_0_cp}\\
& & K^D_{0} =\frac{1}{2r^2} \bigg[ a_{\al}^{'2}r^2 + b_{\al}^{'2}r^2 + c_{\al}^{'2}r^2+ d_{\al}^{'2}r^2 +\al^{'2}+ \be^{'2}\bigg], \label{e:K^D_0_cp} \\
&  &K^G_{0} =\frac{1}{2r^2} \bigg[ \frac{1}{4r^2}(\al^{2}+\be^{2} -2)^2 \nonumber \\ 
 & & \hspace{1.0cm}+\frac{1}{4}(a_{\al}^{2}+b_{\al}^{2}+c_{\al}^{2}+d_{\al}^{2}) (\al^{2}+\be^{2} +2)  \nonumber \\
& & \hspace{1.0cm}  +\frac{\sqrt{2}\be }{2}(a_{\al}^{2}+b_{\al}^{2}-c_{\al}^{2}-d_{\al}^{2})
- \sqrt{2}\al (a_{\al}c_{\al}+b_{\al}d_{\al}) \bigg], \label{e:K^G_0_cp} \\
K_{1} &  = \frac{1}{2r^2}&  \bigg[ (a_{\al}^{'}d_{\al}^{'} - b_{\al}^{'}c_{\al}^{'})r^2
+\frac{1}{4}(a_{\al}d_{\al}+b_{\al}c_{\al}) (\al^{2}+\be^{2} -2)\bigg], \label{e:K_1_cp} \\
V_{0}& =\frac{\upsilon^{2}}{16M_{W}^{2}}&\bigg[(\lambda_{1}+\lambda_{3})\left(a_{1}^{2}+b_{1}^{2}+c_{1}^{2}+d_{1}^{2}-4\cos^{2}\beta\right)^{2}\nonumber\\
  & &+(\lambda_{2}+\lambda_{3})\left(a_{2}^{2}+b_{2}^{2}+c_{2}^{2}+d_{2}^{2}-4\sin^{2}\beta\right)^{2} \nonumber\\
  & &+2\lambda_{3}\left(a_{1}^{2}+b_{1}^{2}+c_{1}^{2}+d_{1}^{2}-4\cos^{2}\beta\right)
                                   \left(a_{2}^{2}+b_{2}^{2}+c_{2}^{2}+d_{2}^{2}-4\sin^{2}\beta\right) \nonumber\\
  & &+\lambda_{4}\left(\left(a_{1}c_{2}-a_{2}c_{1}+b_{1}d_{2}-b_{2}d_{1}\right)^{2}\right.\nonumber\\
  & &\hspace{0.9cm}\left.+\left(a_{1}d_{2}+a_{2}d_{1}-b_{1}c_{2}-b_{2}c_{1}\right)^{2}
                            -4\left(a_{1}d_{1}-b_{1}c_{1}\right)\left(a_{2}d_{2}-b_{2}c_{2}\right)\right) \nonumber\\
  & &+(\lambda_{+}+\chi_{1})\left(a_{1}a_{2}+b_{1}b_{2}+c_{1}c_{2}+d_{1}d_{2}-4\cos\beta\sin\beta\right)^{2} \nonumber\\
  & &+(\lambda_{+}-\chi_{1})\left(a_{1}b_{2}-a_{2}b_{1}+c_{1}d_{2}-c_{2}d_{1}\right)^{2} \\
  & &+2\chi_{2}\left(a_{1}a_{2}+b_{1}b_{2}+c_{1}c_{2}+d_{1}d_{2}-4\cos\beta\sin\beta\right)\left(a_{1}b_{2}-a_{2}b_{1}+c_{1}d_{2}-c_{2}d_{1}\right) \bigg],\nonumber \\
V_{1}& =\frac{\upsilon^{2}}{16M_{W}^{2}}&\bigg[4(\lambda_{1}+\lambda_{3})\left(a_{1}^{2}+b_{1}^{2}+c_{1}^{2}+d_{1}^{2}-4\cos^{2}\beta\right)\left(a_{1}d_{1}-b_{1}c_{1}\right)\label{e:pot_1_cp}\nonumber\\
  & &+4(\lambda_{2}+\lambda_{3})\left(a_{2}^{2}+b_{2}^{2}+c_{2}^{2}+d_{2}^{2}-4\sin^{2}\beta\right)\left(a_{2}d_{2}-b_{2}c_{2}\right) \nonumber\\
  & &+4\lambda_{3}\left(\left(a_{1}^{2}+b_{1}^{2}+c_{1}^{2}+d_{1}^{2}-4\cos^{2}\beta\right)\left(a_{2}d_{2}-b_{2}c_{2}\right)\right. \nonumber\\
  & & \hspace{0.9cm}+\left.\left(a_{2}^{2}+b_{2}^{2}+c_{2}^{2}+d_{2}^{2}-4\sin^{2}\beta\right)\left(a_{1}d_{1}-b_{1}c_{1}\right)\right) \nonumber\\
  & &+2(\lambda_{+}+\chi_{1})\left(a_{1}a_{2}+b_{1}b_{2}+c_{1}c_{2}+d_{1}d_{2}-4\cos\beta\sin\beta\right)\left(a_{1}d_{2}+a_{2}d_{1}-b_{1}c_{2}-b_{2}c_{1}\right) \nonumber\\
  & &-2(\lambda_{+}-\chi_{1})\left(a_{1}b_{2}-a_{2}b_{1}+c_{1}d_{2}-c_{2}d_{1}\right)\left(a_{1}c_{2}-a_{2}c_{1}+b_{1}d_{2}-b_{2}d_{1}\right) \nonumber\\
  & &+2\chi_{2}\left[\left(a_{1}a_{2}+b_{1}b_{2}+c_{1}c_{2}+d_{1}d_{2}-4\cos\beta\sin\beta\right)\left(a_{1}c_{2}-a_{2}c_{1}+b_{1}d_{2}-b_{2}d_{1}\right)\right.\nonumber\\
  & &\hspace{0.9cm}-\left.\left(a_{1}b_{2}-a_{2}b_{1}+c_{1}d_{2}-c_{2}d_{1}\right)\left(a_{1}d_{2}+a_{2}d_{1}-b_{1}c_{2}-b_{2}c_{1}\right)\right] \bigg], \\
V_{2}& =\frac{\upsilon^{2}}{16M_{W}^{2}}&\bigg[4(\lambda_{1}+\lambda_{3})\left(a_{1}d_{1}-b_{1}c_{1}\right)^{2}\label{e:pot_2_cp}\nonumber\\
  & &+4(\lambda_{2}+\lambda_{3})\left(a_{2}d_{2}-b_{2}c_{2}\right)^{2}\nonumber\\
  & &+8\lambda_{3}\left(a_{1}d_{1}-b_{1}c_{1}\right)\left(a_{2}d_{2}-b_{2}c_{2}\right) \nonumber\\
  & &-\lambda_{4}\left(\left(a_{1}c_{2}-a_{2}c_{1}+b_{1}d_{2}-b_{2}d_{1}\right)^{2}\right.\nonumber\\
  & &\hspace{0.9cm}\left.+\left(a_{1}d_{2}+a_{2}d_{1}-b_{1}c_{2}-b_{2}c_{1}\right)^{2}
                            -4\left(a_{1}d_{1}-b_{1}c_{1}\right)\left(a_{2}d_{2}-b_{2}c_{2}\right)\right) \nonumber\\
  & &+(\lambda_{+}+\chi_{1})\left(a_{1}d_{2}+a_{2}d_{1}-b_{1}c_{2}-b_{2}c_{1}\right)^{2} \nonumber\\
  & &+(\lambda_{+}-\chi_{1})\left(a_{1}c_{2}-a_{2}c_{1}+b_{1}d_{2}-b_{2}d_{1}\right)^{2}  \nonumber\\
  & &-2\chi_{2}\left(a_{1}d_{2}+a_{2}d_{1}-b_{1}c_{2}-b_{2}c_{1}\right)\left(a_{1}c_{2}-a_{2}c_{1}+b_{1}d_{2}-b_{2}d_{1}\right) \bigg].
\eea

\section{Numerical scheme}
\label{a:numerics}
%%%%%%%%%%%%%%%%%%%%%%%%%%%%%%%%%%%%%%%%%%%%%%%%%%%%%%%%%%%%%%%%%%%%%%%%%%%%%%%%%%%%%
%%                                                                                 %%
%%                                                                                 %%
%%  	NUMERICAL SCHEME						           %%
%%                                                                                 %%
%%%%%%%%%%%%%%%%%%%%%%%%%%%%%%%%%%%%%%%%%%%%%%%%%%%%%%%%%%%%%%%%%%%%%%%%%%%%%%%%%%%%%

To implement the scheme numerically, we discretise the $n$ fields
into $N$ values $f_{Ai}$ in the range $0 \le r \le R$.  The values at the
boundaries $f_{A0}$ and $f_{A(N-1)}$ are determined by the boundary conditions
in a way which we specify below. Hence 
${\cal
E}^{''}$ is a $n(N-2) \times n(N-2)$ matrix, and 
$\delta f$ and ${\cal E}^{'}$ are $n(N-2)$ column vectors. 
 
To increase the accuracy of the solution while minimising 
the number of points $N$ we use a rescaled co-ordinate $s$, where
\ben
\label{e:s_def}
s  =  \frac{1}{\ln |C|}\ln\left[\frac{1 + \mu r}{1 + r}\right], \quad
\mu  =  \frac{M_{max}}{M_w}, \quad
C=  \frac{1+ \mu R}{1+ R}.
\een
Here, $M_{max}$
is the maximum of [$M_h$, $M_H$, $M_A$, $M_{H^\pm}$], and for $M_{max} = M_w$ we used 
$M_{max} = 1.01 \times M_w$.
We took $R= 20 M_w^{-1}$ and 
used $N=51$ points throughout.
It is also convenient to define two new functions $X(s)$, $Y(s)$ through 
\bea
X(s) \equiv \frac{ds}{dr} &=& \frac{1}{\ln | C|}\frac{1}{(\mu - 1)} \frac{ (C^s - \mu)^2 }{ C^s}, \\
Y(s) \equiv \frac{dX}{ds} &=& \frac{1}{\mu - 1} \frac{ (C^s - \mu) (C^s +\mu)}{ C^s}. 
\eea
The first derivative of the energy ${\cal E}^{'}$ may be split into Higgs and
gauge parts 
\bea
\label{e:en_der_higgs}
{\cal E^{'}}_{H} & = & -(Yr^2 +2r)\frac{d f_{H} }{ds} - Xr^2\frac{d^2 f_{H}}{ds^2} + \frac{1}{X}\frac{d~}{d f_{H}}
\left[ K^G_0 +V_0 +\frac{1}{3} V_2 \right], \\
\label{e:en_der_gauge}
{\cal E^{'}}_{G} & = & -Y\frac{d f_{G}}{ds} - X \frac{d^2 f_{G}}{ds^2} + \frac{1}{X}\frac{d~}{d f_{G}}
 \left[ K^G_0 +V_0 +\frac{1}{3} V_2 \right].
\eea
We use symmetric second-order accurate differencing for the derivatives, and so
\bea
\label{e:en_der_higgs_dis}
{\cal E}'_{Hi} & = & -(Y_{i}r_{i}^2 +2r_{i})\frac{(f_{Hi+1}- f_{Hi-1})}{2 h_s} 
- X_{i}r_{i}^2\frac{(f_{Hi+1}-2f_{Hi}+f_{Hi-1})}{h_s^2},  \nonumber\\
& & + \frac{1}{X_{i}}\frac{d~}{d f_{Hi}}
\left[ K^G_{0i} +V_{0i} +\frac{1}{3} V_{2i} \right] \\
\label{e:en_der_gauge_dis}
{\cal E}'_{Gi}
& = & -Y_{i}\frac{(f_{Gi+1}- f_{Gi-1})}{h_s}
- X_{i} \frac{(f_{Gi+1}-2f_{Gi-1}+f_{Gi-1})}{h_s^2}   \nonumber\\
& &+ \frac{1}{X_{i}}\frac{d~}{d f_{Gi}}
 \left[ K^G_{0i} +V_{0i} +\frac{1}{3} V_{2i} \right], 
\eea
where the index $i=1,...,(N-2)$, runs over the rescaled co-ordinate $s$, excluding the first and last points, 
and $h_s = (N-1)^{-1}$, is the separation between each adjacent rescaled co-ordinate. We did not use 
$(f_{Hi+2}-2f_{Hi}+f_{Hi-2})/(2 h_s)^2$ for the second order derivative, as this would have produced two systems 
independent in derivative terms, 
one seeing the even points and one seeing the odd points.

The matrix ${\cal
E}^{''}$ is a block tridiagonal $n(N-2)\times n(N-2) $ matrix of the form
\bea
\hspace{0.2cm}\makebox[2.0cm]{\rule[-0.75cm]{0.0cm}{1.5cm} $0$}
\hspace{0.2cm}\framebox[2.0cm]{\rule[-0.75cm]{0.0cm}{1.5cm} $D^-_{i-1,i-2}$}
\hspace{0.2cm}\framebox[2.0cm]{\rule[-0.75cm]{0.0cm}{1.5cm} $D^0_{i-1,i-1}$} 
\hspace{0.2cm}\framebox[2.0cm]{\rule[-0.75cm]{0.0cm}{1.5cm} $D^+_{i-1,i}$}
\hspace{0.2cm}\makebox[2.0cm]{\rule[-0.75cm]{0.0cm}{1.5cm} $0$}
\hspace{0.2cm}\makebox[2.0cm]{\rule[-0.75cm]{0.0cm}{1.5cm} $\cdots$} \nonumber \\
\hspace{0.2cm}\makebox[2.0cm]{\rule[-0.75cm]{0.0cm}{1.5cm} $\cdots$}
\hspace{0.2cm}\makebox[2.0cm]{\rule[-0.75cm]{0.0cm}{1.5cm} $0$}
\hspace{0.2cm}\framebox[2.0cm]{\rule[-0.75cm]{0.0cm}{1.5cm} $D^-_{i,i-1}$}
\hspace{0.2cm}\framebox[2.0cm]{\rule[-0.75cm]{0.0cm}{1.5cm} $D^0_{i,i}$} 
\hspace{0.2cm}\framebox[2.0cm]{\rule[-0.75cm]{0.0cm}{1.5cm} $D^+_{i,i+1}$}
\hspace{0.2cm}\makebox[2.0cm]{\rule[-0.75cm]{0.0cm}{1.5cm} $0$} \nonumber \\
\hspace{0.2cm}\makebox[2.0cm]{\rule[-0.75cm]{0.0cm}{1.5cm} $\cdots$}
\hspace{0.2cm}\makebox[2.0cm]{\rule[-0.75cm]{0.0cm}{1.5cm} $\cdots$}
\hspace{0.2cm}\makebox[2.0cm]{\rule[-0.75cm]{0.0cm}{1.5cm} $0$}
\hspace{0.2cm}\framebox[2.0cm]{\rule[-0.75cm]{0.0cm}{1.5cm} $D^-_{i+1,i}$}
\hspace{0.2cm}\framebox[2.0cm]{\rule[-0.75cm]{0.0cm}{1.5cm} $D^0_{i+1,i+1}$} 
\hspace{0.2cm}\framebox[2.0cm]{\rule[-0.75cm]{0.0cm}{1.5cm} $D^+_{i+1,i+2}$}\nonumber
\eea
where each of these boxes are $n \times n$ matrices, 
and there are $(N-2) \times (N-2) $ such boxes.
The only non zero terms are the $D^-_{i,i-1}$, $D^0_{i,i}$, and $D^+_{i,i+1}$. 
Note that $D^-_{i,i-1}$ and $D^+_{i,i+1}$ are themselves diagonal, with 
entries  
\bea
 D^-_{i,i-1}  & = &
\frac{1}{2 h_s}Y_{i} - \frac{1}{h_s^2}X_{i}, \label{e:g_D-}\\
 D^+_{i,i+1}&=& -\frac{1}{2 h_s}Y_{i} - \frac{1}{h_s^2}X_{i},  \label{e:g_D+}
\eea
for the two gauge fields, and 
\bea
D^-_{i,i-1} & = &
\frac{1}{2 h_s}
(2r_{i}+r_{i}^2 Y_{i}) - \frac{1}{h_s^2} r_{i}^2 X_{i},\label{e:h_D-}  \\
 D^+_{i,i+1} & = &-\frac{1}{2 h_s} (2r_{i}+r_{i}^2 Y_{i}) - \frac{1}{h_s^2} r_{i}^2 X_{i}, \label{e:h_D+} 
\eea
for the remaining Higgs fields. If we write 
\ben 
\label{e:D0}
D^0_{Ai,Bi} \equiv D^{0der}_{Ai,Bi} + D^{0mat}_{Ai,Bi},
\een
then  $D^{0der}_{i,i} $ are diagonal in $A$, $B$ with 
\bea\label{e:D0der}
D^{0der}_{i,i}  =&   \frac{2}{h_s^2}X_{i}&\quad
\mbox{(gauge fields),} \label{e:h_D0}\\
D^{0der}_{i,i} =&\frac{2}{h_s^2}r_{i}^2
X_{i} &\quad \mbox{(higgs fields).} \label{e:g_D0}
\eea
The non-diagonal elements are symmetric in $A$, $B$ with 
\ben\label{e:D0mat}
 D^{0mat}_{Ai,Bi}  =\frac{d^2}{d f_{Ai} d f_{Bi}} \left[ K^G_{0i} + V_{0i} + \frac{1}{3} V_{2i} \right].\label{e:h_D0mat}
\een

We have to be careful about the form of ${\cal E}^{''}$ at the top left corner of the matrix, corresponding to the $i=1$ point, affecting the
$D^0_{1,1}$, and the $D^+_{1,2}$ terms.
Also the bottom right corner, corresponding to the $i=(N-2)$ point, affecting the $D^-_{N-2,N-3}$, and the $D^0_{N-2,N-2}$ 
since these must implement the boundary conditions.
\bea
\hspace{0.2cm}\framebox[2.0cm]{\rule[-0.75cm]{0.0cm}{1.5cm} $D^0_{1,1}$} 
\hspace{0.2cm}\framebox[2.0cm]{\rule[-0.75cm]{0.0cm}{1.5cm} $D^+_{1,2}$}
\hspace{0.2cm}\makebox[2.0cm]{\rule[-0.75cm]{0.0cm}{1.5cm} $0$}  
\hspace{0.2cm}\makebox[2.0cm]{\rule[-0.75cm]{0.0cm}{1.5cm} $0$}  
\hspace{0.2cm}\makebox[2.0cm]{\rule[-0.75cm]{0.0cm}{1.5cm} $\cdots$}   
\hspace{0.2cm}\makebox[2.0cm]{\rule[-0.75cm]{0.0cm}{1.5cm} }\nonumber \\
\hspace{0.2cm}\framebox[2.0cm]{\rule[-0.75cm]{0.0cm}{1.5cm} $D^-_{2,1}$}
\hspace{0.2cm}\framebox[2.0cm]{\rule[-0.75cm]{0.0cm}{1.5cm} $D^0_{2,2}$} 
\hspace{0.2cm}\framebox[2.0cm]{\rule[-0.75cm]{0.0cm}{1.5cm} $D^+_{2,3}$}
\hspace{0.2cm}\makebox[2.0cm]{\rule[-0.75cm]{0.0cm}{1.5cm} $0$}  
\hspace{0.2cm}\makebox[2.0cm]{\rule[-0.75cm]{0.0cm}{1.5cm}  $\cdots$}  
\hspace{0.2cm}\makebox[2.0cm]{\rule[-0.75cm]{0.0cm}{1.5cm}  } \nonumber \\
\hspace{0.2cm}\makebox[2.0cm]{\rule[-0.25cm]{0.0cm}{0.5cm} $\ddots$ } 
\hspace{0.2cm}\makebox[2.0cm]{\rule[-0.25cm]{0.0cm}{0.5cm}$\ddots$ } 
\hspace{0.2cm}\makebox[2.0cm]{\rule[-0.25cm]{0.0cm}{0.5cm}$\ddots$ }
\hspace{0.2cm}\makebox[2.0cm]{\rule[-0.25cm]{0.0cm}{0.5cm}  } 
\hspace{0.2cm}\makebox[1.0cm]{\rule[-0.25cm]{0.0cm}{0.5cm}}   \nonumber \\ 
\hspace{0.2cm}\makebox[2.0cm]{\rule[-0.75cm]{0.0cm}{1.5cm} }  
\hspace{0.2cm}\makebox[2.0cm]{\rule[-0.75cm]{0.0cm}{1.5cm} $\cdots$ }  
\hspace{0.2cm}\makebox[2.0cm]{\rule[-0.75cm]{0.0cm}{1.5cm} $0$} 
\hspace{0.2cm}\framebox[2.0cm]{\rule[-0.75cm]{0.0cm}{1.5cm} $D^-_{N-3,N-4}$}
\hspace{0.2cm}\framebox[2.0cm]{\rule[-0.75cm]{0.0cm}{1.5cm} $D^0_{N-3,N-3}$} 
\hspace{0.2cm}\framebox[2.0cm]{\rule[-0.75cm]{0.0cm}{1.5cm} $D^+_{N-3,N-2}$} \nonumber \\ 
\hspace{0.2cm}\makebox[2.0cm]{\rule[-0.75cm]{0.0cm}{1.5cm} }  
\hspace{0.2cm}\makebox[2.0cm]{\rule[-0.75cm]{0.0cm}{1.5cm} $\cdots$}  
\hspace{0.2cm}\makebox[2.0cm]{\rule[-0.75cm]{0.0cm}{1.5cm} $0$ } 
\hspace{0.2cm}\makebox[2.0cm]{\rule[-0.75cm]{0.0cm}{1.5cm} $0$}
\hspace{0.2cm}\framebox[2.0cm]{\rule[-0.75cm]{0.0cm}{1.5cm} $D^-_{N-2,N-3}$} 
\hspace{0.2cm}\framebox[2.0cm]{\rule[-0.75cm]{0.0cm}{1.5cm} $D^0_{N-2,N-2}$}\nonumber
\eea

Because for the sphaleron the boundary conditions at the origin are never updated, $D^0_{1,1}$, and $D^+_{1,2}$ for the sphaleron are as Eq.\ \ref{e:D0}. 
For the RWS, and bisphalerons at the origin we use for the gauge 
fields
\bea
\label{e:F1F2bc}
f_G^{'}|_{r=0}= 0& \rightarrow & f_{G}|_{i=0}=f_{G}|_{i=1},
\eea
from this we are able to calculate 
\bea\label{e:Psibc}
\Psi|_{i=0}= \arctan( -\al /\be )|_{i=0}.
\eea
For the Higgs fields we use Tables \ref{t:BCbisphal} and \ref{t:BCRWsphal} to give
\bea
\label{e:def_theta}
c_\al|_{i=0} & = & a_\al|_{i=0} \tan \Theta_\al|_{i=0},  \\
d_\al|_{i=0} & = & b_\al|_{i=0} \tan \Theta_\al|_{i=0},
\eea
where the $\Theta_\al|_{i=0}$ are calculated from $\Psi|_{i=0}$ of Eq.\ \ref{e:Psibc}, and using Tables \ref{t:BCbisphal} and \ref{t:BCRWsphal} 
according to whether we are looking for the bisphalerons or RW sphalerons. Further 
imposing smoothness of $\phi_\al^\dag\phi_{\be}$ at the origin gives boundary conditions 
\bea
a_\al|_{i=0} & = & a_\al|_{i=1} \cos^2 \Theta_\al|_{i=0} + c_\al|_{i=1} \sin \Theta_\al|_{i=0} \cos \Theta_\al|_{i=0},\label{e:abc}  \\
b_\al|_{i=0} & = & b_\al|_{i=1} \cos^2 \Theta_\al|_{i=0} + d_\al|_{i=1} \sin \Theta_\al|_{i=0} \cos \Theta_\al|_{i=0},\label{e:bbc} \\
c_\al|_{i=0} & = & c_\al|_{i=1} \sin^2 \Theta_\al|_{i=0} + a_\al|_{i=1} \cos \Theta_\al|_{i=0} \sin\Theta_\al|_{i=0},\label{e:cbc} \\  
d_\al|_{i=0} & = & d_\al|_{i=1} \sin^2 \Theta_\al|_{i=0} + b_\al|_{i=1} \cos \Theta_\al|_{i=0} \sin\Theta_\al|_{i=0}.\label{e:dbc}
\eea  

To update the origin after each Newton Raphson iteration we use Eqs.\ \ref{e:F1F2bc}, \ref{e:abc}-\ref{e:dbc}. We also use these
to give us the form of $D^0_{1,1}$ and $D^+_{1,2}$ when looking for the bisphalerons or RW sphalerons. 
We did this by first writing, for $a_{\al}$:
\bea 
\label{e:higgsfi}
 &&-(Yr^2 +2r)\frac{d a_{\al} }{ds}|_{1}-Xr^2\frac{d^2 a_{\al}}{ds^2}|_{1} =\nonumber \\ 
&& -(Y_{1}r_{1}^2 +2r_{1})\frac{1}{2 hs}
 (a_{\al}|_{2}-a_{\al}|_{1}\cos^2\Theta_{\al}|_{1}-c_{\al}|_{1}\cos\Theta_{\al}|_{1}\sin\Theta_{\al}|_{1}) \nonumber \\ 
& & - X_{1}r_{1}^2\frac{1}{hs^2}(a_{\al}|_{2}-2a_{\al}|_{1}
+a_{\al}|_{1}\cos^2\Theta_{\al}|_{1}+c_{\al}|_{1}\cos\Theta_{\al}|_{1}\sin\Theta_{\al}|_{1}),
\eea
with the equivalent expression for the other Higgs fields; and using Eq.\ \ref{e:F1F2bc}, for the gauge fields, we write
\bea 
\label{e:gaugefi}
-Y\frac{d f_{G} }{ds}|_{1}-Xr^2\frac{d^2 f_{G}}{ds^2}|_{1} 
= -Y_{1}\frac{1}{2 hs}(f_{G}|_{2}-f_{G}|_{1})- X_{1}\frac{1}{hs^2}(f_{G}|_{2}-f_{G}|_{1}).
\eea 
We then, after functional differentiation of Eqs.\ \ref{e:higgsfi} and \ref{e:gaugefi}, 
get a form of $D^0_{1,1}$ and $D^+_{1,2}$ that sees 
the boundary conditions. 

The $\Theta_{\al}$ throughout are zero if we are looking for sphaleron solutions, and are determined from either 
Tables \ref{t:BCbisphal} or \ref{t:BCRWsphal} with Eq.\ \ref{e:Psibc} according to whether we are looking for bisphalerons or RWS.

We now turn
to the boundary conditions at infinity. The last point is never updated since this boundary does not evolve, and
$D^{0}_{N-2,N-2}$ is as Eqs.\ \ref{e:D0}-\ref{e:D0mat}. 
We did not use $f^{'}_G |_{r\rightarrow \infty}=(f_G|_{N-1}-f_G|_{N-2})/hs=0$ as the boundary condition
since rescaling 
the radial co-ordinate to allow greater accuracy at the origin reduces the number of points at large distances. This meant that
the form of the first and second derivative were not very accurate at the last few points. 

The form of $\cal{E^{'}}$ of Eqs.\ \ref{e:en_der_higgs_dis} and \ref{e:en_der_gauge_dis} was not affected by the boundary conditions.
Because $\cal{E^{'}}$ is only defined for $i=1,...,(N-2)$ and first and second derivatives 
at $i=1$, and $i=N-2$ are obtained from the 
already updated fields $f_{A}|_0$ and $f_{A}|_{N-1}$.

Also recalling that $\cal{E^{''}}$ of Eqs.\ \ref{e:eignevals} and \ref{e:2nd_der} used in the evaluation of the curvature eigenvalues is functionally differentiated 
with repect to $f_G$ and $r f_H$, and not $f_G$ and $f_H$. The form of $D^0_{1,1}$ and $D^+_{1,2}$
for evaluating the curvature eigenvalues is for the Higgs fields components as 
Eqs.\ \ref{e:D0} and \ref{e:D0mat} since $\delta (r f_H)|_{0}=0$. We again use Eq.\ \ref{e:F1F2bc} for the gauge fields.

To find solutions other than the original sphaleron 
we first find the sphaleron and determine the curvature eigenvalues and eigenfunctions of the configuration. If there is
more than one negative curvature eigenvalue, we succesively add a fraction of the eigenfunction 
of the second (or third) negative eigenvalue to the sphaleron field configuration, measuring the energy at each step. 
If we chose this 
fraction small enough (typically between $0.01$ and $0.1$) the energy at each step will decreases until it reaches a minimum. 
When the energy after a step is larger than the energy measured after the previous step, we 
multiply the fraction by $-0.1$ and continue until the fraction is $-10^{-9}$ times its original value.

This configuration is then used as the initial configuration 
for the Newton Raphson minimisation routine to find the RW sphalerons (or bisphalerons).

Sliding down the most negative eigenfunction of a sphaleron configuration reaches the vacuum. Sliding down the second 
most negative eigenfunction reaches the lowest energy branch of sphaleron like solutions, a third negative eigenfunction will reach 
the second lowest energy branch and so on. In this way we were able to find bisphalerons and RW sphalerons of 
the theory.

We use BLAS fortran subroutines {\tt dgbco} and {\tt dgbsl} to solve for 
$\delta f_{\alpha}$ of Eq.\ \ref{e:eignevals} and subroutine 
{\tt dgeev} to evaluate the curvature eigenvalues and eigenfunctions.


\begin{thebibliography}{99}
%%%%%%%%%%%%%%%%%%%%%%%%%%%%%%%%%%%%%%%%%%%%%%%%%%%%%%%%%%%%%%%%%%%%%%%%%%%%%%%%%%%%%
%%                                                                                 %%
%%                                                                                 %%
%%  	Refererences							           %%
%%                                                                                 %%
%%%%%%%%%%%%%%%%%%%%%%%%%%%%%%%%%%%%%%%%%%%%%%%%%%%%%%%%%%%%%%%%%%%%%%%%%%%%%%%%%%%%%

%\cite{Sakharov:1967dj}
\bibitem{Sakharov:1967dj}
A.~D.~Sakharov,
%``Violation Of CP Invariance, C Asymmetry, And Baryon Asymmetry Of The Universe,''
Pisma Zh.\ Eksp.\ Teor.\ Fiz.\  {\bf 5} (1967) 32. JETP Lett. {\bf 5} (1967) 24.
%%CITATION = ZFPRA,5,32;%%

%\cite{'tHooft:1976up}
\bibitem{'tHooft:1976up}
G.~'t Hooft,
%``Symmetry breaking through Bell-Jackiw anomalies,''
Phys.\ Rev.\ Lett.\  {\bf 37} (1976) 8.
%%CITATION = PRLTA,37,8;%%

%\cite{Kuzmin:1985mm}
\bibitem{Kuzmin:1985mm}
V.~A.~Kuzmin, V.~A.~Rubakov and M.~E.~Shaposhnikov,
%``On The Anomalous Electroweak Baryon Number Nonconservation In The Early Universe,''
Phys.\ Lett.\  {\bf B155} (1985) 36.
%%CITATION = PHLTA,B155,36;%%

%\cite{Kajantie:1996kf}
\bibitem{Kajantie:1996kf}
K.~Kajantie, M.~Laine, K.~Rummukainen and M.~Shaposhnikov,
%``The Electroweak Phase Transition: A Non-Perturbative Analysis,''
Nucl.\ Phys.\  {\bf B466} (1996) 189
[hep-lat/9510020].
%%CITATION = HEP-LAT 9510020;%%

%\cite{Kajantie:1997qd}
\bibitem{Kajantie:1997qd}
K.~Kajantie, M.~Laine, K.~Rummukainen and M.~Shaposhnikov,
%``A non-perturbative analysis of the finite T phase transition in  SU(2) x U(1) electroweak theory,''
Nucl.\ Phys.\  {\bf B493}, 413 (1997)
[hep-lat/9612006].
%%CITATION = HEP-LAT 9612006;%%

%\cite{Huet:1996sh}
\bibitem{Huet:1996sh}
P.~Huet and A.~E.~Nelson,
%``Electroweak baryogenesis in supersymmetric models,''
Phys.\ Rev.\  {\bf D53} (1996) 4578
[hep-ph/9506477].
%%CITATION = HEP-PH 9506477;%%

%\cite{Carena:1997gx}
\bibitem{Carena:1997gx}
M.~Carena, M.~Quiros, A.~Riotto, I.~Vilja and C.~E.~Wagner,
%``Electroweak baryogenesis and low energy supersymmetry,''
Nucl.\ Phys.\  {\bf B503} (1997) 387
[hep-ph/9702409].
%%CITATION = HEP-PH 9702409;%%

%\cite{Cline:2000nw}
\bibitem{Cline:2000nw}
J.~M.~Cline, M.~Joyce and K.~Kainulainen,
%``Supersymmetric electroweak baryogenesis,''
JHEP {\bf 0007} (2000) 018
[hep-ph/0006119].
%%CITATION = HEP-PH 0006119;%%

%\cite{deCarlos:1997ru}
\bibitem{deCarlos:1997ru}
B.~de Carlos and J.~R.~Espinosa,
%``The baryogenesis window in the MSSM,''
Nucl.\ Phys.\  {\bf B503} (1997) 24
[hep-ph/9703212].
%%CITATION = HEP-PH 9703212;%%

%\cite{Laine:1998qk}
\bibitem{Laine:1998qk}
M.~Laine and K.~Rummukainen,
%``The MSSM electroweak phase transition on the lattice,''
Nucl.\ Phys.\  {\bf B535} (1998) 423
[hep-lat/9804019].
%%CITATION = HEP-LAT 9804019;%%

%\cite{Cline:1998hy}
\bibitem{Cline:1998hy}
J.~M.~Cline and G.~D.~Moore,
%``Supersymmetric electroweak phase transition: Baryogenesis versus  experimental constraints,''
Phys.\ Rev.\ Lett.\  {\bf 81} (1998) 3315
[hep-ph/9806354].
%%CITATION = HEP-PH 9806354;%%

%\cite{Cohen:1991iu}
\bibitem{Cohen:1991iu}
A.~G.~Cohen, D.~B.~Kaplan and A.~E.~Nelson,
%``Spontaneous baryogenesis at the weak phase transition,''
Phys.\ Lett.\  {\bf B263} (1991) 86.
%%CITATION = PHLTA,B263,86;%%

%\cite{Rubakov:1997sr}
\bibitem{Rubakov:1997sr}
V.~A.~Rubakov and M.~E.~Shaposhnikov,
%``Electroweak baryon number non-conservation in the early universe and  in high energy collisions,''
{\it Prepared for La Plata Meeting on Trends in Theoretical Physics, La Plata, Argentina, 28 Apr - 6 May 1997}.
%TO BE ADDED 
Phys. Usp. {\bf 39}, 
461 (1996),  {[hep-ph/9603208]}

%\cite{Riotto:1999yt}:
\bibitem{Riotto:1999yt}
A.~Riotto and M.~Trodden,
%``Recent progress in baryogenesis,''
Ann.\ Rev.\ Nucl.\ Part.\ Sci.\  {\bf 49} (1999) 35
[hep-ph/9901362].
%%CITATION = HEP-PH 9901362;%%

%\cite{Trodden:1999ym}
\bibitem{Trodden:1999ym}
M.~Trodden,
%``Electroweak baryogenesis,''
Rev.\ Mod.\ Phys.\  {\bf 71}, 1463 (1999)
[hep-ph/9803479].
%%CITATION = HEP-PH 9803479;%%

%\cite{Shaposhnikov:1986jp}:
\bibitem{Shaposhnikov:1986jp}
M.~E.~Shaposhnikov,
%``Possible Appearance Of The Baryon Asymmetry Of The Universe In An Electroweak Theory,''
JETP Lett.\  {\bf 44} (1986) 465.
%%CITATION = JTPLA,44,465;%%

%\cite{Shaposhnikov:1987tw}:
\bibitem{Shaposhnikov:1987tw}
M.~E.~Shaposhnikov,
%``Baryon Asymmetry Of The Universe In Standard Electroweak Theory,''
Nucl.\ Phys.\  {\bf B287} (1987) 757.
%%CITATION = NUPHA,B287,757;%%

%\cite{Bochkarev:1990fx}
\bibitem{Bochkarev:1990fx}
A.~I.~Bochkarev, S.~V.~Kuzmin and M.~E.~Shaposhnikov,
%``Electroweak Baryogenesis And The Higgs Boson Mass Problem,''
Phys.\ Lett.\  {\bf B244} (1990) 275.
%%CITATION = PHLTA,B244,275;%%

%\cite{Dashen:1975xh}
\bibitem{Dashen:1975xh}
R.~F.~Dashen, B.~Hasslacher and A.~Neveu,
%``Semiclassical Bound States In An Asymptotically Free Theory,''
Phys.\ Rev.\  {\bf D12} (1975) 2443.
%%CITATION = PHRVA,D12,2443;%%

%\cite{Soni:1980ps}
\bibitem{Soni:1980ps}
V.~Soni,
%``Possible Classical Solutions In The Weinberg-Salam Model,''
Phys.\ Lett.\  {\bf B93} (1980) 101.
%%CITATION = PHLTA,B93,101;%%

%\cite{Burzlaff:1984jb}
\bibitem{Burzlaff:1984jb}
J.~Burzlaff,
%``A Classical Lump In SU(2) Gauge Theory With A Higgs Doublet,''
Nucl.\ Phys.\  {\bf B233} (1984) 262.
%%CITATION = NUPHA,B233,262;%%

%\cite{Klinkhamer:1984di}
\bibitem{Klinkhamer:1984di}
F.~R.~Klinkhamer and N.~S.~Manton,
%``A Saddle Point Solution In The Weinberg-Salam Theory,''
Phys.\ Rev.\  {\bf D30} (1984) 2212.
%%CITATION = PHRVA,D30,2212;%%

%\cite{Kunz:1989sx}
\bibitem{Kunz:1989sx}
J.~Kunz and Y.~Brihaye,
%``New Sphalerons In The Weinberg-Salam Theory,''
Phys.\ Lett.\  {\bf B216} (1989) 353.
%%CITATION = PHLTA,B216,353;%%

%\cite{Yaffe:1989ms}
\bibitem{Yaffe:1989ms}
L.~G.~Yaffe,
%``Static Solutions Of SU(2) Higgs Theory,''
Phys.\ Rev.\  {\bf D40} (1989) 3463.
%%CITATION = PHRVA,D40,3463;%%

%\cite{Kastening:1991nw}
\bibitem{Kastening:1991nw}
B.~Kastening, R.~D.~Peccei and X.~Zhang,
%``Sphalerons in the two doublet Higgs model,''
Phys.\ Lett.\  {\bf B266} (1991) 413.
%%CITATION = PHLTA,B266,413;%%

%\cite{Bachas:1996ap}
\bibitem{Bachas:1996ap}
C.~Bachas, P.~Tinyakov and T.~N.~Tomaras,
%``On spherically-symmetric solutions in the two Higgs standard model,''
Phys.\ Lett.\  {\bf B385} (1996) 237
[hep-ph/9606348].
%%CITATION = HEP-PH 9606348;%%

%\cite{Kleihaus:1999bh}
\bibitem{Kleihaus:1999bh}
B.~Kleihaus,
%``Energy barrier in the two-Higgs model,''
Mod.\ Phys.\ Lett.\  {\bf A14} (1999) 1431
[hep-ph/9808295].
%%CITATION = HEP-PH 9808295;%%

%\cite{Moreno:1997zm}
\bibitem{Moreno:1997zm}
J.~M.~Moreno, D.~H.~Oaknin and M.~Quiros,
%``Sphalerons in the MSSM,''
Nucl.\ Phys.\  {\bf B483} (1997) 267
[hep-ph/9605387].
%%CITATION = HEP-PH 9605387;%%

%\cite{Kleihaus:1991ks}
\bibitem{Kleihaus:1991ks}
B.~Kleihaus, J.~Kunz and Y.~Brihaye,
%``The Electroweak sphaleron at physical mixing angle,''
Phys.\ Lett.\  {\bf B273} (1991) 100.
%%CITATION = PHLTA,B273,100;%%

%\cite{James:1992re}
\bibitem{James:1992re}
M.~E.~James,
%``The Sphaleron at nonzero Weinberg angle,''
Z.\ Phys.\  {\bf C55} (1992) 515.
%%CITATION = ZEPYA,C55,515;%%

%\cite{Klinkhamer:1992fi}
\bibitem{Klinkhamer:1992fi}
F.~R.~Klinkhamer and R.~Laterveer,
%``The Sphaleron at finite mixing angle,''
Z.\ Phys.\  {\bf C53} (1992) 247.
%%CITATION = ZEPYA,C53,247;%%

%\cite{Saffin:1998ae}:
\bibitem{Saffin:1998ae}
P.~M.~Saffin and E.~J.~Copeland,
%``Electrically charged sphalerons,''
Phys.\ Rev.\  {\bf D57} (1998) 5064
[hep-ph/9710343].
%%CITATION = HEP-PH 9710343;%%

%\cite{Grant:1999ci}
\bibitem{Grant:1999ci}
J.~Grant and M.~Hindmarsh,
%``Sphalerons with CP-violating Higgs potentials,''
Phys.\ Rev.\  {\bf D59} (1999) 116014
[hep-ph/9811289].
%%CITATION = HEP-PH 9811289;%%

%\cite{Arnold:1987mh}
\bibitem{Arnold:1987mh}
P.~Arnold and L.~McLerran,
%``Sphalerons, Small Fluctuations And Baryon Number Violation In Electroweak Theory,''
Phys.\ Rev.\  {\bf D36} (1987) 581.
%%CITATION = PHRVA,D36,581;%%

%\cite{Carson:1990rf}
\bibitem{Carson:1990rf}
L.~Carson and L.~McLerran,
%``Approximate Computation Of The Small Fluctuation Determinant Around A Sphaleron,''
Phys.\ Rev.\  {\bf D41}, 647 (1990).
%%CITATION = PHRVA,D41,647;%%

%\cite{Carson:1990jm}
\bibitem{Carson:1990jm}
L.~Carson, X.~Li, L.~McLerran and R.~Wang,
%``Exact Computation Of The Small Fluctuation Determinant Around A Sphaleron,''
Phys.\ Rev.\  {\bf D42}, 2127 (1990).
%%CITATION = PHRVA,D42,2127;%%

%\cite{Baacke:1994aj}
\bibitem{Baacke:1994aj}
J.~Baacke and S.~Junker,
%``Quantum fluctuations around the electroweak sphaleron,''
Phys.\ Rev.\  {\bf D49}, 2055 (1994)
[hep-ph/9308310];
%%CITATION = HEP-PH 9308310;%%
%\cite{Baacke:1994ix}
%\bibitem{Baacke:1994ix}
J.~Baacke and S.~Junker,
%``Quantum fluctuations of the electroweak sphaleron: Erratum and addendum,''
Phys.\ Rev.\  {\bf D50}, 4227 (1994)
[hep-th/9402078].
%%CITATION = HEP-TH 9402078;%%

%\cite{Diakonov:1996xz}
\bibitem{Diakonov:1996xz}
D.~Diakonov, M.~Polyakov, P.~Sieber, J.~Schaldach and K.~Goeke,
%``Sphaleron transitions in the minimal standard model and the upper bound for the Higgs mass,''
Phys.\ Rev.\  {\bf D53} (1996) 3366
[hep-ph/9502245].
%%CITATION = HEP-PH 9502245;%%

%\cite{Moore:1996jv}
\bibitem{Moore:1996jv}
G.~D.~Moore,
%``Fermion determinant and the sphaleron bound,''
Phys.\ Rev.\  {\bf D53} (1996) 5906
[hep-ph/9508405].
%%CITATION = HEP-PH 9508405;%%

%\cite{Jarlskog:1989bm}
\bibitem{Jarlskog:1989bm}
C.~Jarlskog,
%``CP Violation,''
{\it Singapore: World Scientific (1989) 723 P. (Advanced Series on Directions in High Energy Physics, 3)}.
	
%\cite{Gunion:1989we}
\bibitem{Gunion:1989we}
J.~F.~Gunion, H.~E.~Haber, G.~L.~Kane and S.~Dawson,
``The Higgs Hunter's Guide,''
SCIPP-89/13;
%\cite{Gunion:1992hs}
%\bibitem{Gunion:1992hs}
J.~F.~Gunion, H.~E.~Haber, G.~L.~Kane and S.~Dawson,
``Errata for the Higgs hunter's guide,''
hep-ph/9302272.
%%CITATION = HEP-PH 9302272;%%
	
%\cite{Ratra:1988dp}
\bibitem{Ratra:1988dp}
B.~Ratra and L.~G.~Yaffe,
%``Spherically Symmetric Classical Solutions In SU(2) Gauge Theory With A Higgs Field,''
Phys.\ Lett.\  {\bf B205} (1988) 57.
%%CITATION = PHLTA,B205,57;%%

%\cite{Carena:2000yi}
\bibitem{Carena:2000yi}
M.~Carena, J.~Ellis, A.~Pilaftsis and C.~E.~Wagner,
%``Renormalization-group-improved effective potential for the MSSM Higgs  sector with explicit CP violation,''
[hep-ph/0003180].
%%CITATION = HEP-PH 0003180;%%

%\cite{Bock:2000gk}
\bibitem{Bock:2000gk}
P.~Bock {\it et al.}  [ALEPH, DELPHI, L3 and OPAL Collaborations],
%``Searches for Higgs bosons: Preliminary combined results using LEP data  collected at energies up to 202-GeV,''
CERN-EP-2000-055.



%\cite{Nolte:1995pz}
\bibitem{Nolte:1995pz}
G.~Nolte and J.~Kunz,
%``Gradient approach to the sphaleron barrier,''
Phys.\ Rev.\  {\bf D51}, 3061 (1995)
[hep-ph/9409445].
%%CITATION = HEP-PH 9409445;%%
%\cite{Brihaye:1994ib}
\bibitem{Brihaye:1994ib}
Y.~Brihaye and J.~Kunz,
%``On axially symmetric solutions in the electroweak theory,''
Phys.\ Rev.\  {\bf D50} (1994) 4175
[hep-ph/9403392].
%%CITATION = HEP-PH 9403392;%%

\end{thebibliography}
\end{document}